%% file: main.tex
\documentclass[final,5p,times,twocolumn]{elsarticle}

\usepackage{lineno,hyperref}
\usepackage{amsmath}
\usepackage{amssymb}
\usepackage{csquotes}
\usepackage{textcomp}
\usepackage{dirtree}
\usepackage{dirtytalk}
\usepackage{xcolor}
\usepackage{caption}
\usepackage{subcaption}
\usepackage[linesnumbered,ruled,vlined]{algorithm2e}

\journal{}

\newcommand*\dd{\mathrm{d}}

\begin{document}
\begin{frontmatter}

\title{Interactive Geometry Modification of High Performance Finite Element Simulations}

\author[mechAddress]{Corey Wetterer-Nelson \corref{mycorrespondingauthor}}
\ead{c.wetterer-nelson@kitware.com}

\author[aeroAddress]{Kenneth E. Jansen}
\ead{ kenneth.jansen@colorado.edu}

\author[aeroAddress]{John A. Evans}
\ead{ john.a.evans@colorado.edu}
\cortext[mycorrespondingauthor]{Corresponding author}

\address[mechAddress]{Kitware, Inc.
1712 Route 9
Suite 300
Clifton Park, New York
12065 USA}
\address[aeroAddress]{Ann and H.J. Smead Department of Aerospace Engineering Sciences. 3775 Discovery Drive
Boulder, CO 80303}

\begin{abstract}
In the context of high performance finite element analysis, the cost of iteratively modifying a computational domain via re-meshing and restarting the analysis becomes time prohibitive as the size of simulations increases. In this paper, we demonstrate a new interactive simulation pipeline targeting high performance finite element simulations where the computational domain is modifiable \textit{in situ}, that is, while the simulation is ongoing. This pipeline is designed to be modular so that it may interface with any existing finite element simulation framework. A server-client architecture is employed to manage simulation mesh data existing on a high performance computing resource while user-prescribed freeform geometric modifications take place on a separate workstation. We employ existing \textit{in situ} visualization techniques to rapidly inform the user of simulation progression, enabling computational steering. By expressing the simulation domain in a reduced fashion on the client application, this pipeline manages highly refined finite element simulation domains on the server while maintaining good performance on the client application.
\end{abstract}

\begin{keyword}
Computational steering, immersive simulation, design space exploration, interactive shape deformation, skeletal rigging, high performance finite element analysis
\end{keyword}

\end{frontmatter}

\section{Introduction}
Modern massively parallel computational resources have revolutionized the ways engineers and scientists can simulate and study complex physical phenomena. The capability of partial differential equation solvers has exploded, enabling the study of detailed three-dimensional systems with evermore complicated setups. Not only are simulations growing in size, but smaller problems are growing in quantity with rapid access to data and near realtime performance enabling massive ensembles of low-fidelity simulations. However, two key difficulties still plague our capability for gaining insight from simulations or using large simulations for engineering design. First, the proliferation of high performance simulation capabilities has outpaced the storage capabilities required to house their massive data output. This leads to a bottleneck for practitioners needing to gain insight from their simulations. Saving all analysis for a post-processing procedure becomes progressively more expensive and time intensive as the size and quantity of simulations increase. Many authors have noted this bottleneck, particularly punctuated by the irreverent Sandia National Laboratories report written by David Thompson et al. which makes a strong case for techniques which mitigate the need to write data to disk \cite{thompson2009design}. The second severe difficulty is the setup and initialization process of large-scale simulations which need to be repeated at each pass of iterative design workflows. Modifying and exploring simulation input parameters and their effects on simulation output is difficult and non-intuitive due to the difficulty required in setting up sophisticated massively parallel simulations. This problem is exacerbated when the geometry of the simulation domain is the variable parameter of interest. Typically, modifying the geometry of a finite-element domain requires expensive re-meshing and re-partitioning. Studying the sensitivity of a system with respect to geometry is tedious at best and prohibitive typically. Hardwick and Clay, in an internal report from Sandia National Laboratories, described the design to analysis pipeline as a 10 step process \cite{boggs2005dart}. Of that process, approximately $72\%$ of the time a practitioner spends working through the process is spent on mesh generation and geometry manipulation. That percentage is inflated by the high iteration probability and \textit{rework fraction}, or the percentage of work that must be redone or iterated upon in order to progress the analysis pipeline. The report rigorously demonstrated that the greatest time requirement of the design to analysis pipeline is redoing work and regenerating data to shuttle through future steps. Geometry modification and re-meshing make up the majority of this rework fraction.

Our work presented in this paper addresses both of these challenges, with particular focus on the latter. To do so, we turn to \textit{computational steering}, or broadly, enabling users to interact with their simulation. We demonstrate a new paradigm of \textit{freeform computational steering}, which brings elements of geometric design into the simulation runtime, empowering practitioners to develop and iteratively manipulate the design of their simulated systems \textit{in situ}, while their simulation is running, circumventing expensive re-meshing and simulation re-initialization.  To realize freeform computational steering, we fuse handle-based polyharmonic surface manipulation with computationally efficient volume mesh deformation schemes, and we introduce a new skeleton-based surface manipulation strategy based on biharmonic deformation of curve-skeletons.  We utilize state of the art co-processing visualization and data extraction systems to combat data bloat and provide a window into the details of the ongoing simulation through which the user can steer the simulation, freely manipulating the domain and watching the simulation respond as it runs. We have implemented this capability in the open-source software suite \texttt{Shoreline} \cite{wetterernelson2021}. Our implementation targets finite element based simulation solvers, especially those which simulate dynamic and unsteady processes with domain modification performed over time or between time steps. Further, our implementation could accommodate existing finite volume solvers with little to no modification. In this article, we demonstrate these capabilities with the high performance computational fluid dynamics (CFD) software \texttt{PHASTA} \cite{whiting2001stabilized}.

Existing computational steering implementations generally rely on exposing scalar parameters to the user which must be set ahead of time. As will be detailed in Section \ref{sec:background}, these parameters must be exposed to the user \textit{a priori} and range from variables affecting solver stability to boundary condition values and only recently have allowed for geometric domain modification. With freeform computational steering, our system takes this further by empowering a user to manipulate the geometry of their simulated system with no pre-defined parameters. That is, the user may freely \textit{discover} a design space while the simulation is ongoing, rather than just explore a previously defined design space. This becomes especially pertinent in the context of biomedical systems such as surgical implants. We explore this in Section \ref{sec:bypass} where straightforward design space parametrizations are difficult or even impossible to define.

The remainder of this paper is organized as follows. In Section \ref{sec:background}, we provide a review of existing computational steering and interactive simulations systems. In Section \ref{sec:steeringOverview}, we present the geometry deformation steering system and describe the software components requisite to its functionality. In Section \ref{sec:srfManipulation}, we detail our approach to user-interactive geometry modification, taking cues from computer animation and discrete differential geometry. In Section \ref{sec:volumeMeshDef}, we provide details on how user-defined geometry modification is propagated across a finite element mesh distributed on multiple partitions. In Section \ref{sec:nozzle}, we demonstrate our interactive computational steering tools with a simulation of flow through a retangular channel which is then modified into flow through a converging/diverging nozzle while the simulation is ongoing. Next, we demonstrate our workflow further with a more complex flow of blood through an arterial stenosis bypass in Section \ref{sec:bypass}. Then, in Section \ref{sec:meshQuality}, we characterize the robustness of mesh quality undergoing freeform deformation. In Section \ref{sec:solCvg}, we study the response of solution convergence to freeform deformation under mesh refinement. In Section \ref{sec:interactiveScaling}, we explore how incorporating our computational steering workflow into an existing simulation workflow may affect the scalability of the workflow. Finally, in Sections \ref{sec:conclusions} and \ref{sec:futureWork}, we provide concluding remarks and discuss future directions for this work.

\input background.tex

\section{Geometry Deformation Steering System} \label{sec:steeringOverview}
In this section, we detail the software components which comprise our freeform computational steering system. Throughout this paper, we consider any geometric manipulation to a simulation domain made by a user to be a geometric deformation from an initial geometry. As such, much of the presented software infrastructure is involved with defining geometric deformations and propagating deformation fields across distributed finite element meshes. As such, we will frequently refer to this collection of software components as the \textit{geometry deformation steering system}.

Our geometry deformation steering system is architected as a server-client system. The server application is responsible for managing the full computational domain and triggering the physical simulation to step in time. The client application runs on a user's workstation and is responsible for handling user interaction with the simulation domain. Communication between the server and the client is performed via a low-level transmission control protocol (TCP) connection directly between the two applications. Visualization and interaction take place in \texttt{ParaView} where we have developed a suite of custom plugins to perform mesh deformation actions from the user's workstation.

The flow of information through the system is illustrated in Figure \ref{fig:Flowchart}. First, the server application loads each part of a pre-partitioned finite element mesh onto processes in the server's message passing interface (MPI) communicator. We employ the \texttt{PUMI} mesh infrastructure \cite{ibanez2016pumi} to handle the mesh data. Then, the server extracts the surface mesh from the volume in parallel and gathers that surface mesh on to process zero. Next, the server waits for communication with a client application. Once a connection is established, the server sends the surface mesh to the client application and begins the finite element simulation. The client application then packages the surface mesh into a Visualization Toolkit (\texttt{VTK}) file \cite{schroeder2004visualization} which is then loaded into \texttt{ParaView}. From \texttt{ParaView}, the user can manipulate and deform the mesh using a stack of custom mesh deformation plugins. Upon completing the deformation, \texttt{ParaView} exports that deformation field as a file which the client application parses and then transmits to the server. Upon receiving the surface deformation field, it is scattered across all processes on the MPI communicator and the server linearly interpolates the surface deformation in order to apply the deformation of the computational volume over a series of steps. It is noted that linear interpolation is problematic in general \cite{von2015real}, but for the well behaved deformations we are targeting, there are no problems with linear interpolation, and linear interpolation is simply unbeatable in terms of computational cost. Now ready with a series of surface deformations, the server can sequentially apply the series of deformations to the simulation domain while running the simulation for a prescribed number of time-steps between deformations. The deformation of the volume mesh takes place using the same MPI processes on the same mesh partition as the simulation. Generally, the goal is to run the unsteady simulation to a statistically stationary state between deformation steps. Upon completing the full series of deformation steps, the server can accept a new deformation order from the client application, thus ensuring the server and client stay synced without transmitting mesh data unnecessarily.

For our demonstration, we use \texttt{PHASTA} \cite{whiting2001stabilized} as a high performance incompressible fluid dynamics simulation system. As \texttt{PHASTA} has been demonstrated to scale to very large computational resources, its capability is representative of the computational scale we wish to achieve.

\subsection{Surface Mesh Extraction}
\begin{figure}[t!]
    \centering
    \begin{subfigure}[b]{\linewidth}
         \centering
          \includegraphics[width=\linewidth]{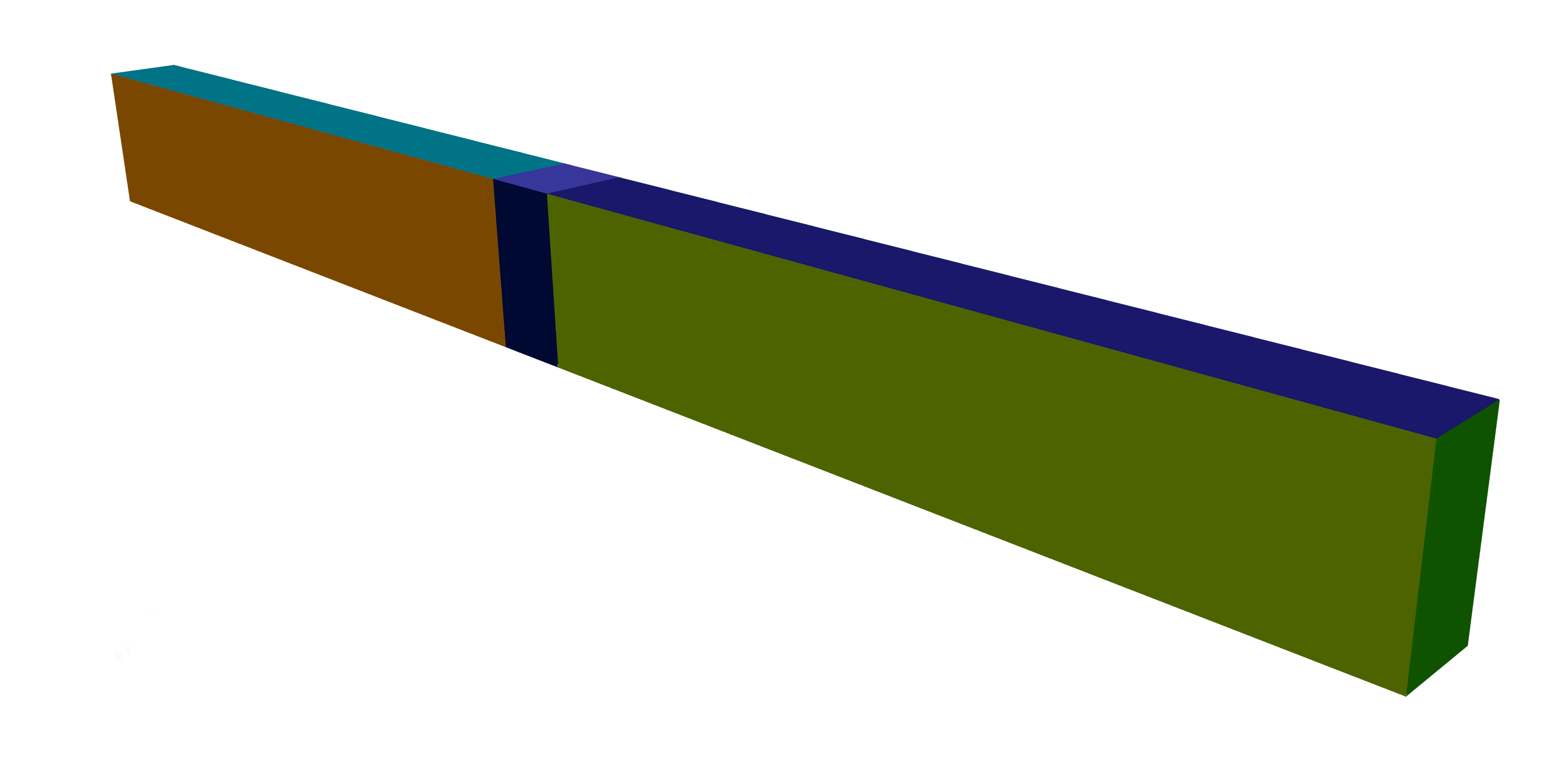}
          \caption{The channel geometry is colored by geometric feature to provide a user with visual indication as to how the geometry can be deformed. Note the region in the middle of separate geometric features. }
          \label{fig:featureColors}
    \end{subfigure}
    
    \begin{subfigure}[b]{\linewidth}
        \centering
        \includegraphics[width=\linewidth]{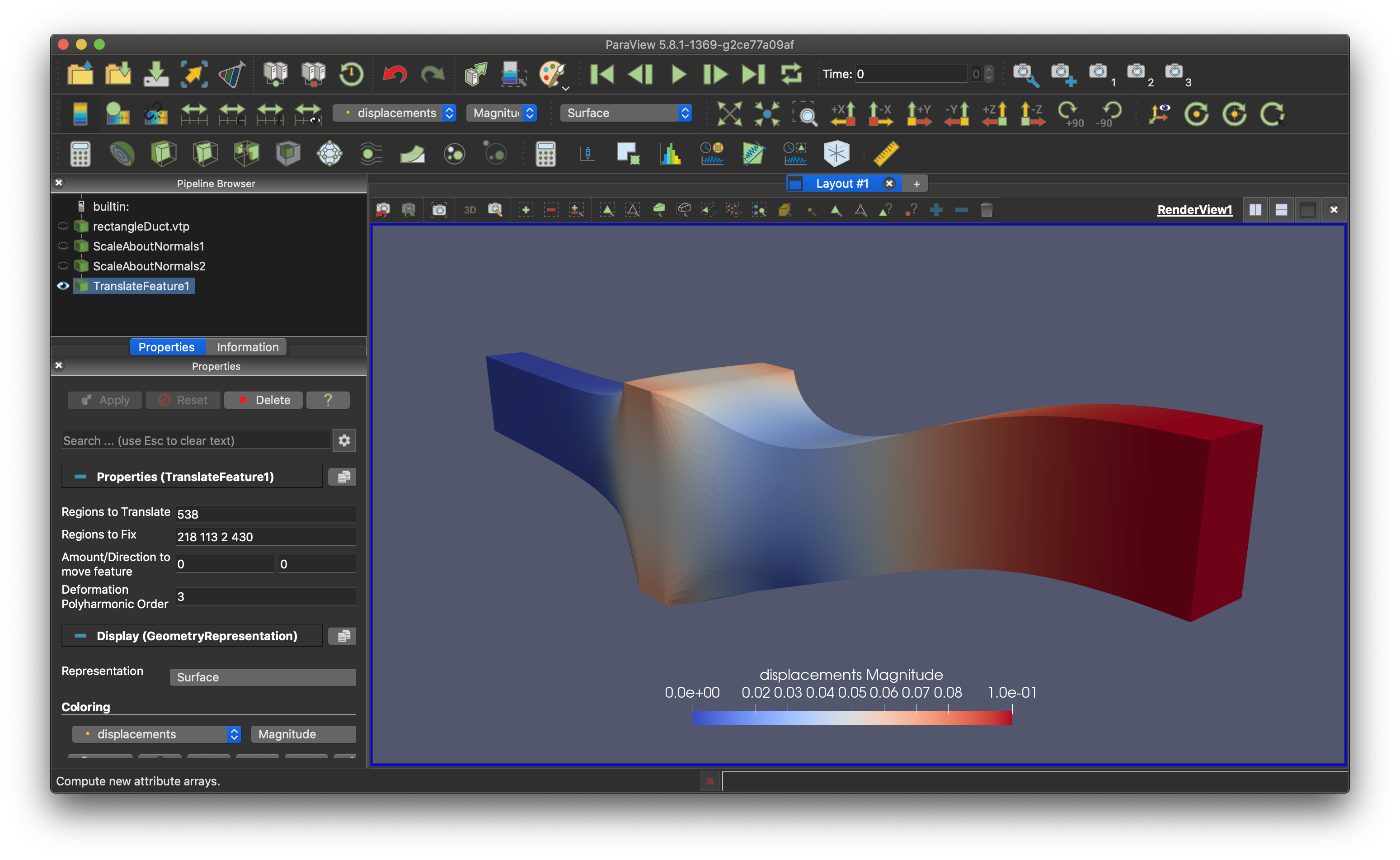}
        \caption{Here we see an (admittedly pathological) example of using our suite of mesh modification actions within \texttt{ParaView}. The Scale by Normals action is applied twice to widen the central region of the channel, then a Translate Feature action is applied with a triharmonic deformation field to move the end of the channel upward.}
        \label{fig:pvGUI}
    \end{subfigure}
    
    \caption{A simple channel geometry undergoes significant geometry modification.}
    \label{}
\end{figure}

After loading the parallel partitioned mesh data, the server application extracts a surface mesh from the volume. The \texttt{PUMI} mesh infrastructure makes this a straightforward task of looping over every element face and adding faces to the surface mesh if they lie on the boundary of the volume mesh. We employ standard C++ library data structures for the surface mesh representation as simple data structures ease the process of serializing the data and passing it between processes on the server and over the TCP connection between the server and the client applications. During this surface extraction process, elements in the surface mesh are tagged by their corresponding geometric feature. These tags are generally dictated by features of the computer aided design (CAD) model progenitor of the mesh. These feature tags will be used by the interactive surface mesh deformation system to represent geometric handles which the user may interact with. In Figure \ref{fig:featureColors}, we see a sparse set of individual features called out, despite the surface being made up of 33,530 elements.

\subsection{Surface Mesh Deformation}
The tagged surface mesh sent to the user's Client device is exported to a form readable by \texttt{ParaView}. From here, we provide a suite of mesh manipulation actions in the form of filter plugins. Currently, the suite is comprised of three direct surface manipulation actions: Scale by Normals, Scale by Direction and Translate Feature. These actions allow a user to select a collection of features tagged on their geometry and scale these features about their surface-normals, scale them in the direction of a vector, or translate them by a specified vector respectively. In Subsection \ref{subsec:SurfHandles}, we provide details on the geometry deformation algorithms used in these plugins. The suite also contains a system for automatic skeletal rigging and freeform manipulation via the Create Mean Curvature Flow Skeleton action and the Interactive Skeleton Widget which are detailed in Subsection \ref{sec:skelManip}. These plugins are stackable, akin to layers in programs like Adobe \texttt{Photoshop}, allowing users to creatively combine actions to produce complex geometric modifications as shown in Figure \ref{fig:pvGUI}.

\begin{figure}[t!]
  \centering
  \includegraphics[width=\linewidth]{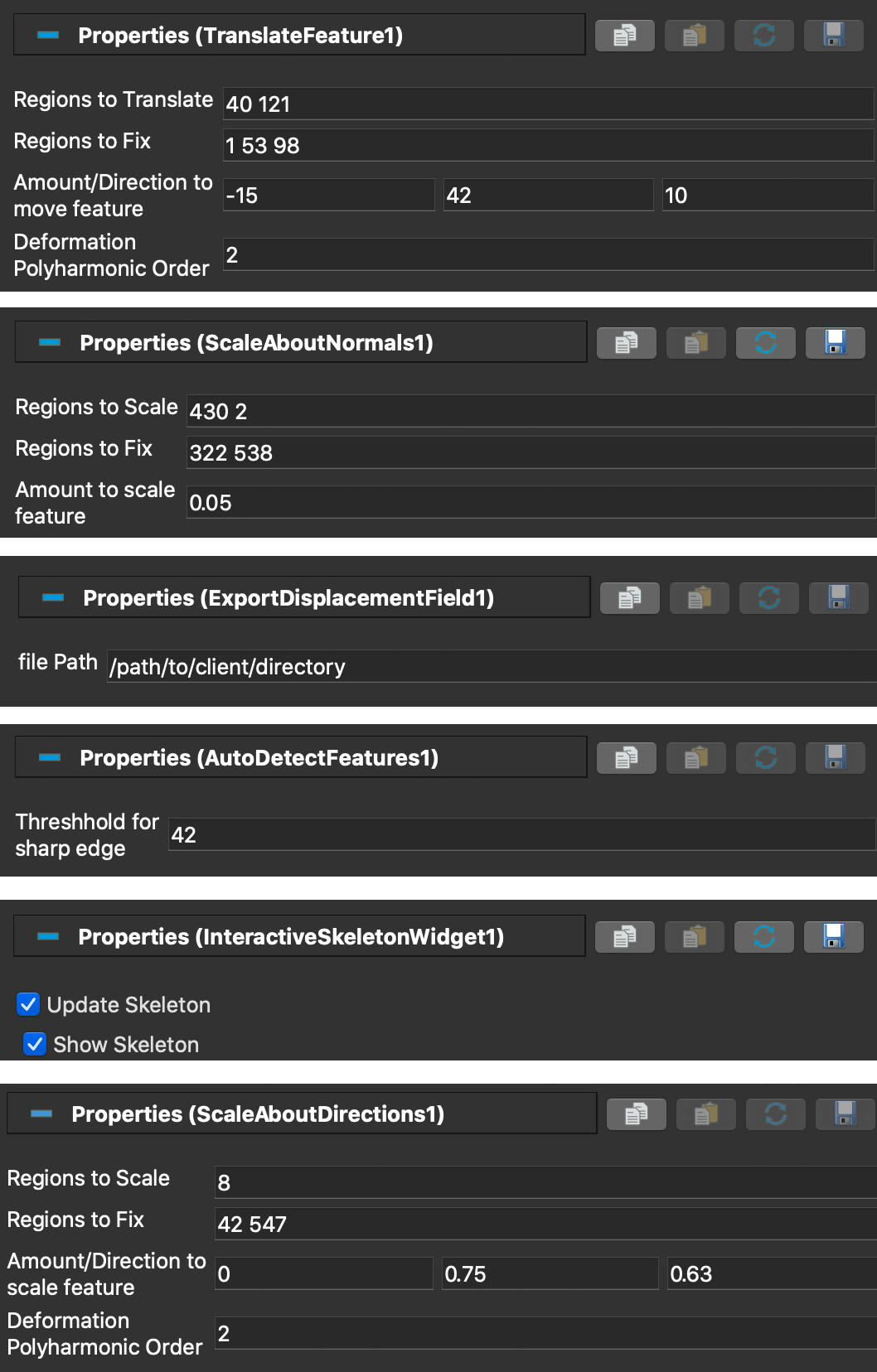}
  \caption{Several GUI interfaces are presented here. Some of the actions expose no controls to the GUI such as the Create Mean Curvature Flow Skeleton action. The GUI interface for the Scale by Direction action is the same as the Scale by Normals action, though requesting three scalars for manual directional scaling, rather than one.}
  \label{fig:GUIExamples}
\end{figure}

Further, the suite provides utility plugins. The Auto Detect Features action tags a surface mesh with features via hard-edge segmentation when no CAD features are available. The Export Displacement Field action manages exporting of user-defined modifications to their geometry. Each plugin (with the exception of the Create Mean Curvature Flow Skeleton action which exposes no user-editable parameters) exposes custom graphical user interface (GUI) elements to the \texttt{ParaView} properties panel. These controls are shown in Figure \ref{fig:GUIExamples} for each of the above mentioned plugins.

\subsection{Volume Mesh Deformation}

When the Server is notified that a user-generated surface geometry deformation is available, a volume deformation system scatters that surface displacement field across the MPI processes where a volumetric deformation algorithm interprets this field as boundary conditions for deforming the simulation mesh. The server application performs this volumetric deformation in between simulation time steps \textit{in situ}, that is, utilizing the same processes and mesh data as the simulation. Currently, two linear volume deformation algorithms are available in the framework. These are a linear elasticity solver with Jacobian-stiffening and a method based on the concept of harmonic maps. Details of these solvers are provided in Section \ref{sec:volumeMeshDef}. 

An instantaneous change in geometry is a non-physical process. In Section \ref{sec:solCvg}, we demonstrate that a severe pressure spike may occur immediately following even a modest geometry deformation. As such, the deformation may be split into a sequence of intermediate deformations which can be spread out over multiple simulation time steps, easing the effect of the deformation on the simulation solution. We refer to this procedure as \textit{deformation scheduling} and demonstrate its use in Section \ref{sec:nozzle}.

\subsection{Communication Protocol}
Communicating data between the server and the client is a critical component to ensure smooth synchronous operation of the interaction loop and the simulation loop. To serve this workflow, we developed a network layer application programming interface (API) built on low-level TCP networking protocols. This API is constructed for communication between a single server and a single client application, and developed specifically for Unix based operating systems. For reference, this system targeted a client application running on macOS with the server application running on Linux.

We utilize C++ standard library vector data structures to package data for transfer between client and server applications. This was chosen to ensure portability of the API, minimizing external dependencies. Also, via templating, this API can transmit any set of serialized data stored in a standard vector data structures.

In order to reduce the complexity of the client application communication management, the communication API is designed to send and receive data from a single process. As such, the distributed server application must gather all data set for transmission to the client on to a single process. Then that single process can communicate data to and from the client. This gather operation must only occur once at the beginning of the workflow, as the surface mesh topology will not change throughout the lifetime of the simulation. The communicator process will need to scatter data back to the rest of the MPI communicator received from the client any time the client sends a message. As this process will likely happen infrequently compared to the total computation happening on the server, this adds little internal overhead to the server application. Currently, the communicator process on the server is the zeroth process in the global communicator used by the simulation. This system will scale well to fairly large problems as message data size is only on the order of the size of the surface mesh, but for extreme sized problems, data reduction techniques such as surface mesh decimation and simplification \cite{lindstrom1998fast, lindstrom1999evaluation, garland1997surface} may need to be employed. In future iterations of this system, we would like to separate the server's communication process from the simulation processes as a separate MPI communicator so that the server may operate in a truly asynchronous fashion.

\subsection{Visualization System}
\texttt{ParaView} was chosen for the surface mesh deformation system precisely for its robust visualization infrastructure. \texttt{ParaView} already has a tight link with \texttt{PHASTA} as its preferred visualization software for high performance CFD computations, and we leverage that relationship in this work. As our system time-evolves the \texttt{PHASTA} flow simulation, visualization data is written out at a user-specified cadence. This can then be readily loaded into \texttt{ParaView} for visualization as the simulation is ongoing. However, with the goal of targeting significantly larger simulations, the pairing of \texttt{PHASTA} and \texttt{ParaView Catalyst} has already demonstrated \textit{in situ} visualizations of simulations on millions of MPI processes \cite{rasquin2014scalable} where visualization data is streamed to the user in a compressed fashion, immediately reducing the data bloat associated with large-scale simulations. In our \texttt{ParaView Catalyst} workflow, the user may link their workstation's instance of \texttt{ParaView} to the \texttt{Catalyst} system so that visualization of the simulation occurs in the same application as user interaction. We deem this streamlined workflow an immersive computational steering system as the user can visualize their simulation and directly manipulate its domain from a graphical user interface on their workstation.

\section{Surface Mesh Manipulation in \texttt{ParaView}} \label{sec:srfManipulation}
Upon receiving the surface mesh, the client application immediately passes the mesh onward to \texttt{ParaView}. In \texttt{ParaView}, we leverage the custom plugin system to provide a suite of geometry modifications which can be stacked to build up a desired mesh deformation. Two paradigms have been implemented for user manipulation of surface meshes. First, a series of plugins utilizing a polyharmonic deformation field algorithm based on Laplacian surface editing \cite{sorkine2004laplacian} allows a user to move individual surface features rigidly, while smoothly deforming the free portions of the surface. We rely on the open source \texttt{libIGL} library for the implementation of this algorithm within our plugins \cite{libigl}. Second, a wildly freeform method of surface editing inspired by graphics animation schemes presents the user with a polyline skeleton of their surface mesh which can be freely manipulated in a smooth fashion.

\subsection{Surface Handle Manipulation}\label{subsec:SurfHandles}
In the context of computer graphics and animation, geometry is typically represented by its boundary, generally parametrized by a triangulated surface mesh. In order for an artist to manipulate said geometry, the goal is generally posed to allow the artist to move vertices or handles on the mesh, and then propagate that prescribed deformation smoothly across the rest of the mesh. Early work revolved around defining a biharmonic energy functional over the mesh, endowing the propagated deformation with appealing, smooth (approximate) $C^1$ continuity of the deformation field at the artist-manipulated handle \cite{botsch2004intuitive, sorkine2004laplacian, jacobson2010mixed, botsch2007linear}.

\begin{figure}[!t]
  \centering
  \includegraphics[width=0.6\linewidth]{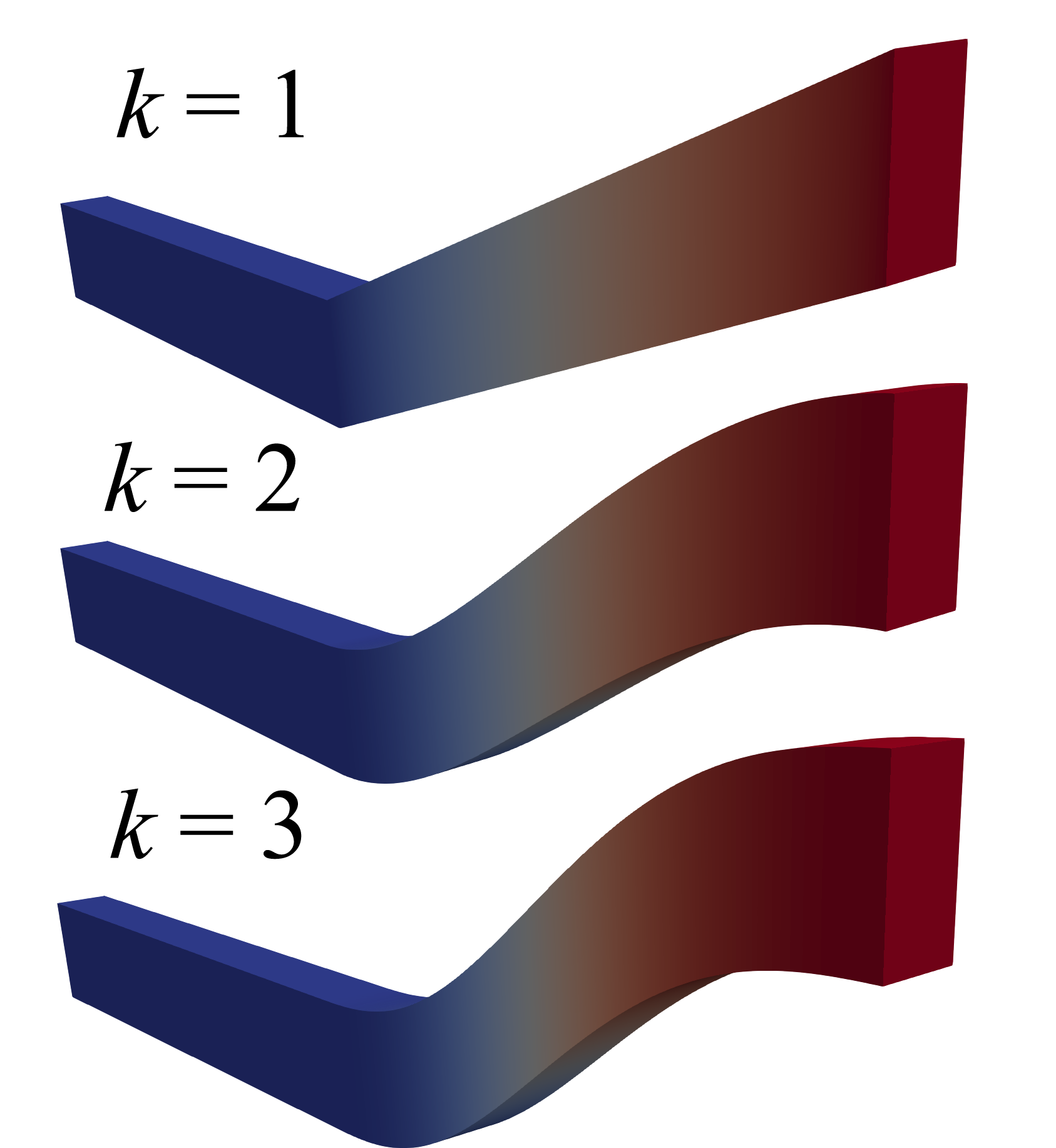}
  \caption{The outlet face of the channel geometry from Figure \ref{fig:featureColors} is moved upward, and the propagated deformation follows solutions to the $k$-Laplacian equation of various $k$. The three modified channels are colored by displacement magnitude.}
  \label{fig:harmonicDefs}
\end{figure}

Formally, a surface mesh $\mathcal{M}^S$ can be defined as a collection of vertices $V$ and simplices $F$ such that $\mathcal{M}^S = \{V, F\}$. We typically require that $\mathcal{M}^S$ is a closed manifold of codimension-1.  The basic formulation of these techniques involves defining an energy over the surface mesh in terms of a displacement field $\mathbf{d}$, starting with the Laplacian energy
\begin{equation}
  E(\mathbf{d}) = \int_{\mathcal{M}^S}\left\lVert (\Delta \mathbf{d}) \right\rVert^2 \dd\Omega.
\end{equation}
Here, $\Delta$ refers to the Laplace-Beltrami operator, a Laplacian with derivatives taken along the surface. Minimizing this energy functional with applied boundary conditions results in a smooth deformation field over the surface, and is equivalent to solving the biharmonic equation with respect to the displacement field
\begin{equation}
  \Delta^2 \mathbf{d} = \mathbf{0}.
\end{equation}
This concept is extendable to variable order on the Laplacian operator. Often referred to as the $k$-Laplacian or polyharmonic equation, this operator grants control over the smoothness of the resultant deformation
\begin{equation}
  \Delta^k \mathbf{d} = \mathbf{0}.
\end{equation}
Figure \ref{fig:harmonicDefs} depicts just such smoothness control in the deformation of the simple channel geometry from Figure \ref{fig:featureColors}. For $k=1$, the deformation field is minimal, and thus leads in the depicted case to a linear deformation. For $k=2$, the deformation field is biharmonic, and thus the tangency at the fixed regions is controlled, here set to a zero tangency constraint. Finally, for $k=3$, the deformation field is triharmonic, leading to control of the curvature at the fixed regions as well as tangency. In the depicted case, tangency and curvature are set to zero at the fixed regions, leading to an extremely smooth transition from the fixed region to the manipulated region. Smoothness of the deformation is a desirable feature for intuitive geometry manipulation, but moreover, manipulating $k$ varies the requisite boundary conditions across the fixed regions. Increasing $k$ requires that more and more derivatives of the deformation field be set at the boundaries, which provides increasingly more powerful control over the types of deformations that are available to the user.

\begin{figure} [!t]
  \centering
  \includegraphics[width=0.5\linewidth]{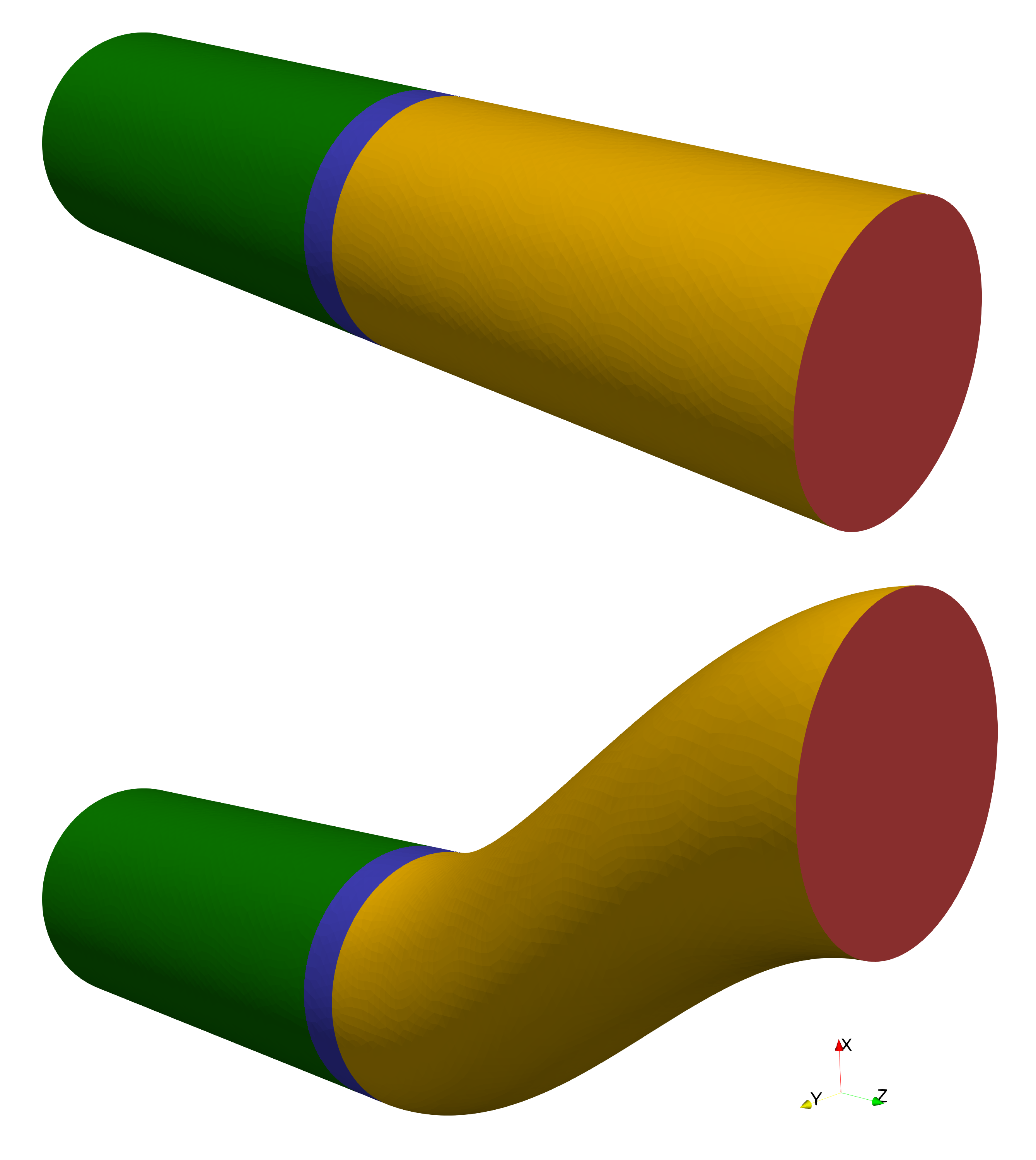}
  \caption{Here, the outlet face of a pipe was translated using our surface modification system in \texttt{ParaView}.}
  \label{fig:cylindertr}
\end{figure}

We use a polyharmonic deformation field with user-selectable harmonic order to provide control over constraints at the boundary of fixed and moved features. Each plugin works by allowing the user to select moveable handle features and fixed features. Features on a mesh are specified by a scalar field in the \texttt{VTK} data labeled \enquote*{features}. These features are either assigned from the initial CAD description of the geometry used to generate the mesh as described previously or can be automatically assigned via feature detection. Our suite includes a \texttt{ParaView} plugin which implements the \texttt{CGAL} shape detection system \cite{cgal:ovja-pssd-20a} for automatically assigning features based on sharp edges in the mesh.

\begin{figure}
  \centering
  \includegraphics[width=\linewidth]{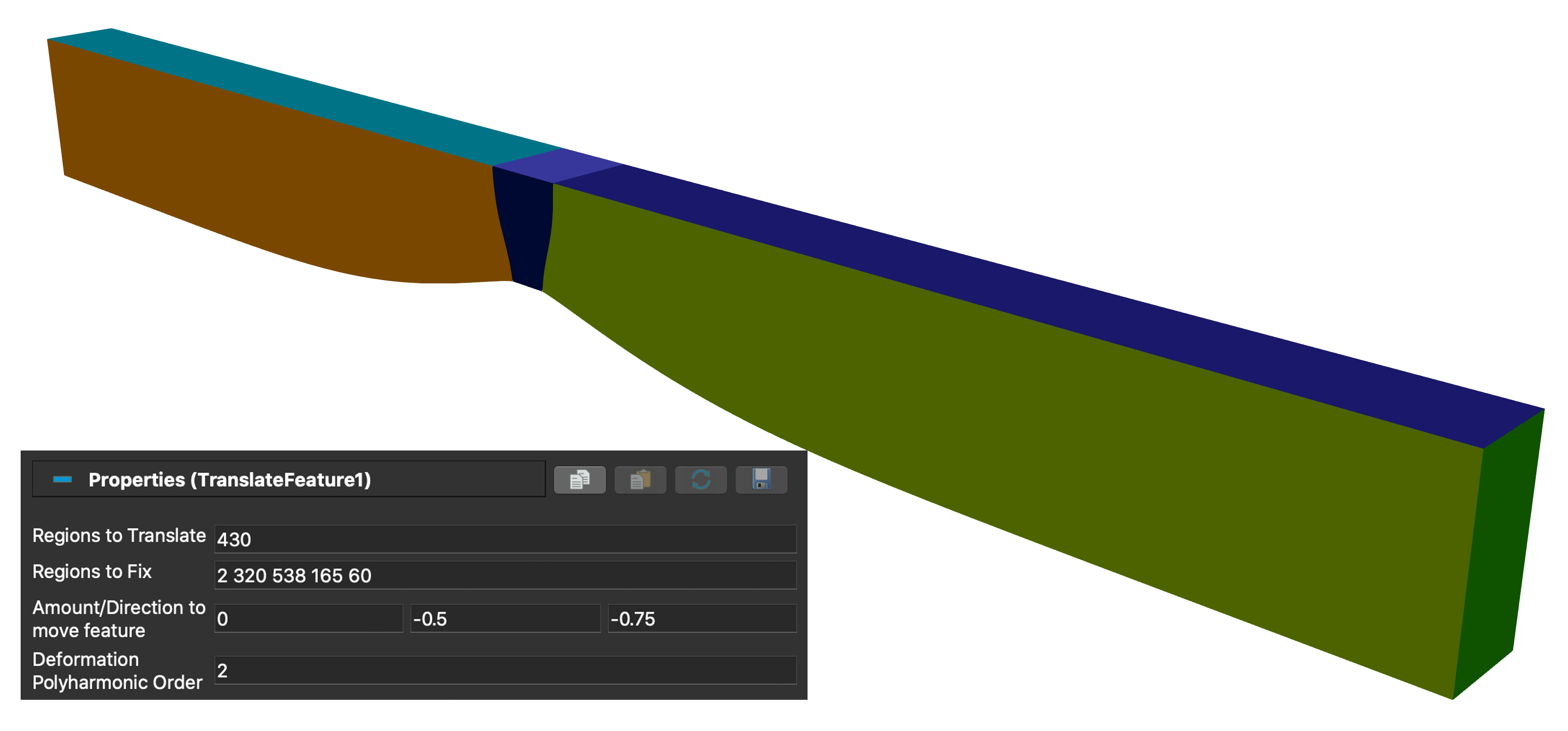}
  \caption{Here, we have scaled a feature in the $y$ and $z$ directions to create a bump within the channel. Also pictured is the user interface for the scaling plugin.}
  \label{fig:scaleBump}
\end{figure}

Currently, we have three direct surface mesh deformation actions implemented in this \texttt{ParaView} plugin suite. First, the Translate Features allows the user to perform arbitrary translation of a collection of features. Figure \ref{fig:cylindertr} depicts the end of a pipe translated upward with the back half of the pipe held fixed, demonstrating how this tool can be used to explore routing of pipes and ducts or more organic geometries such as segments of cardiovascular networks. Second, the Scale by Direction action enables scaling of a collection of features in the direction of a user-specified vector. Figure \ref{fig:scaleBump} depicts a scale deformation operating on a single feature on the lower face of the channel. In this example, the top faces of the channel and both ends are held fixed, and the biharmonic deformation field is computed over the unprescribed faces. Third, the Scale about Normals plugin enables the scaling of a feature about its surface-normal vector. This plugin is used in Section \ref{sec:interactiveScaling} to increase the diameter of an arterial bypass which, owing to its intrinsically organic shape, has no single convenient vector about which to scale the feature.

The deformation field computed by a plugin's operation is passed through the plugin so that operations may be stacked as illustrated in Figure \ref{fig:pvGUI}. Upon finishing the desired deformation procedure, the user can activate an export plugin to pass the computed displacement field from \texttt{ParaView} back to the client application via writing to a plain-text file. The advantage of this method is that the prescribed displacement field is saved and can be loaded for subsequent simulations without needing to repeat interactions with \texttt{ParaView}.

\subsection{Skeleton-Based Surface Manipulation} \label{sec:skelManip}
\begin{figure}[t!]
  \centering
  \includegraphics[width=\linewidth]{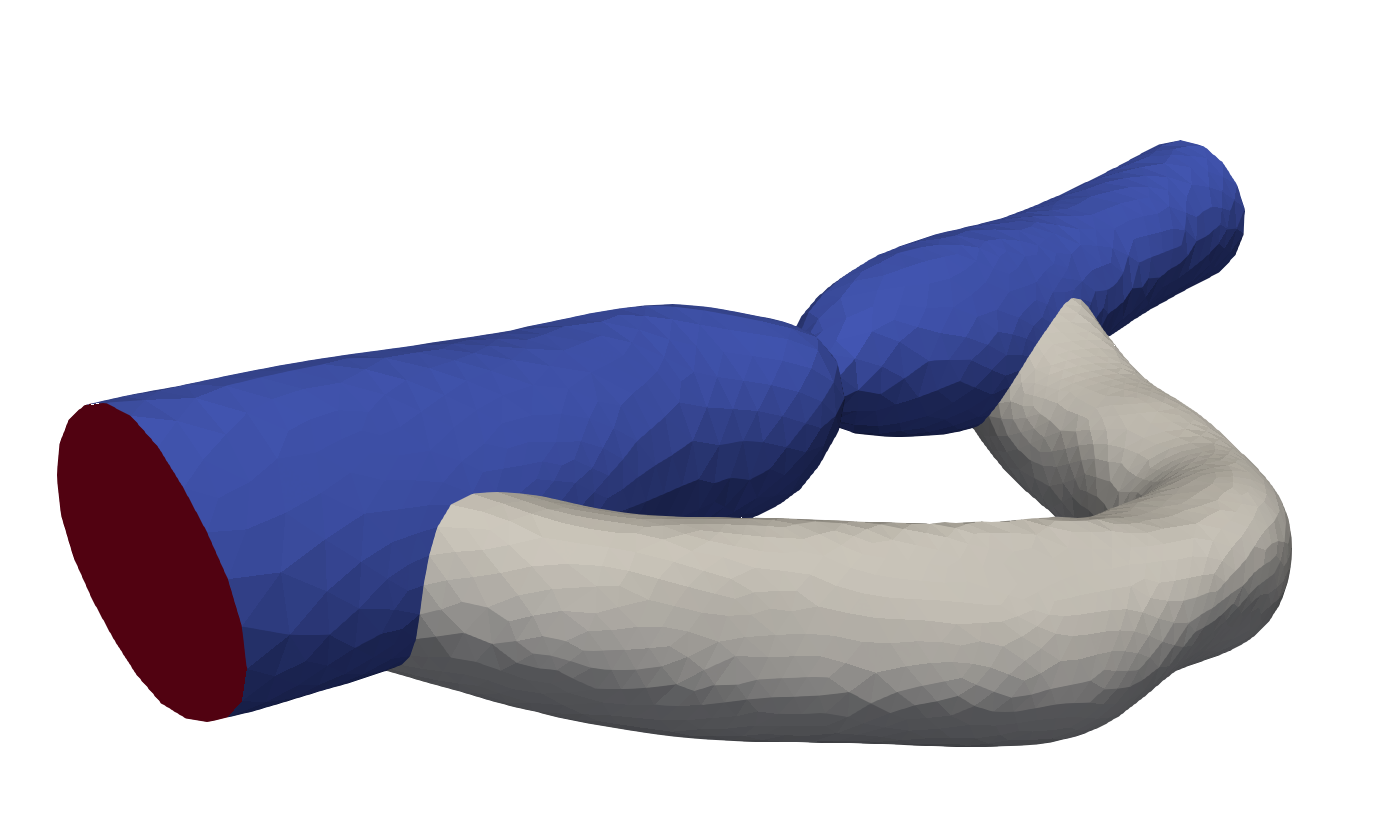}
  \caption{The surface mesh of the bypass only contains four features: the inlet face (red), the outlet (occluded), the artery boundary (blue) and the bypass boundary (white) for which to define handles.}
  \label{fig:bypassFeatures}
\end{figure}

In the previous subsection, we described a system whereby the user interacts with the surface mesh directly. That is, the handles are defined by features of the surface mesh, and the space of realizable manipulations is restricted to manipulations where these features undergo affine transformations. There is a critical drawback to that system. Specifically, while defining manipulation handles based on surface features may provide an excellent interface for manipulating bulk shapes of a simulation domain, it is often the case that a surface mesh is quite barren of features for which to define interactive handles. Often, a user may desire to manipulate the shape of a single feature. Take, for example, the arterial bypass geometry as shown in Figure \ref{fig:bypassFeatures}. The geometry inherited only has four features during the mesh generation process for which to define handles. Further, in the case of this arterial bypass, the interesting geometric considerations primarily revolve around the shape of the bypass. As such, we require a tool which allows a user to manipulate individual features in a freeform and intuitive manner that does not rely on CAD-inherited feature data or sharp edge segmentation. In this section, we present a novel tool for just such user interaction inspired by tools for posing and animating used in the computer graphics and animation field. Our strategy involves first automatically generating a \textit{skeleton} of the surface mesh represented by a collection of polylines, then enabling smooth interactive manipulation of the skeleton, followed by casting that skeletal deformation back onto the surface mesh.

\subsubsection{Automatic Skeleton Generation} \label{sec:skelGen}
In order to actualize our intent to manipulate surface features in a freeform and intuitive manner, we desire to generate a \textit{rig} or control system which simplifies the user's interface to manipulating the mesh, while maintaining as much control over the shape of the surface as possible. This is to say, we do not seek to require the user to manipulate every single vertex in the surface mesh individually by hand, but rather provide intuitive handles that the user may manipulate that in turn manipulate bulk shapes of the surface mesh in an intuitive fashion. For our aims, we have developed a system for skeletal manipulation of surface meshes.

In typical skeleton rigging workflows, a user is tasked with manually designing a skeleton which fits inside their geometry. That is, they must define every bone and joint, placing them within their surface mesh geometry for intuitive manipulation of the geometry later in the pipeline. This process is labor intensive at best, and generally demands a non-zero level of artistic experience of the user.

Automatic skeleton generation systems seek to reduce this burden placed on the user by programmatically generating a skeleton rig within the geometry. However, replacing artistic experience with an algorithm is no easy feat, and many of the proposed algorithms for automatic skeleton generation are complex and each are typically only successful in a small subset of geometric situations. For instance, the heavily cited Pinocchio automatic rigging system \cite{baran2007automatic} is an example of a skeleton generation algorithm which begins with producing an approximation of the input surface's medial axis and then trimming down to a usable skeleton rig. This algorithm works fantastically well for the cartoon character model geometries on which this algorithm was first demonstrated. However, when we implemented this algorithm, we found it does not generalize to even simple genus-1 surfaces such as the arterial bypass geometry in Figure \ref{fig:bypassFeatures}. Next, a class of automatic rigging systems rely on the presence of a volume mesh or discretization such as a volumetric Voronoi diagram or voxelization \cite{bloomenthal1999skeletal, wang2008curve}. These methods rely on the resolution of their internal volumetric discretization, and their generated skeletons are subject to artifacts caused by lack of volumetric resolution. Further for our purposes, employing such algorithms would be a step backward, adding a volume discretization to our surface mesh representation which is already associated with a volume mesh, though one that is engaged in parallel simulation computations on another compute resource. As our goal is to provide a light-weight, performant solution to mesh manipulation, we seek to avoid re-discretizing the volume of the surface mesh. Finally, we turn to curve-skeletonization. These algorithms typically attempt to represent a medial axis with some level of resolution rather than minimizing the number of joints and bones as is typical in skeleton rigs for animation. This choice does lead to the challenge of manipulating a large quantity of skeleton joints simultaneously which we address with a new curve manipulation algorithm presented in Subsubsection \ref{sec:skelInteractivity}.

Automatic generation of curve-skeletons takes many forms dependent on the desired characteristics of the skeleton and the description of the base geometry. Broadly, these generation techniques can be classified as either thinning methods, distance field methods or geometric methods \cite{saha2016survey}. Thinning methods work directly on the given surface mesh by reducing a representation of the surface geometry in the direction opposite the surface's normal vector. This can be done in a number of ways, such as is done elegantly through mean curvature flow  \cite{wang2008curve, au2008skeleton, zhou2020curve}. Field based methods vary widely, though are broadly categorized to defining an ambient field in terms of the surface geometry where that the field minima approximates the medial axis of the geometry. Examples of such techniques include employing geodesic distance fields \cite{dey2006defining}, or repulsive force-fields \cite{cornea2005computing, pantuwong2010skeleton}. Geometric methods involve prescribing other geometric data on which to compute a medial surface or curved-skeleton. For instance, Voronoi diagrams of the volume can be employed to compute distance from the surface to find the medial axis on which to produce a skeleton \cite{brandt1992continuous, attali2001delaunay, yan2018voxel}. These techniques compute the Voronoi diagram of the boundary vertices and then take the intersection of that with the polygonal shape of the geometry. This intersection is wildly sensitive to the polygonal shape, and as such, an approximation of that shape can be employed. However this still requires a fairly well resolved approximation, leading still to skeletons which are sensitive to geometric perturbations and exhibit frequent spurious skeletal branches \cite{saha2016survey}.

In order to generate skeletons in our geometry deformation steering system, we employ a mean curvature flow skeletonization algorithm developed by Tagliasacchi et al. \cite{tagliasacchi2012mean}. This algorithm is robust to arbitrary genus geometry and does not rely on a tetrahedral mesh describing the volume of the geometry. We found that this algorithm produces acceptable skeletons for our application.

\begin{figure}[t!]
  \centering
  \includegraphics[width=\linewidth]{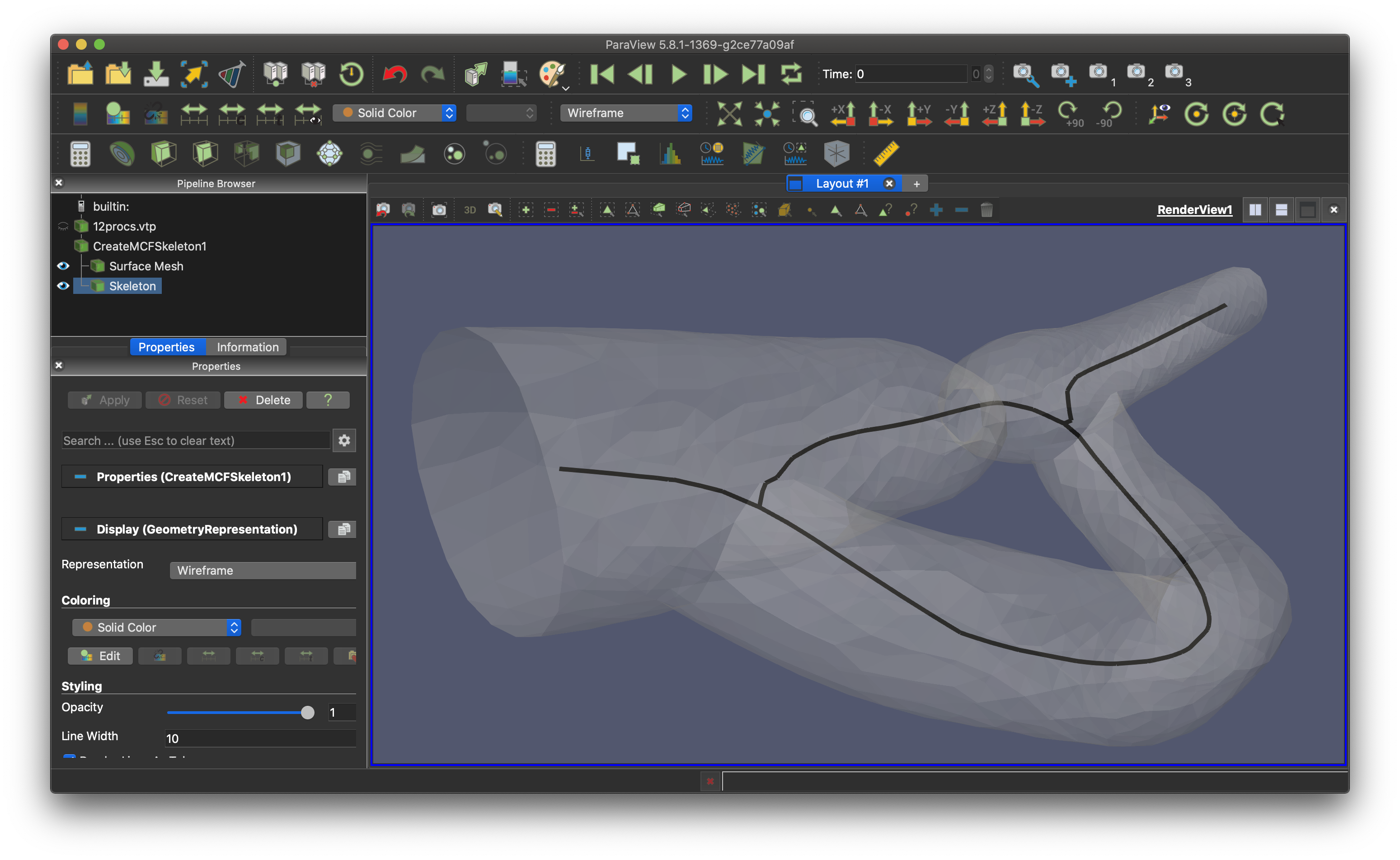}
  \caption{The implementation of the skeleton generation occurs in a \texttt{ParaView} plugin which, when activated, automatically generates a skeleton of the input geometry.}
  \label{fig:bypassSkeletonScreenshot}
\end{figure}

We implemented the mean curvature flow skeletonization algorithm in a \texttt{ParaView} filter plugin, enabling a user to automatically generate a skeleton with which to perform remarkably freeform manipulation of their simulation geometry. Figure \ref{fig:bypassSkeletonScreenshot} illustrates such a mean curvature skeleton computed for the arterial bypass geometry generated by our plugin in \texttt{ParaView}. In the next subsubsection, we describe the details of how such interaction is made possible in the visualization software \texttt{ParaView}, employing a novel curve-skeleton manipulation technique.

\subsubsection{Skeleton Interactivity} \label{sec:skelInteractivity}
Armed with a curve-skeleton plotted through the medial axis of the surface mesh, the next challenge is enabling manipulation of the skeleton. In typical skeleton based animation and geometry manipulation, the skeleton is extremely sparse, meaning allowing the user to manipulate every joint in the skeleton is a feasible workflow. However, with the curve-skeletons defined above, the user is presented with hundreds of joints and bones, making it completely infeasible to expect a user to manipulate each joint or bone individually. This challenge has been addressed previously by defining rigid bones by chains of line segments in a curve-skeleton with user-defined joints \cite{barbieri2016interactive}. This method however lacks the flexibility required for truly freeform design manipulation.

Thus, a new formulation was devised enabling a user to drag and deform a curve-skeleton as if it were a collection of metal wires. Our algorithm shares intent with as-rigid-as-possible deformation algorithms \cite{lan2017medial, yang2018dmat, lan2020medial}, though is more directly inspired by biharmonic surface deformation formulations discussed above. Here, we define a biharmonic displacement field over a polyline curve.

To begin, we must define a Laplace operator $\underline{\underline{Lc}}$ over a curve. This is done via a central difference approximation about each joint $\mathbf{x}_i$ on the interior of each curve in the skeleton as
\begin{equation}
  Lc_{ij} = \begin{cases}
    j = i \pm 1 & 1/\|\mathbf{x}_i - \mathbf{x}_j\|,\\
    j \notin [i-1,i+1] & 0,\\
    i = j & -\sum_{k \neq i} Lc_{ik}.
\end{cases}
\end{equation}
For our purposes, we explicitly treat any point with greater than or less than two neighbors as boundary points. This choice leads to intuitive user interaction where modifying a given curve within the skeleton only modifies that selected curve, pinning all other curves potentially branching about the skeleton.

We define the mass matrix $\underline{\underline{Mc}}$ with the mass of a vertex as the sum of the half-distance to each of its neighboring points. For points on the interior of a curve, this is
\begin{equation}
  Mc_{ii} = \left(\|\mathbf{x}_i - \mathbf{x}_{i+1}\| + \|\mathbf{x}_i - \mathbf{x}_{i-1}\|\right) / 2.
\end{equation}
For points on the boundary, the mass is explicitly set to one to facilitate boundary condition enforcement.

We then combine these to form our biharmonic operator for a curve $\underline{\underline{Bc}}$, inspired by the construction of the surface biharmonic operator in Section \ref{subsec:SurfHandles}:
\begin{equation}
  \underline{\underline{Bc}} = \underline{\underline{Lc Mc}}^{-1} \underline{\underline{Lc}}.
\end{equation}

Rows of $\underline{\underline{Bc}}$ corresponding to boundary nodes are zeroed out and a 1 is placed on the diagonal element of that row. We solve for the biharmonic deformation $\underline{d}$ field via
\begin{equation}
  \underline{\underline{Bc}} \underline{d} = \underline{0}.
\end{equation}
This completes our formulation for the curve-skeleton manipulation. Formation and assembly of the matrix system, as well as the system solve, are all performed during user interaction at interactive rates using the \texttt{Eigen} linear algebra library \cite{eigenweb}.

\begin{figure}
     \centering
     \begin{subfigure}[b]{\linewidth}
        \centering
        \includegraphics[width=\linewidth]{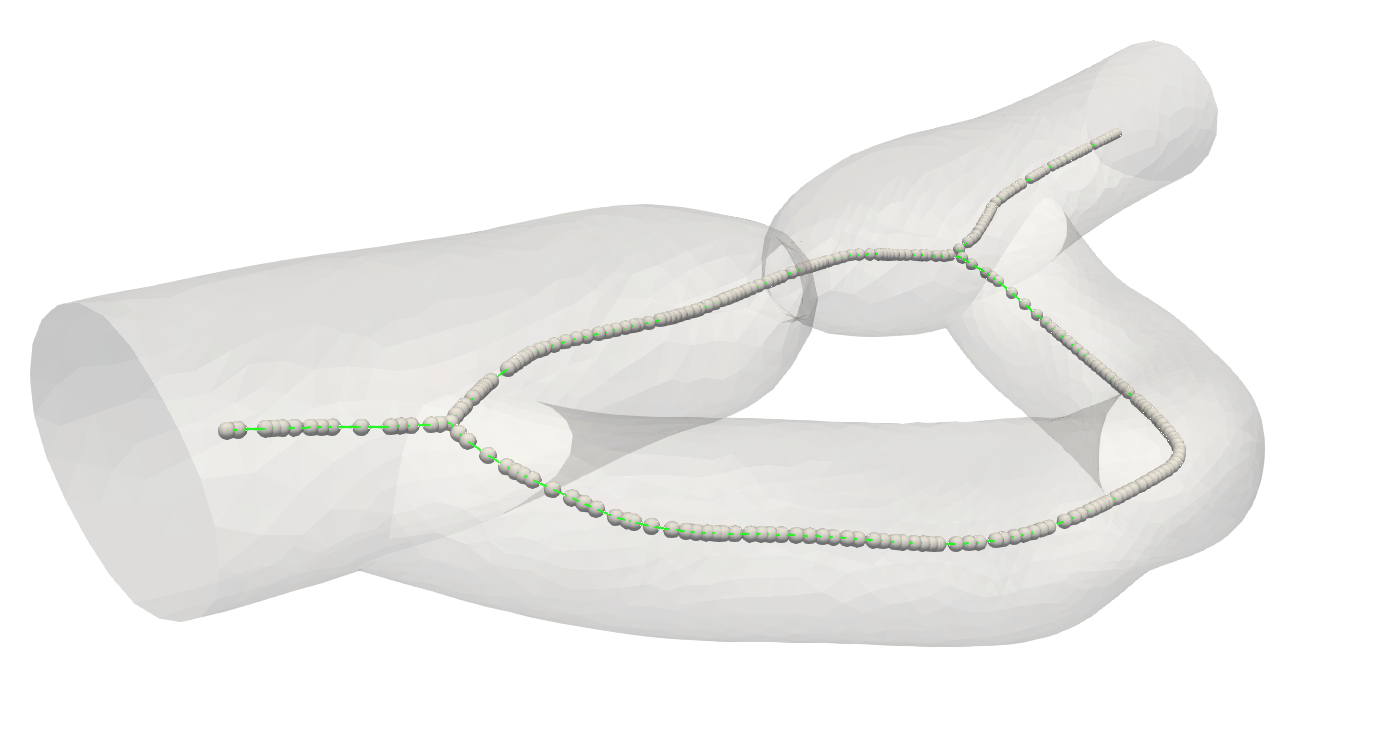}
        \caption{Once the skeleton is generated, it is passed along with the surface mesh to another plugin where the skeleton becomes an interactive widget which the user can freely manipulate by simply dragging around joints with their cursor.}
        \label{fig:bypassSkeletonWidget}
     \end{subfigure}
     
    \begin{subfigure}[b]{\linewidth}
        \centering
        \includegraphics[width=\linewidth]{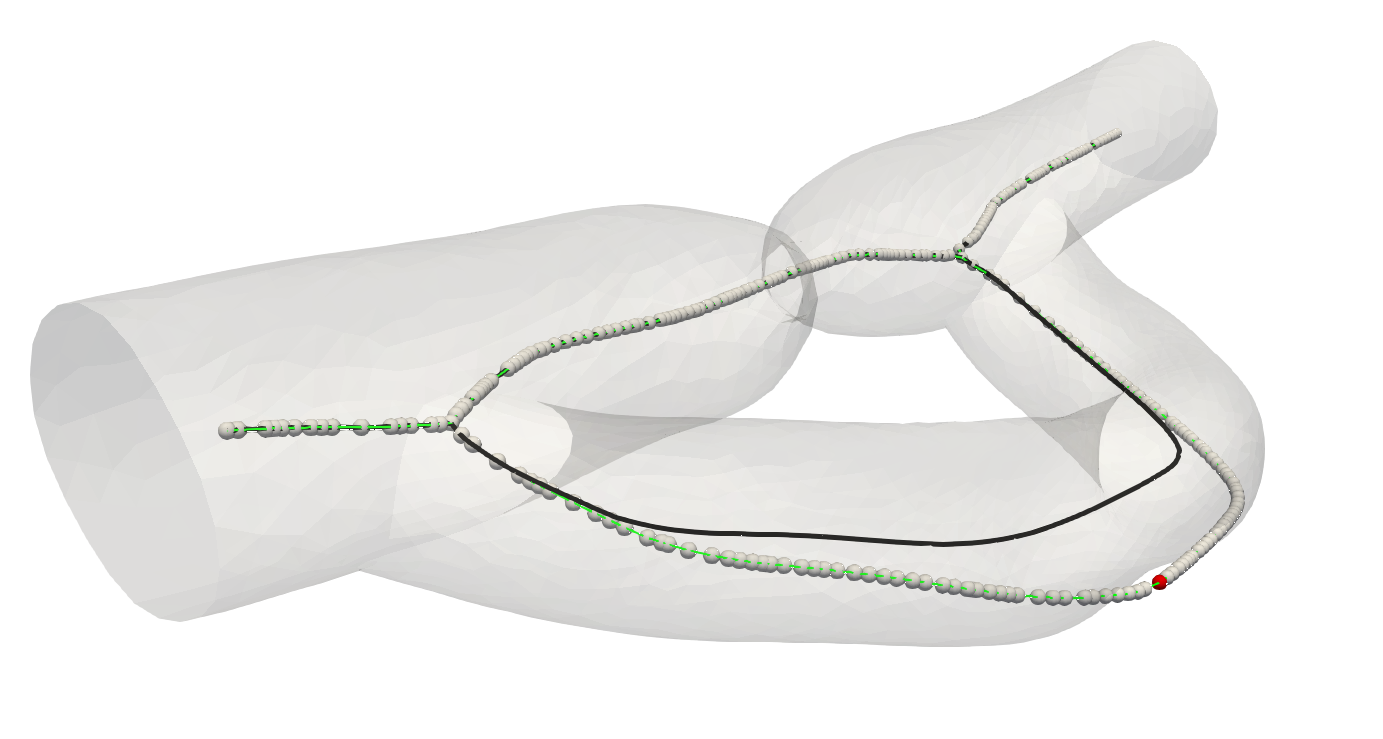}
        \caption{When the user pulls on a handle (as was done here, the handle marked in red was manipulated), the entire skeleton responds by bending according to a biharmonic deformation field. Boundary conditions are set such that only the bypass section of the skeleton moves, leaving the main artery channel unmoved.}
        \label{fig:skeletonWidgetMoved}
    \end{subfigure}
     
    \begin{subfigure}[b]{\linewidth}
        \centering
        \includegraphics[width=\linewidth]{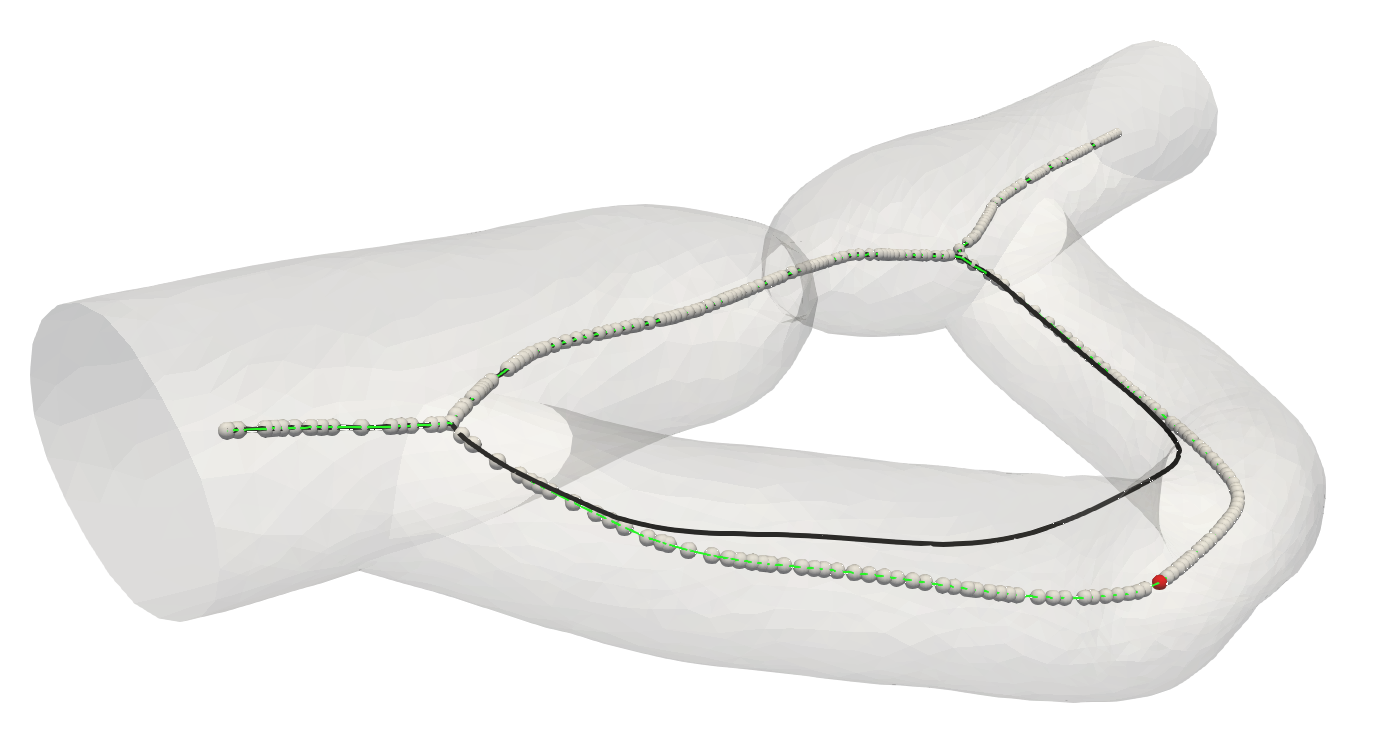}
        \caption{Upon completing manipulation of the skeleton, the user can apply the manipulation to the surface mesh. The black poly-curve here still represents the original skeleton of the undeformed surface mesh.}
        \label{fig:bypassDeformed}
    \end{subfigure}
    \caption{Progression of user interaction with the interactive skeleton widget.}
    \label{fig:three graphs}
\end{figure}

To actualize curve-skeleton manipulation, we designed an interactive widget which exposed the skeleton to the user as an intuitive interactive tool. \texttt{ParaView}  provides a small collection of interactive widgets which are exposed to several visualization filters to provide interactive reconfiguration of those filters. With that basic infrastructure, we developed a custom interactive skeleton \texttt{ParaView} widget which ingests a previously generated curve-skeleton and its associated surface mesh and enables direct, real-time interactivity within the \texttt{ParaView} viewport. Figure \ref{fig:bypassSkeletonWidget} showcases the final product, a collection of spheres are rendered, representing joints, connected by green line segments representing bones.

The deformation algorithm is run every time the user issues a \texttt{MovePoint} command which occurs when the user clicks on a joint in the skeleton and drags the joint. In Figure \ref{fig:skeletonWidgetMoved}, we see the effect of dragging the point marked in red outward away from the initial configuration of the skeleton drawn in black. Many joints in the skeleton are also dragged in an intuitive manner which deforms the entire segment of the skeleton much like a bendable metal wire. This user interaction takes place directly in the \texttt{ParaView} viewport shown in Figure \ref{fig:bypassSkeletonScreenshot}.

The user may continue manipulating any joint in the skeleton, freely deforming the skeleton. However, this deformation is not propagated to the surface until the user issues the \texttt{Apply} command. This propagation of skeleton modification to the surface mesh is referred to as \textit{skinning}. Though many complex skinning algorithms exist, for our purposes, a lowest order skinning algorithm is sufficient. Our strategy is to associate every vertex in the surface mesh with a joint in the skeleton. When a joint in the skeleton is translated, the collection of vertices on the surface mesh associated with that joint are translated by the same amount. This skinning approach performs acceptably for small modifications, thanks in part to the high density of joints along the skeleton. However, this approach cannot handle rotations of features. In order to enable rotations, each vertex in the surface mesh must be associated with a bone, or collection of bones in the skeleton, and a transformation matrix must be computed for each bone upon modification of the skeleton, which can then be applied to the associated vertices in the surface mesh. Figure \ref{fig:bypassDeformed} illustrates the bypass surface geometry properly adjusted to match the new configuration of the skeleton using our skinning algorithm.

\section{Volume Mesh Deformation} \label{sec:volumeMeshDef}
A vibrant ecosystem of volume mesh deformation techniques pervades the literature. Typically, these techniques fall into one or more of several camps: physically inspired methods based on material deformation \cite{froehle2015nonlinear, tezduyar1992computation, stein2003mesh, stein2004automatic}, and more exotic methods based on intrinsic discrete differential properties of the mesh \cite{wang2003volumetric, helenbrook2003mesh}, or based on interpolation schemes \cite{de2007mesh, luke2012fast}. Nonlinear deformation methods such as \cite{froehle2015nonlinear} provide typically superior deformation quality, especially compared to linear elasicity methods and interpolation methods, though at a significant computational overhead for the requisite nonlinear equation solve.

For large deformations, it is often advantageous to adapt the mesh via local refinement or coarsening along with an expensive interpolation procedure which maps solution fields from the old mesh to the adapted mesh \cite{alauzet2016decade, baker2002mesh, yano2012optimization}. However, for large meshes, this becomes a significant computational cost, and as such, we avoid this process in this work.

\begin{figure}
  \centering
  \includegraphics[width=0.75\linewidth]{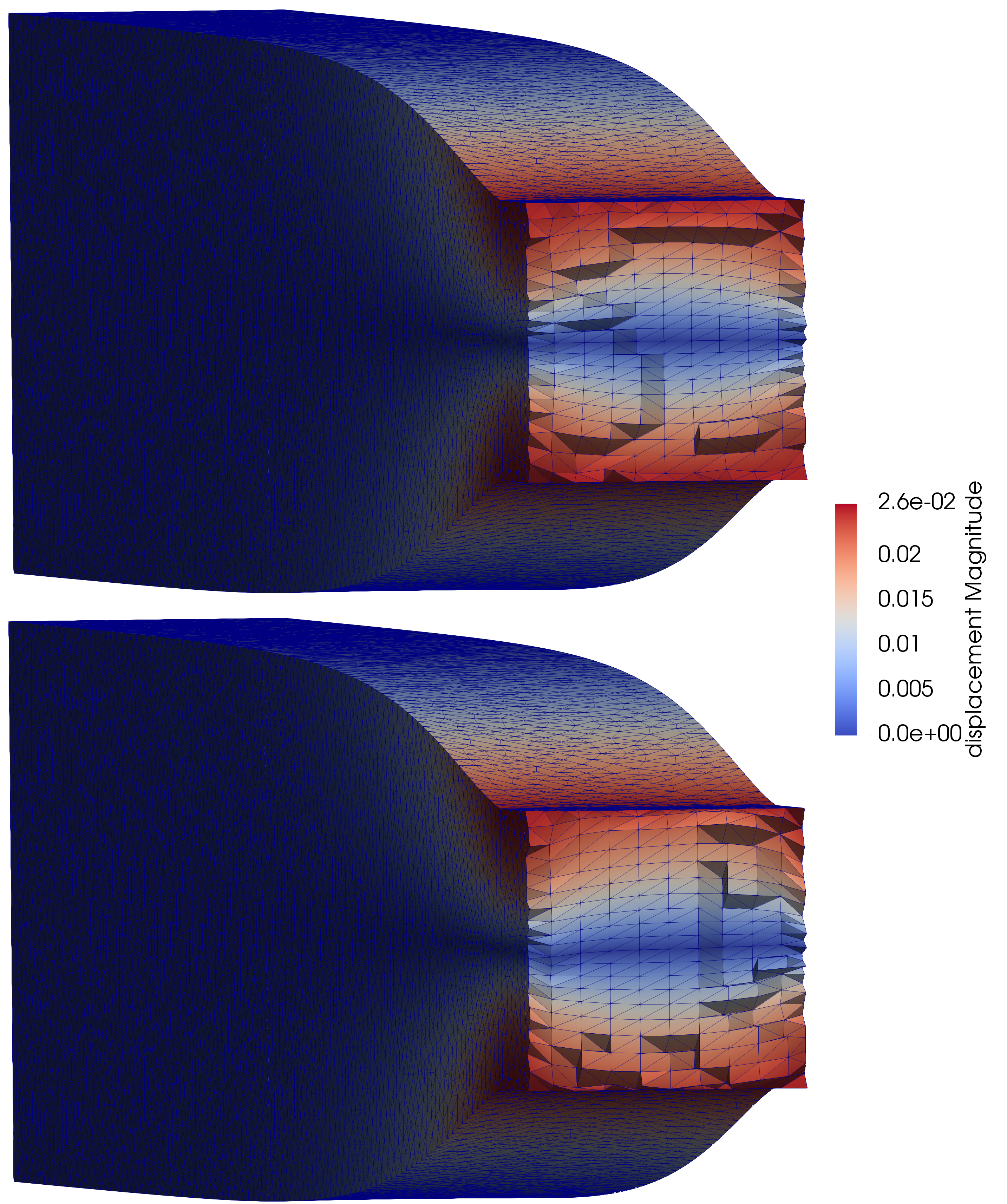}
  \caption{(Top) The channel is deformed using the Jacobian scaled linear elasticity technique. Here, elements near the boundary are less compressed, though elements toward the interior are slightly more compressed. (Bottom) The channel is deformed using the harmonic map technique. Note the slightly more isotropic deformation through the body of the domain. }
  \label{fig:defCompare}
\end{figure}

In our freeform deformation framework, we implemented two linear methods for volume deformation. These are a linear elasticity method with Jacobian based stiffening and a method based on discrete harmonic maps. A comparison of the two methods is given below. These methods were selected primarily for their low computational cost. We desire this workflow to be comparatively computationally inexpensive with respect to the targeted fluid dynamics simulation in order to be unobtrusive to the user. In this workflow, upon receiving the displacement field of the surface, the server resource computes the deformation of the fluid domain volume using the surface displacement field as a boundary condition. In each case, we employ the high performance linear algebra library \texttt{PETSc} to solve the resulting matrix systems \cite{petsc-user-ref, petsc-efficient}. For both volume deformation strategies, the deformation is computed in parallel, utilizing the same mesh partitioning used by the CFD solver.

First, we implemented a linear elasticity based deformation scheme with Jacobian based stiffening \cite{tezduyar1992computation, stein2003mesh, stein2004automatic} where element stiffness is controlled by the Jacobian determinant of the element. This has the effect of making smaller elements stiffer and thus reducing the warping of small elements. This generally leads to better deformation of boundary layer meshes, where very small elements with large aspect ratios are used to resolve near-wall flow structures. 

Second, we implemented a method of generating harmonic maps between a mesh and a deformed mesh. This method bares its origin in volumetric feature identification and registration \cite{wang2003volumetric}. To formulate this method, we follow the derivation of \cite{crane2019n} to produce a \textit{mimetic} definition of the Laplacian operator over tetrahedralized volumes. Then, we directly apply the procedure for Laplacian surface editing \cite{sorkine2004laplacian} to the volume. This results in decoupled displacements in each cardinal direction which are computationally inexpensive to produce. 

For comparison of these two volume deformation techniques, we compressed the inner section of a rectangular channel geometry shown in Figure \ref{fig:flowSetup} using the Jacobian scaled linear elasticity technique and again using the harmonic map technique. In Section \ref{sec:nozzle}, we will demonstrate how this deformation transforms a simple internal flow through a channel into flow through a converging/diverging nozzle. The channel mesh consists of 256,961 tetrahedral elements, partitioned into four parts. Figure \ref{fig:defCompare} shows a cross-section of the deformed mesh at a location 0.05 m from the center of the channel. Both techniques produce acceptable deformation of this mesh, though the boundary layer elements are substantially less compressed under the linear elasticity deformation. Typically, the harmonic map technique will be computationally cheaper for simple deformation operations which deform the mesh in one cardinal direction, as the solver will not need to compute deformation in the directions with no deformation. The linear elasticity solver is designed to be robust for deformations of anisotropic meshes, and thus will out-perform the harmonic map technique when deforming meshes with high aspect ratio boundary layer meshes.

\section{Demonstration: Flow in a Constricting Nozzle} \label{sec:nozzle}
For the first demonstration of our computational steering workflow, we investigate how a simple incompressible internal flow through a duct evolves after a user pinches the central region of the duct. We expect this change of geometry to produce a nozzle which exhibits flow separation downstream of the pinched region. We also use this demonstration to explore the impact of different deformation schedules and investigate whether or not a one step deformation schedule is appropriate. For this demonstration, the simulation and server application operated on a small compute node running the Debian 3.16 operating system with 40 total cores over dual socket Intel Xeon E5-2650 CPUs with 256 GB available RAM. The client application and \texttt{ParaView} operated on a 2018 Apple MacBook Pro running macOS 10.15 with a quad-core Intel i5 CPU and 8 GB of RAM.


\subsection{Simulation Setup}
\begin{figure}[b]
  \centering
  \includegraphics[width=\linewidth]{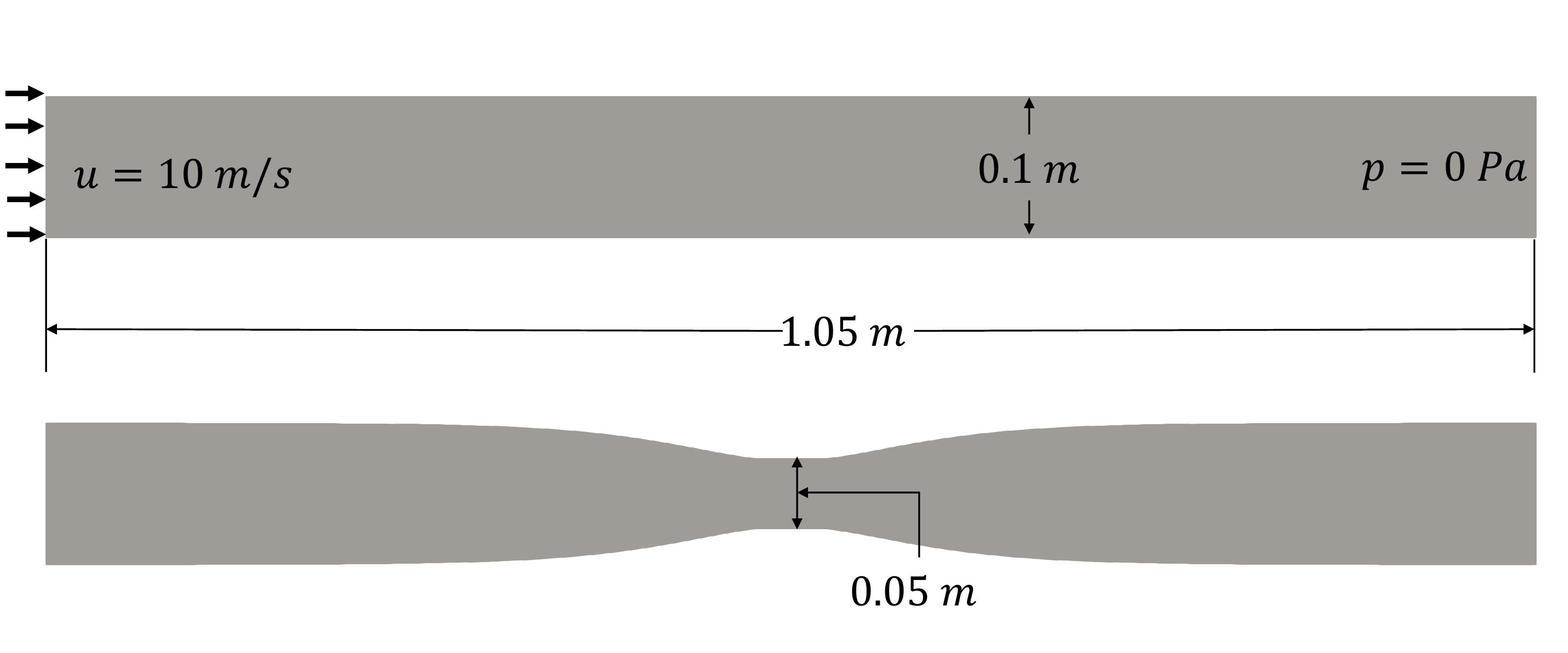}
  \caption{The initial fluid domain is a three dimensional channel $1.05 m$ long in the $y$ direction, $0.1 m$ thick in the $z$ direction, and $0.05 m$ thick in the $x$ direction. The targeted deformation narrows the center region of the domain to $0.05 m$ in the z direction, creating a nozzle which converges upstream and diverges downstream. }
  \label{fig:flowSetup}
\end{figure}
The initial flow configuration for this demonstration is illustrated in Figure \ref{fig:flowSetup}. The domain is setup to be a rectilinear channel $\Omega = \{x,y,z \: |\: x\in [-0.025,0.025] m, y\in[-0.525,0.525] m, z\in[-0.05,0.05] m\}$. Incompressible flow is driven by an inlet velocity at the $y=-0.525 m$  face of $u = 10 m/s$  and an outflow condition is set at the $y=0.525 m$  with outflow pressure $p_o = 0 Pa$. All other faces are set as no-slip, no-passthrough walls. This flow is driven to steady state before any deformation is applied. The domain is meshed isotropically into 256,961 linear tetrahedral elements. Boundary layer meshing was not employed as the near wall behavior of the fluid flow was not of interest, thus the extra resolution would have been an excessive computational expense. A fixed time step size was chosen of $0.001 s$. An initial condition of $u = 0$ was applied.

\subsection{Steering the Computation}
Once presented with the surface mesh on the client application, the throat region of the channel is reduced to 50\% of its thickness. We rely on the biharmonic surface deformation solver to produce smooth transitions from the wide to narrow region of the channel. This deformation field is then immediately passed back to the server. We investigate three deformation schedules:
\begin{enumerate}
  \item 1-step instantaneous deformation at the fluid simulation time $t=0.0 s$,
  \item 1-step instantaneous deformation at the fluid simulation time $t=0.1 s$,
  \item 10-step deformation at the fluid simulation time $t=0.1 s$ with $0.01 s$ simulated between steps.
\end{enumerate}

\begin{figure}[t!]
  \centering
  \includegraphics[width=\linewidth]{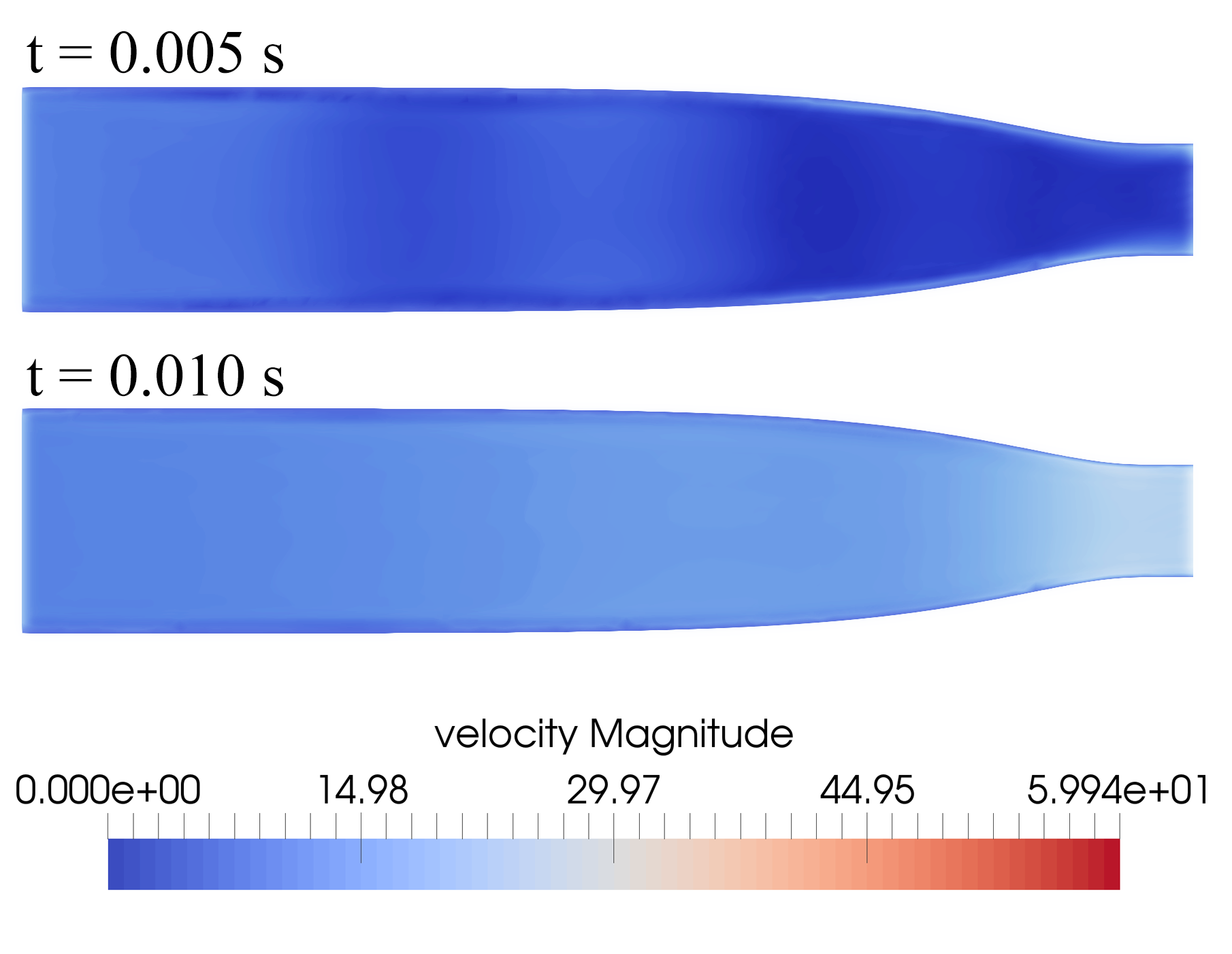}
  \caption{In deformation schedule 1, after initializing the simulation with a zero-flow initial condition and the nozzle deformation fully applied, we see an unphysical velocity oscillation in the region upstream. However, this unphysical behavior quickly dissipates.}
  \label{fig:UnphysicalVSPhysical}
\end{figure}

We selected the Jacobian scaled linear elasticity volume deformation strategy for this demonstration, as we found that the near-wall region of the nozzle was more desirably deformed by this strategy as shown in Figure \ref{fig:defCompare}.

\subsection{Results}
\begin{figure}[t!]
  \centering
  \includegraphics[width=\linewidth]{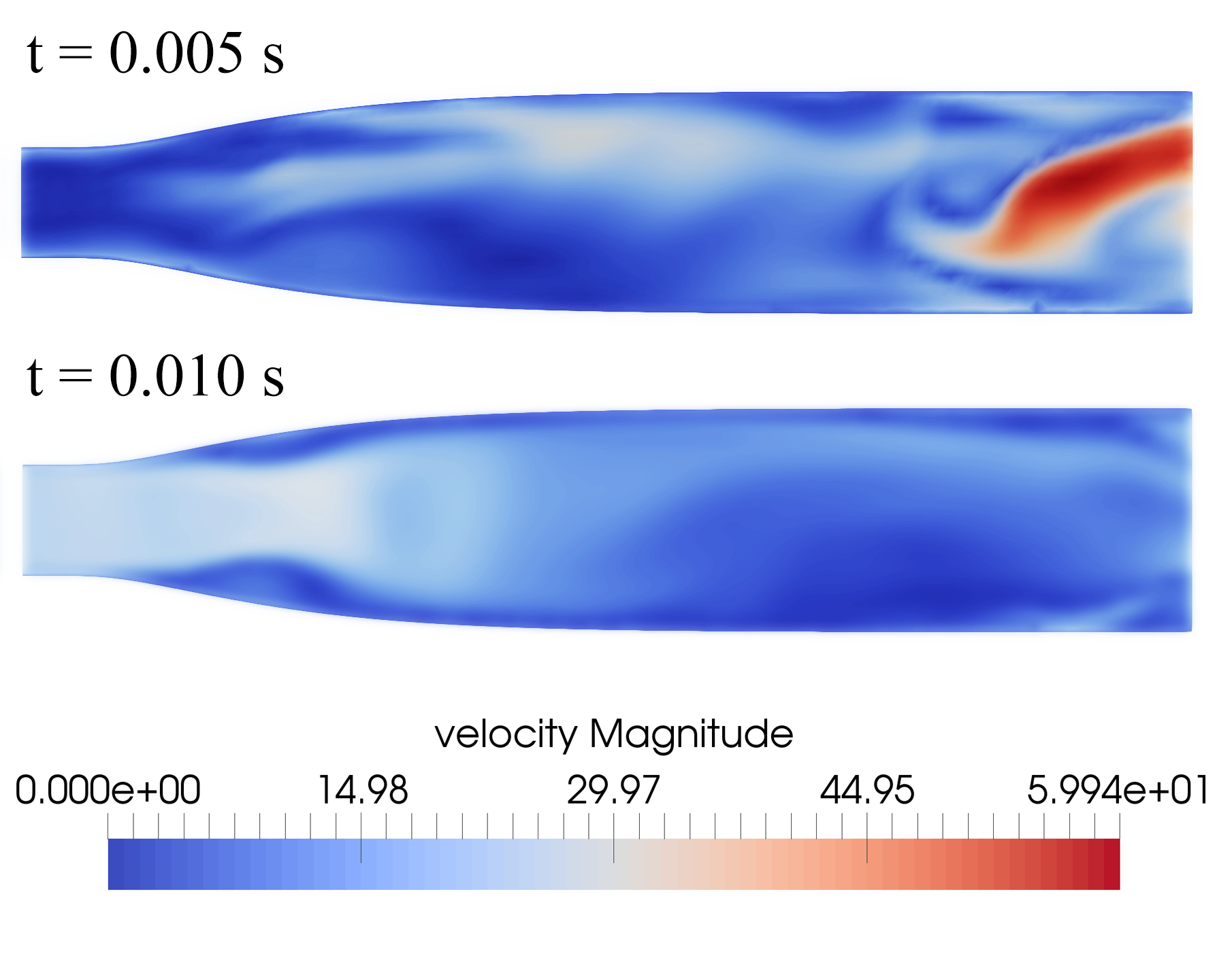}
  \caption{These instantaneous flow profiles are taken from deformation schedule one.}
  \label{fig:UnsteadyJet}
\end{figure}

In Figures \ref{fig:UnphysicalVSPhysical} and \ref{fig:UnsteadyJet}, the velocity magnitudes upstream and downstream of the throat are displayed after the nozzle deformation is applied for schedule one.  An unphysical velocity oscillation appears upstream of the throat after the deformation is applied but disappears within $0.01 s$. A very strong jet of flow is observed downstream of the throat following the deformation, and strong swirling and mixing remain even after $0.01 s$.

\begin{figure}[!t]
  \centering
  \includegraphics[width=\linewidth]{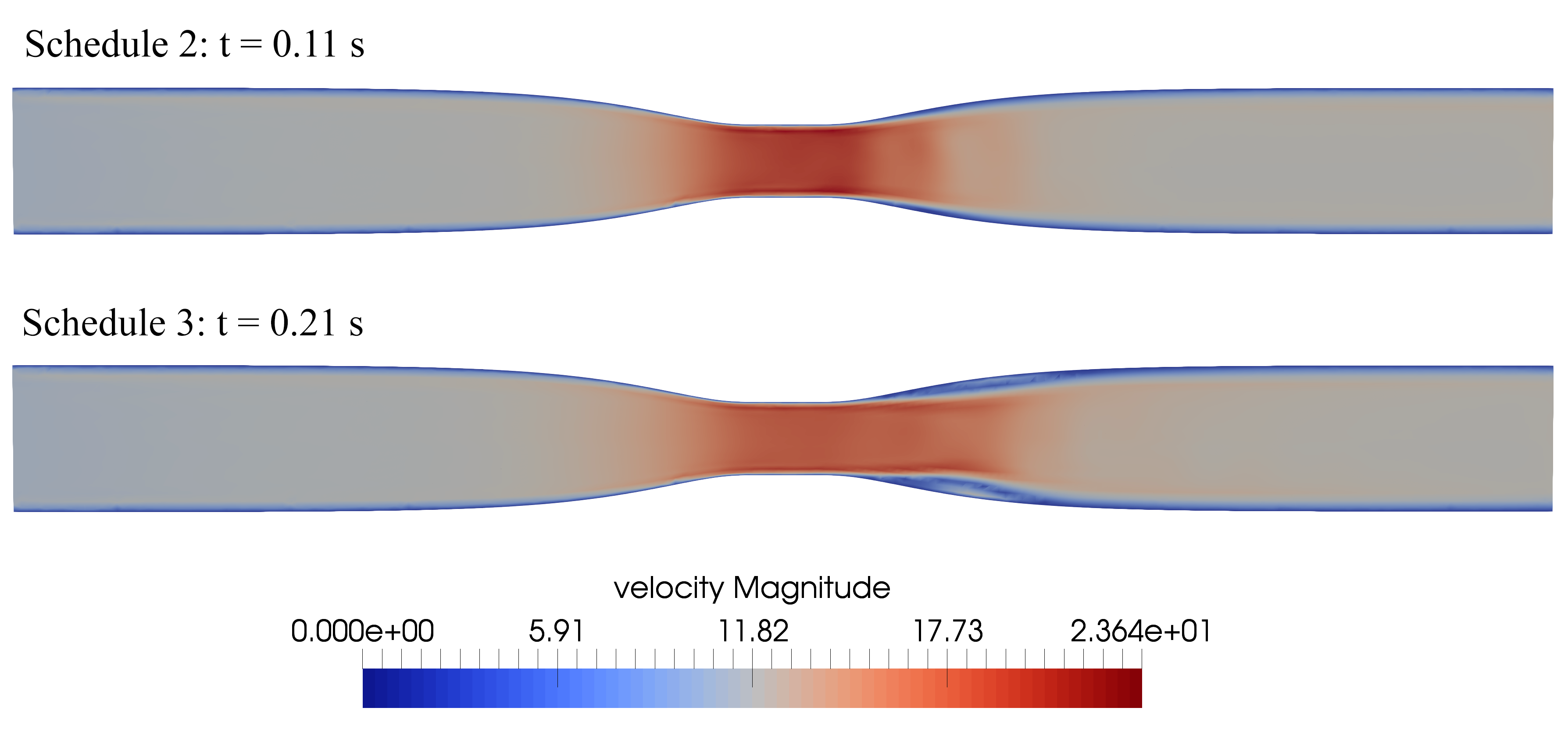}
  \caption{These instantaneous velocity magnitude fields were captured $0.01 s$ after the completion of the deformation schedule. Here, we note that under schedule two, the flow downstream of the nozzle has not had as much time to develop as in schedule three, but no spurious unphysical oscillations are noted upstream of the deformation.}
  \label{fig:configs2And3}
\end{figure}

In Figure \ref{fig:configs2And3}, the velocity magnitude throughout the channel is displayed $0.01 s$ after the deformation is complete for schedules two and three.  The flow is much better behaved than that seen for schedule one.  The schedule two results suggest a one step deformation schedule can be effective if a flow is allowed to reach a stationary state before a deformation is applied.


\section{Demonstration: Arterial Stenosis Bypass \textit{In Situ} Design} \label{sec:bypass}
In this section, we demonstrate our framework applied to the design and analysis of organic simulation domains where straightforward CAD parametrizations are unavailable. Here, we observe the effects of incident angle of an arterial stenosis bypass upstream of the blockage. 

Cardiovascular disease is the leading cause of death and disability worldwide. Regions of artery walls with low oscillating shear stress are known to be at high risk for elevated particle residence times and thus plaque buildup \cite{ku1985pulsatile, zarins1998hemodynamic}. Atherosusceptability, characterizing the likelihood of the development of atherosclerosis and arterial blockage is a patient specific and difficult to define metric affected by many chemical pathways and physical arterial features. For instance, within an artery at regions of recirculation, blood cells and undesirable metabolites can be trapped in recirculating eddies and potentially lead to cell damage, increasing atherosusceptability \cite{davies2013atherosusceptible}. This is a purely geometric pathway to increased atherosusceptability dependent on patient specific orientations of arterial networks. Qualitative characterization of such phenomona is well understood, but direct observation alone makes studying risk and response to adverse conditions difficult to quantify. An existing treatment for arterial blockages involves implanting a bypass to redirect blood flow around the blockage. The performance and longevity of these implanted bypasses are correlated with the blood flow dynamics through the complex organic geometry of the bypass and its interface with the artery. Further, understanding the impact of geometric deviation from a planned implant caused by potential surgical variation or defect would aid a surgeon in the surgical planning process. This all inspires a need for simulation based tools \cite{updegrove2017simvascular}, and our framework is particularly well suited for rapidly characterizing the impact of changes to geometry in these systems. 

\begin{figure}[t!]
  \centering
  \includegraphics[width=0.7\linewidth]{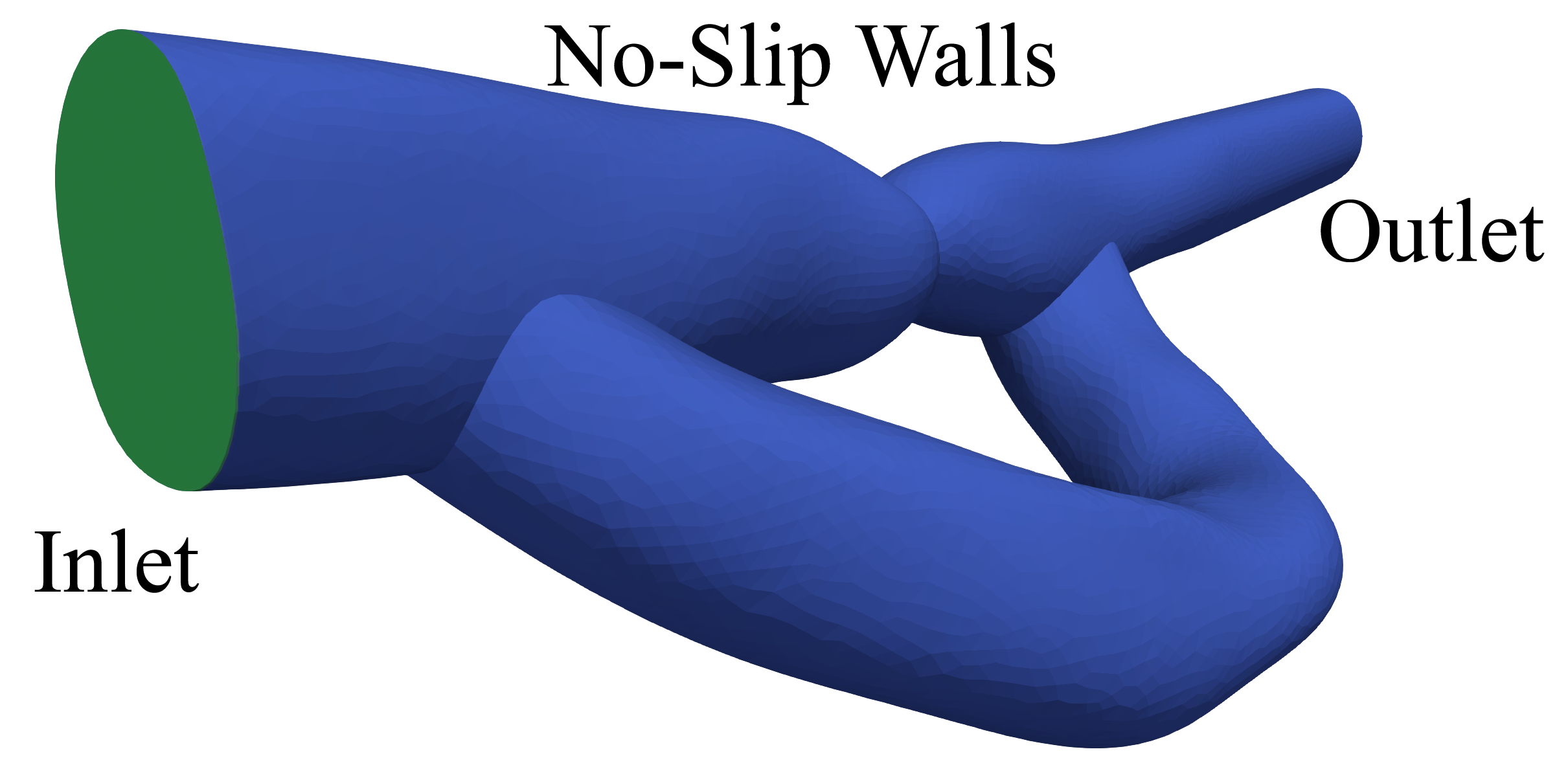}
  \caption{The bypass simulation is characterized by a constant-flow inlet, no-slip walls, and a zero-gauge pressure outlet with zero viscous traction.}
  \label{fig:bypassBCs}
\end{figure}

The novel workflow detailed in this article enables exploration of design spaces of systems in a manner that does not rely on rigid parametrizations. This feature makes it an excellent framework for exploring the behavior of organic physiological flow domains such as those found in human cardiovascular systems. These domains are intrinsically difficult to develop simple, actionable parametrizations of, and efforts to explore patient specific therapies lead to flow simulation domains specified by nothing but point clouds or tetrahedralized meshes. As in Section \ref{sec:nozzle}, the simulation and server application operated on a small compute node running the Debian 3.16 operating system with 40 total cores over dual socket Intel XeonE5-2650 CPUs with 256 GB available RAM. The client application and \texttt{ParaView} operated on a 2018 Apple MacBook Pro running macOS 10.15 with a quad-core Intel i5 CPU and 8 GB of RAM.

\subsection{Simulation Setup}
Figure \ref{fig:bypassBCs} illustrates the bypass geometry colored by boundary conditions used in this simulation. A constant velocity of $0.15 m/s$ was set at the inlet. The walls of the artery and bypass were set as no-slip walls, and the outlet pressure was set to zero-gauge pressure. The mesh of the domain was made of 59,850 tetrahedral elements. The time step size was set to $0.001 s$. Viscosity of the fluid was set to $3.3 \times 10^{-6} m^2/s$. No turbulence model was used in this simulation.

The Server was set to check for user defined mesh modifications every 10 time steps of the simulation. A deformation schedule of one step was used for all deformations.

\subsection{\textit{In Situ} Design Space Exploration}
\begin{figure}[t!]
  \centering
  \includegraphics[width=0.8\linewidth]{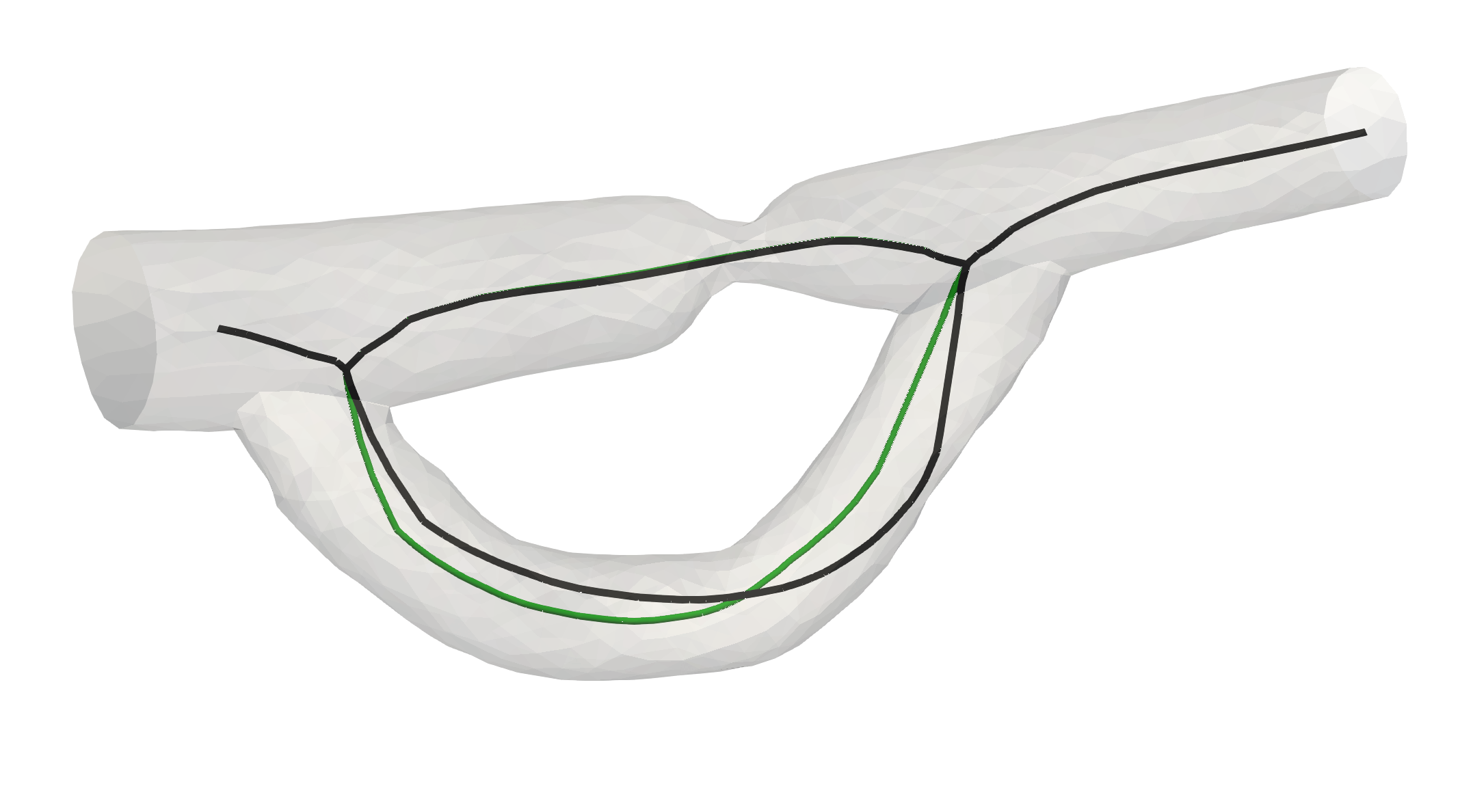}
  \caption{(The initial skeleton is drawn in black. The skeleton was then manipulated to the green curve to decrease the incident angle of the bypass.}
  \label{fig:angleChangeSkeleton}
\end{figure}

\begin{figure}[t!]
  \centering
  \includegraphics[width=\linewidth]{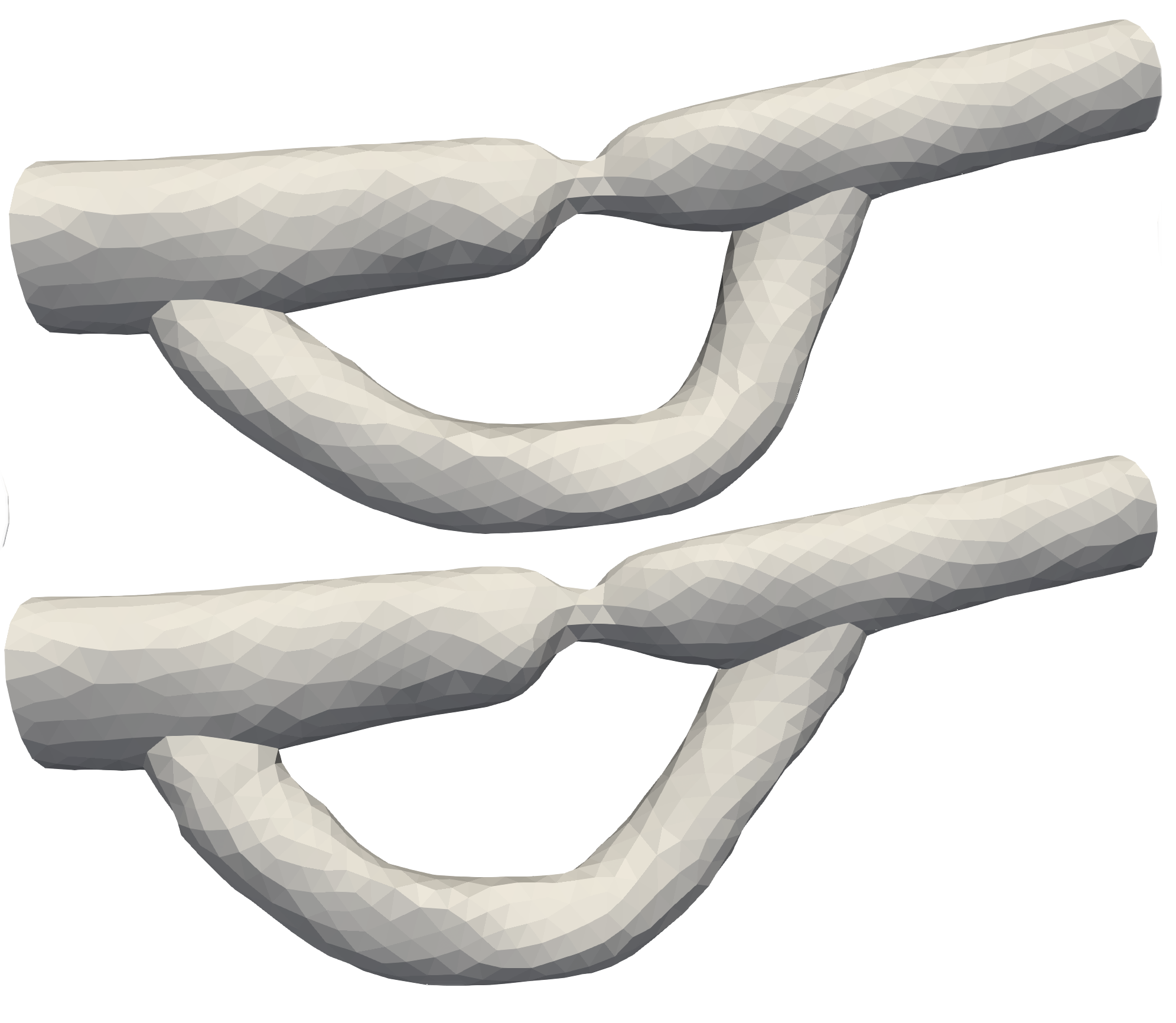}
  \caption{(top) The initial configuration of the bypass geometry was modified using the Interactive Skeleton plugin to reduce the outlet angle (bottom) of the bypass incident with the artery outlet.}
  \label{fig:angleChangeAction}
\end{figure}

In this design consideration, we were interested in how the incident angle of the bypass outlet with the artery affects bulk dynamics of the flow. To study this, we employed the Interactive Skeleton Manipulation action detailed in Section \ref{sec:skelInteractivity} to manipulate that angle. Figure \ref{fig:angleChangeSkeleton} illustrates the specific modification of the skeleton applied. This freeform interaction stretched the bypass slightly upstream of the bypass outlet and decreased the angle at which the  bypass reenters the artery by approximately 15\textdegree. Figure \ref{fig:angleChangeAction} shows the initial and final design of the bypass. The artery geometry only changed marginally around the outlet of the bypass to accommodate the modified bypass.

To properly resolve the effects of the manipulated geometry, the simulation was first run with the initial geometric configuration for 1000 time steps, for a total physically simulated time of 1 second. This brought the flow to a stationary state before we applied the deformation. The simulation was then run again for an additional 1000 time steps to ensure another stationary configuration of the flow was reached, though we should note that the flow did not change considerably after approximately 100 time steps post deformation.

\subsection{Results}
\begin{figure}[t!]
  \centering
  \includegraphics[width=\linewidth]{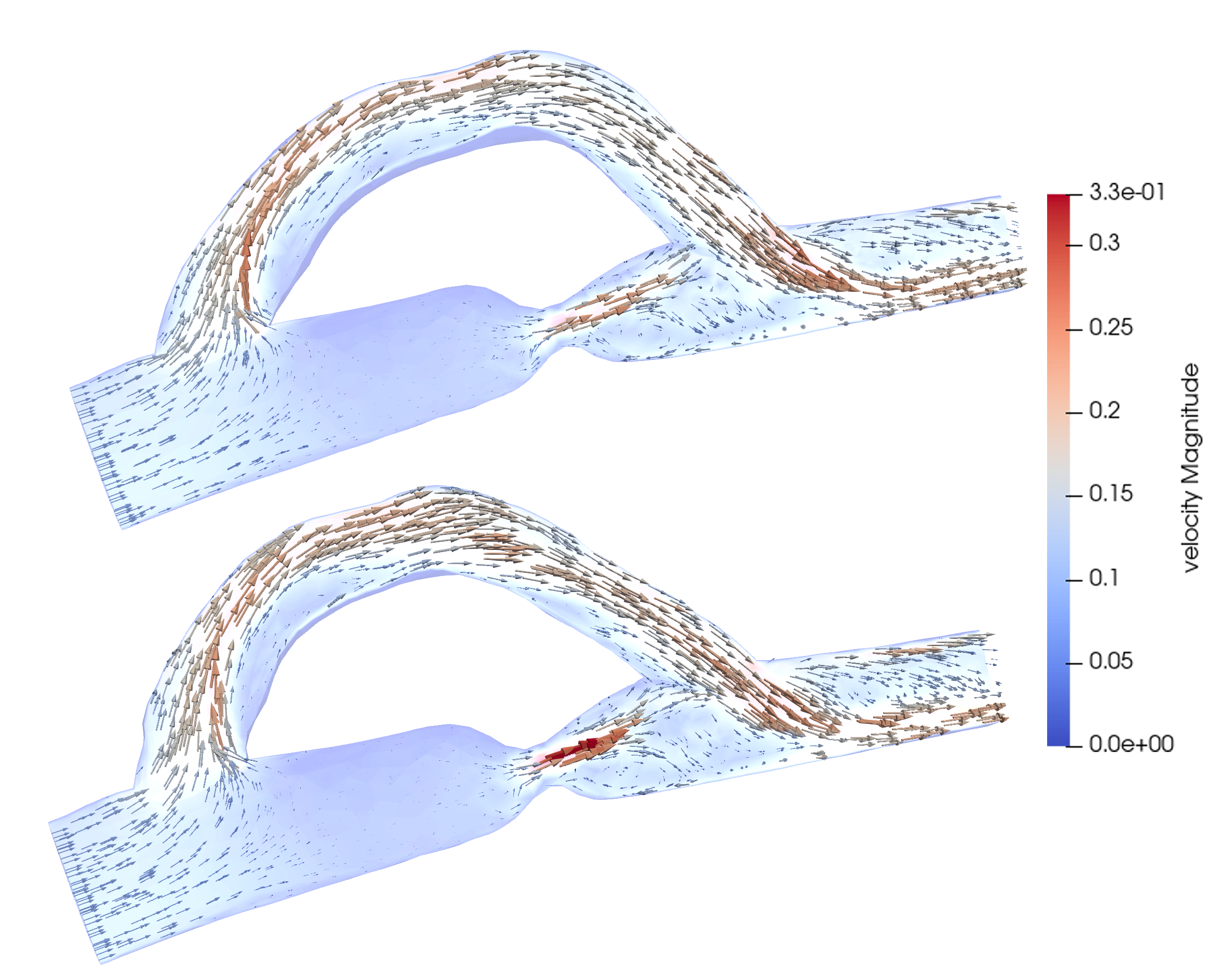}
  \caption{The instantaneous velocity field of the bypass is plotted before (top) and after (bottom) modification of the incident angle of the bypass with the artery downstream of the stenosis.}
  \label{fig:BypassVelocity}
\end{figure}

\begin{figure}[t!]
  \centering
  \includegraphics[width=\linewidth]{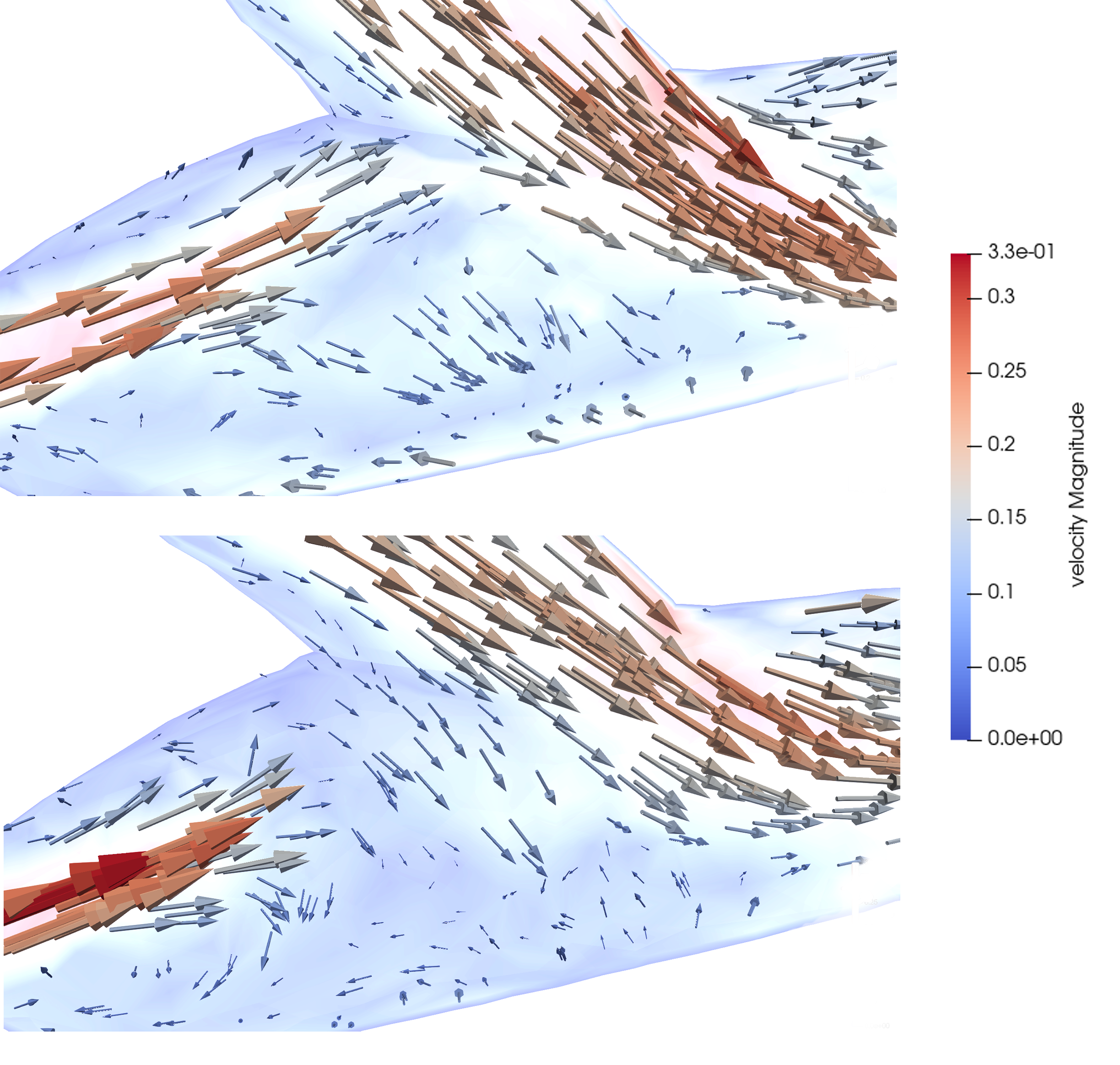}
  \caption{Zooming in on the region where the bypass channel meets the artery downstream of the stenosis, the instantaneous velocity field is plotted before (top) and after (bottom) modification.}
  \label{fig:velocityZoom}
\end{figure}

\begin{figure}[t!]
  \centering
  \includegraphics[width=\linewidth]{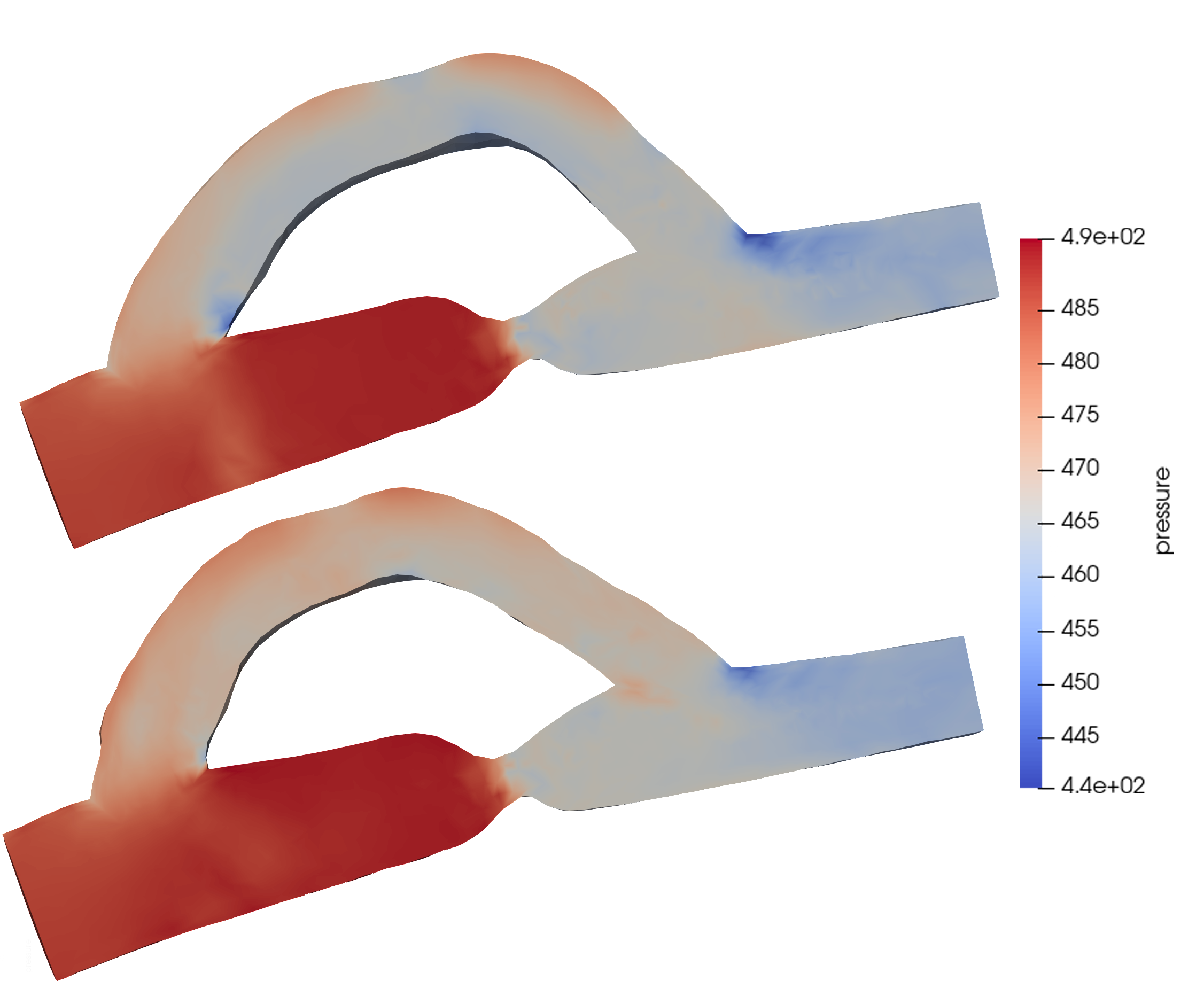}
  \caption{The pressure field of the bypass is plotted before (top) and after (bottom) modification of the incident angle of the bypass with the artery downstream of the stenosis.}
  \label{fig:bypassPressure}
\end{figure}

\begin{figure}[t!]
  \centering
  \includegraphics[width=\linewidth]{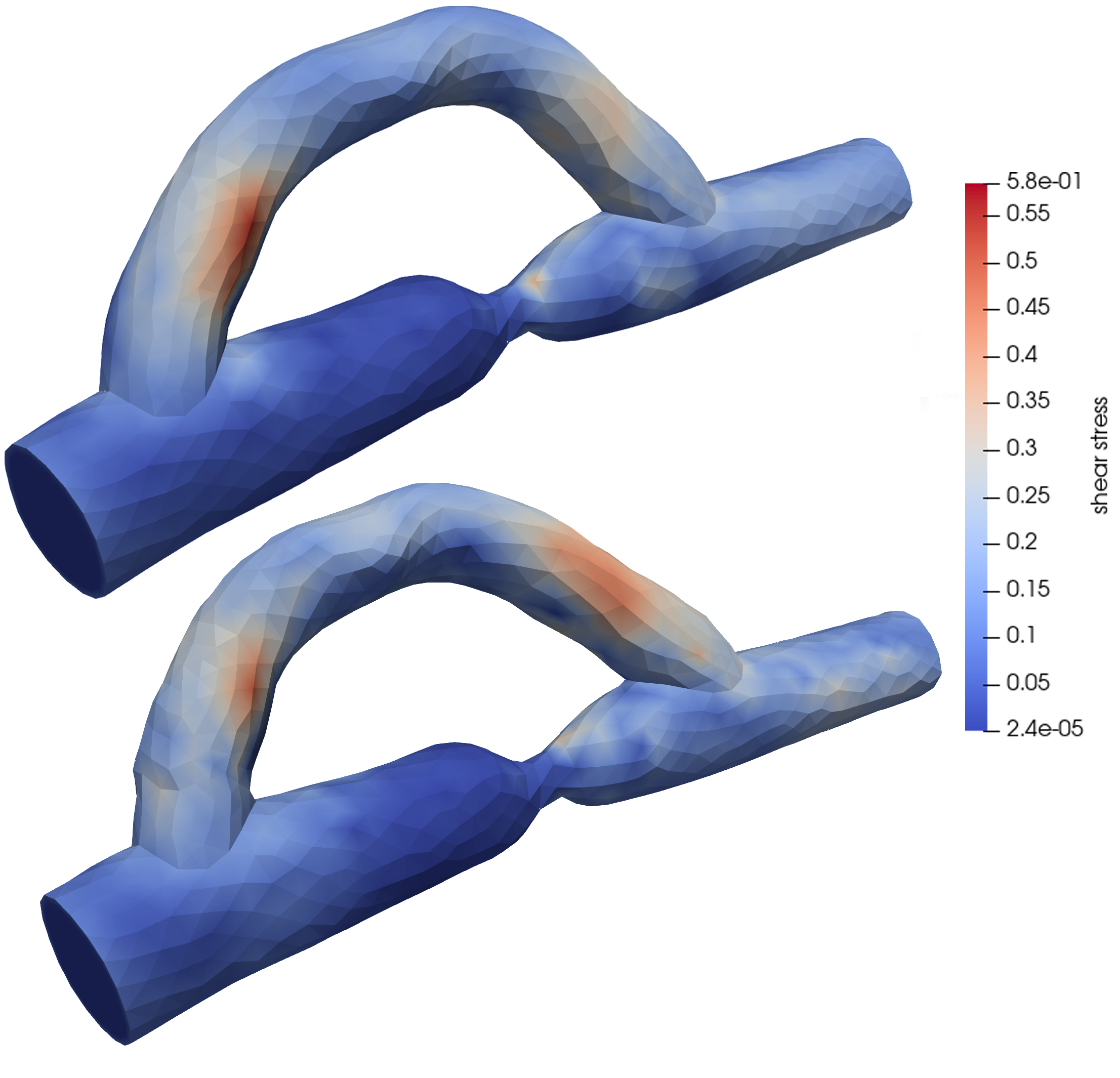}
  \caption{The shear stress field at the walls of the bypass and artery is plotted before (top) and after (bottom) modification of the incident angle of the bypass with the artery downstream of the stenosis.}
  \label{fig:bypassShear}
\end{figure}

In Figure \ref{fig:BypassVelocity}, a representative instantaneous velocity field of the domain is presented from before and after design modification. From this perspective, we note that the an increased flow through the stenosis post-modification. Zooming in on the region between the stenosis and the bypass outlet in Figure \ref{fig:velocityZoom}, we see this increase in flow through the stenosis is coupled with reduced recirculation in the region of the artery just upstream of the bypass outlet. Also from this zoomed perspective, we can see that the velocity at the point where the bypass intersects the artery farthest downstream is reduced after the modification is performed. All figures reported were captured during simulation runtime using the \texttt{Catalyst} co-processing visualization system.

In Figure \ref{fig:bypassPressure}, the pressure field reveals that the downstream reentrant corner of the bypass where the velocity field was noted to have been reduced post-modification corresponds with an increase in pressure around that corner. The pressure field through the bypass is marked by several high and low points before modification. Particularly, the initial design admits a much sharper corner before the outlet of the bypass, where a high pressure region occurs. This high pressure region is reduced significantly post modification.

The shear stress along the walls of the bypass and artery are illustrated in Figure \ref{fig:bypassShear}. Before modification, a localized region of increase shear stress is observed along the upstream side of the bypass. This region is somewhat reduced post modification, though traded for an increased localized region of shear stress at the downstream side bypass. However, it is regions of reduced localized shear stress which are associated with stagnation regions and longer particle residence times that can lead to the buildup of plaque and subsequent blockages \cite{paszkowiak2003arterial}.

\section{Mesh Deformation Robustness} \label{sec:meshQuality}
In this section, we characterize the robustness of our volumetric mesh deformation procedure with respect to a selection of freeform surface deformations. The freeform surface deformation system has few guardrails shielding a user from performing drastic or intuitively unreasonable deformations, and it is certainly possible to produce a surface deformation which leads to an invalid volume mesh exhibiting unusable mesh quality or even inverted elements. In lieu of adaptive mesh refinement, it is critical to illustrate the robustness of our mesh deformation system. In this section, we present two experiments which illustrate how increasingly extreme deformations affect mesh quality. The first experiment in Subsection \ref{sec:sharingMesh} examines deformation of isotropic meshes under severe shearing. The second experiment in Subsection \ref{sec:compressBL} examines compression of a boundary layer mesh. These two experiments cover anisotropic deformation of an isotropic mesh and isotropic deformation of an anisotropic mesh.

We measure mesh quality with the scaled Jacobian metric \cite{shewchuk2002good}. Given a transformation from a reference element to a physical element $E$ encoded in the Jacobian matrix $J$, let $v_j$ be the $j$th column of $J$. The, we can define the scaled Jacobian as
\begin{equation}
  \sigma_{sj}(E) = \frac{|J|}{\Pi_j \lVert v_j \rVert}.
\end{equation}
The scaled Jacobian varies from 0 to 1, where 1 represents an ideal element, and 0 being a flattened element. The measure becomes negative for inverted elements. It is related to the interpolation error of a finite element method. Further, in order to better characterize \textit{change} in quality from an initial mesh, we define the normalized scaled Jacobian of an element $E'$ deformed from an initial configuration $E^0$ as
\begin{equation}
    \sigma_{nj}(E') = \frac{\sigma_{sj}(E')}{\sigma_{sj}(E^0)}.
\end{equation}
We report this normalized quality metric throughout this section. Using this definition, we do observe some elements within deformed meshes with a normalized scaled Jacobian greater than $1.0$ as the deformed element's quality was improved from its initial state. This behavior was observed in every mesh deformation experiment in this section, though the effect was isolated to small collections of elements with little effect on the statistics reported outside the maximum.

In this set of experiments, no simulation solver was attached to the freeform geometry modification framework. The design intent for our framework is a generalizable system which could be readily adapted to any mesh based simulation application. It is reasonable to expect any specific simulation solver to have some threshold for tolerance to poor mesh quality under which the simulation results become unreliable. Because this threshold is not general across solvers and across physical models, this set of experiments illustrates what amount of mesh quality degradation can be expected from our framework, making no assumptions about what any given simulation solver can tolerate.

\subsection{Shearing an Isotropic Mesh} \label{sec:sharingMesh}
\begin{figure}[!b]
  \centering
  \includegraphics[width=0.65\linewidth]{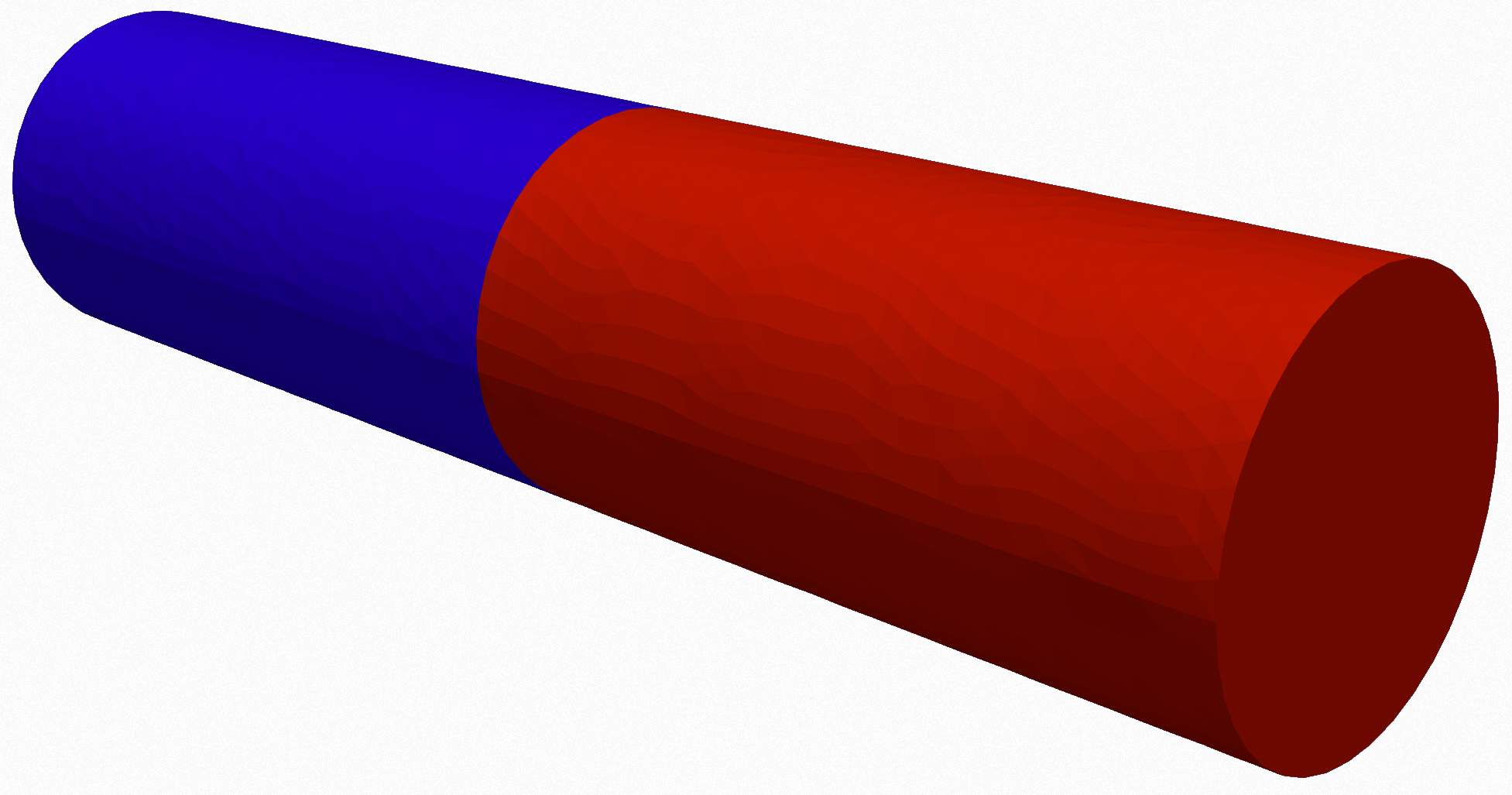}
  \caption{The center of the cylinder is at the origin. The fixed region of the cylinder (blue) spans axially between $[-0.625, 0.025]$, and the shearing region spans axially between $[0.025, 0.625]$.}
  \label{fig:cylinderShearRegions}
\end{figure}

In this subsection, we examine the mesh quality response of an isotropically meshed cylinder undergoing severe shearing. The cylinder was divided into two regions, a fixed region and a shearing region, depicted in Figure \ref{fig:cylinderShearRegions}. The cylinder has a radius of 0.15 units and a length of 1.25 units. Two meshes were generated for this geometry, a coarse mesh consisting of 43,938 elements, and a fine mesh consisting of 182,402 elements.

\begin{figure*}[!t]
  \centering
  \includegraphics[width=0.8\linewidth]{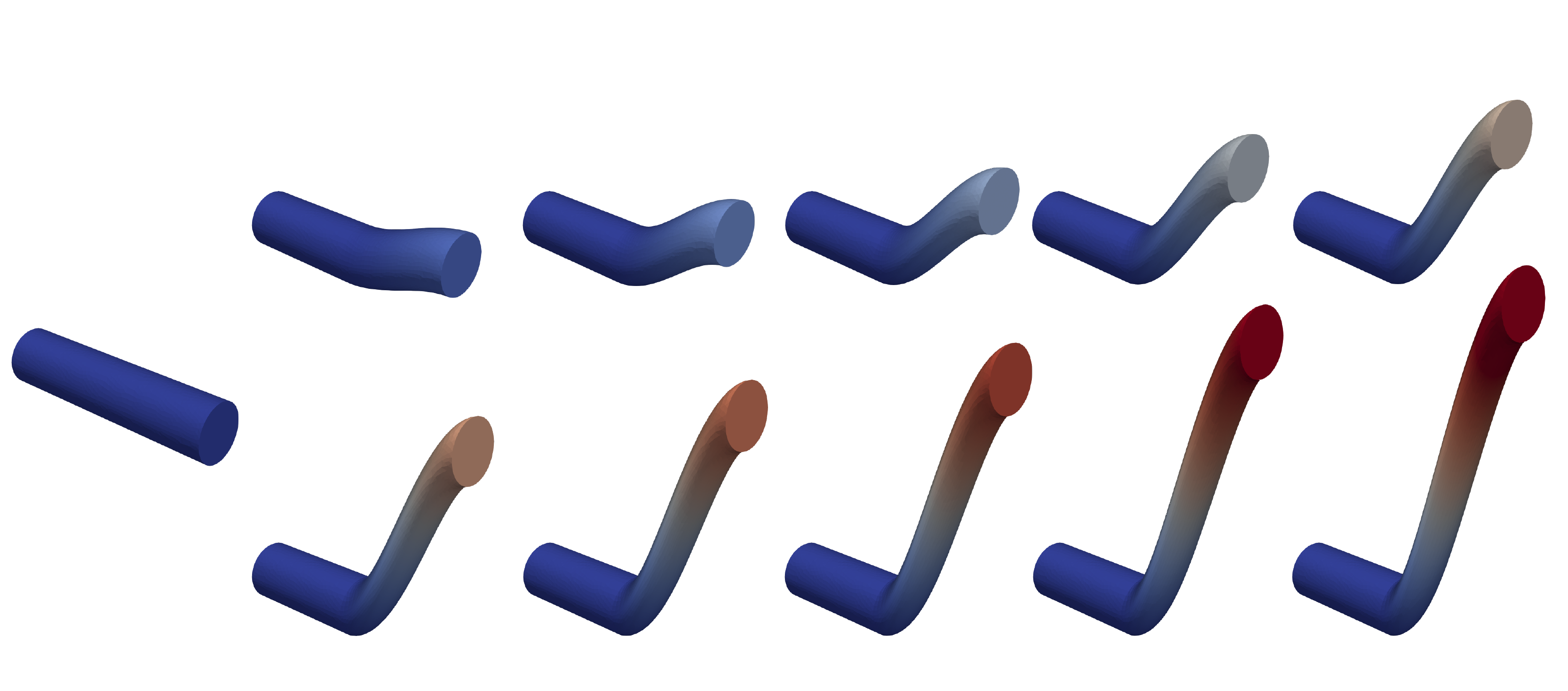}
  \caption{Here, the fine cylinder mesh is shown undeformed and then sheared from one to ten radii.}
  \label{fig:cylinderShearDepiction}
\end{figure*}

In this test, the circular face of the cylinder in the shearing region was translated normal to its central axis by integer multiples of its radius while the fixed region remained fixed. Note that the fixed region only applies to the fixation of boundary vertices during surface deformation, no constraints were placed on volume elements in the fixed region. Each deformation was taken in one step, that is to say, no multi-step deformation schedule was employed in order to measure the limits of single step deformation. The linear elasticity solver was employed for volume deformation. A biharmonic deformation field was used to propagate the translation of the circular face to the rest of the shearing region. Figure \ref{fig:cylinderShearDepiction} depicts each shearing deformation of the cylinders.

\begin{table*}[!t]
\centering
\caption{ Normalized Scaled Jacobian Mesh Quality Statistics for Coarse Sheared Isotropic Cylinder Mesh. }
\begin{tabular}{|| c | c | c | c ||}
  \hline
  Amount Sheared & Mean Normalized Quality & Min Normalized Quality & Max Normalized Quality \\
  \hline
  \hline
   $0r$  & 1.000000  & 1.000000  & 1.000000 \\
   \hline
   $1r$  & 0.942714  & 0.641904  & 1.494435 \\
   \hline
   $2r$  & 0.878471  & 0.408294  & 2.246570 \\
   \hline
   $3r$  & 0.819966  & 0.252435  & 3.208303 \\
   \hline
   $4r$  & 0.771026  & 0.159230  & 3.602791 \\
   \hline
   $5r$  & 0.731101  & 0.103874  & 3.635846 \\
   \hline
   $6r$  & 0.698656  & 0.066737  & 4.248812 \\
   \hline
   $7r$  & 0.672212  & 0.042723  & 3.900661 \\
   \hline
   $8r$  & 0.650334  & 0.026897  & 3.161174 \\
   \hline
   $9r$  & 0.632012  & 0.016207  & 3.231860 \\
   \hline
   $10r$ & 0.616426  & 0.008846  & 3.359668 \\
  \hline
\end{tabular}
\label{table:coarseShearStats}
\end{table*}

\begin{table*}[!t]
\centering
\caption{ Normalized Scaled Jacobian Mesh Quality Statistics for Fine Sheared Isotropic Cylinder Mesh. }
\begin{tabular}{|| c | c | c | c ||}
  \hline
  Amount Sheared & Mean Normalized Quality & Min Normalized Quality & Max Normalized Quality \\
  \hline
  \hline
  $0r$ & 1.000000  & 1.000000 & 1.000000 \\
  \hline
  $1r$ & 0.932491 & 0.621786 & 1.562472 \\
  \hline
  $2r$ & 0.862242 & 0.389323 & 2.285814 \\
  \hline
  $3r$ & 0.801268 & 0.236938 & 2.868266 \\
  \hline
  $4r$ & 0.752120 & 0.144464 & 3.878493 \\
  \hline
  $5r$ & 0.713175 & 0.090750 & 3.937560 \\
  \hline
  $6r$ & 0.682087 & 0.058418 & 3.828411 \\
  \hline
  $7r$ & 0.656880 & 0.038234 & 3.803568 \\
  \hline
  $8r$ & 0.636095 & 0.025181 & 3.967885 \\
  \hline
  $9r$ & 0.618672 & 0.016483 & 4.017845 \\
  \hline
  $10r$ &0.603879 & -0.013790& 4.015214 \\
  \hline
\end{tabular}
\label{table:fineShearStats}
\end{table*}

Table \ref{table:coarseShearStats} summarizes the statistics of the normalized scaled Jacobian metric for the sheared coarse mesh. Here, we see that no elements were inverted even under extreme shearing. Overall, we see a reduction in quality from the original mesh by $38.6\%$ at the most extreme deformation. The most deformed elements were almost completely flattened as evidenced by a reduction in quality of $99.1\%$. A collection of isolated elements actually experienced improved quality as noted by the max normalized quality of over $1.0$.

Table \ref{table:fineShearStats} summarizes the normalized mesh quality statistics for the sheared fine mesh. At a shearing deformation of $10r$, finally some elements inverted in on themselves leading to a minimum quality less than zero. The coarse mesh performed slightly better under severe shearing due to the increased range of motion per element vertex relative to the scale of the geometric volume.

\begin{figure}[!t]
  \centering
  \includegraphics[width=\linewidth]{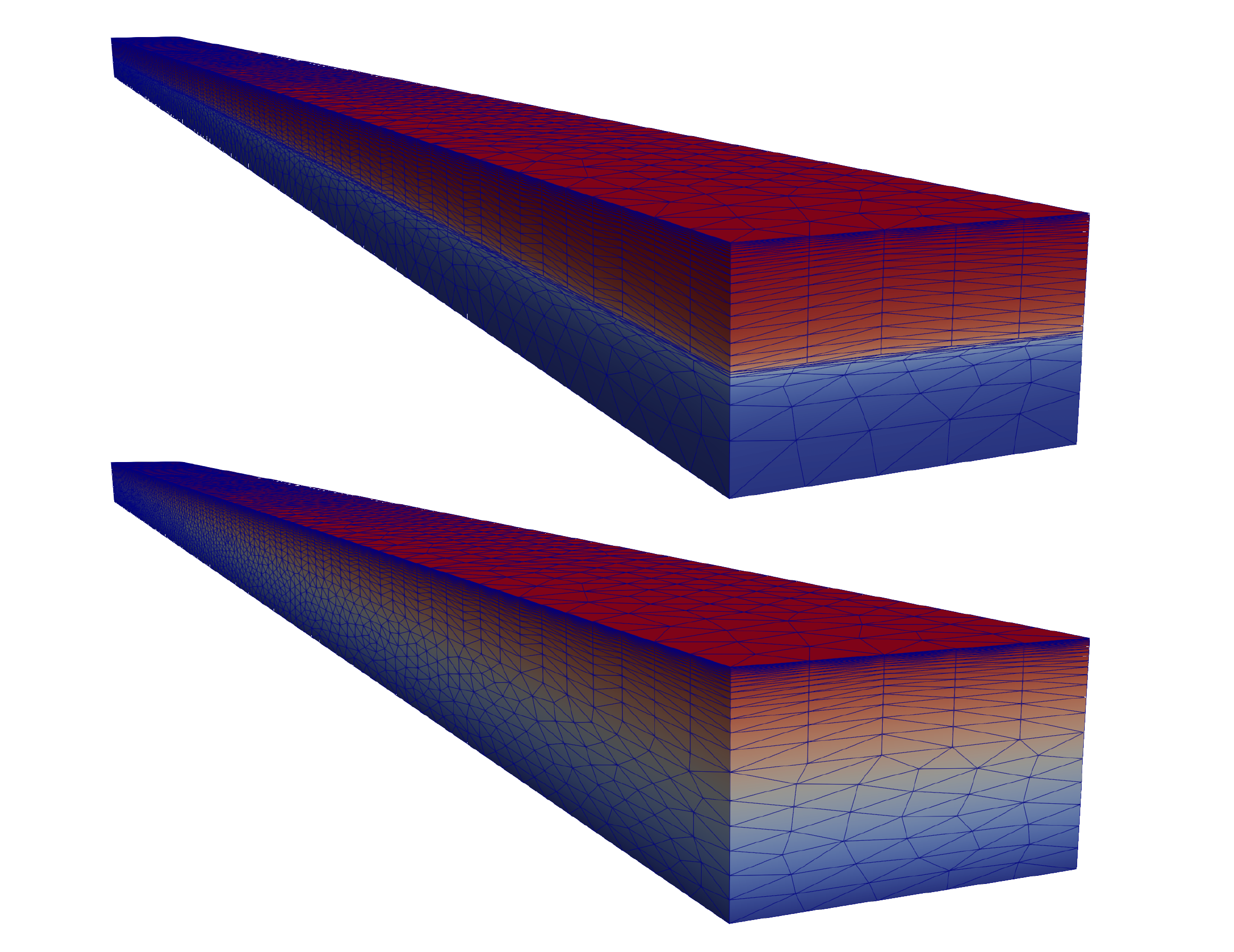}
  \caption{The channel is compressed 70\% using a biharmonic (top) and harmonic (bottom) surface deformation.}
  \label{fig:duct70Compressed}
\end{figure}
\begin{figure}[!t]
  \centering
  \includegraphics[width=\linewidth]{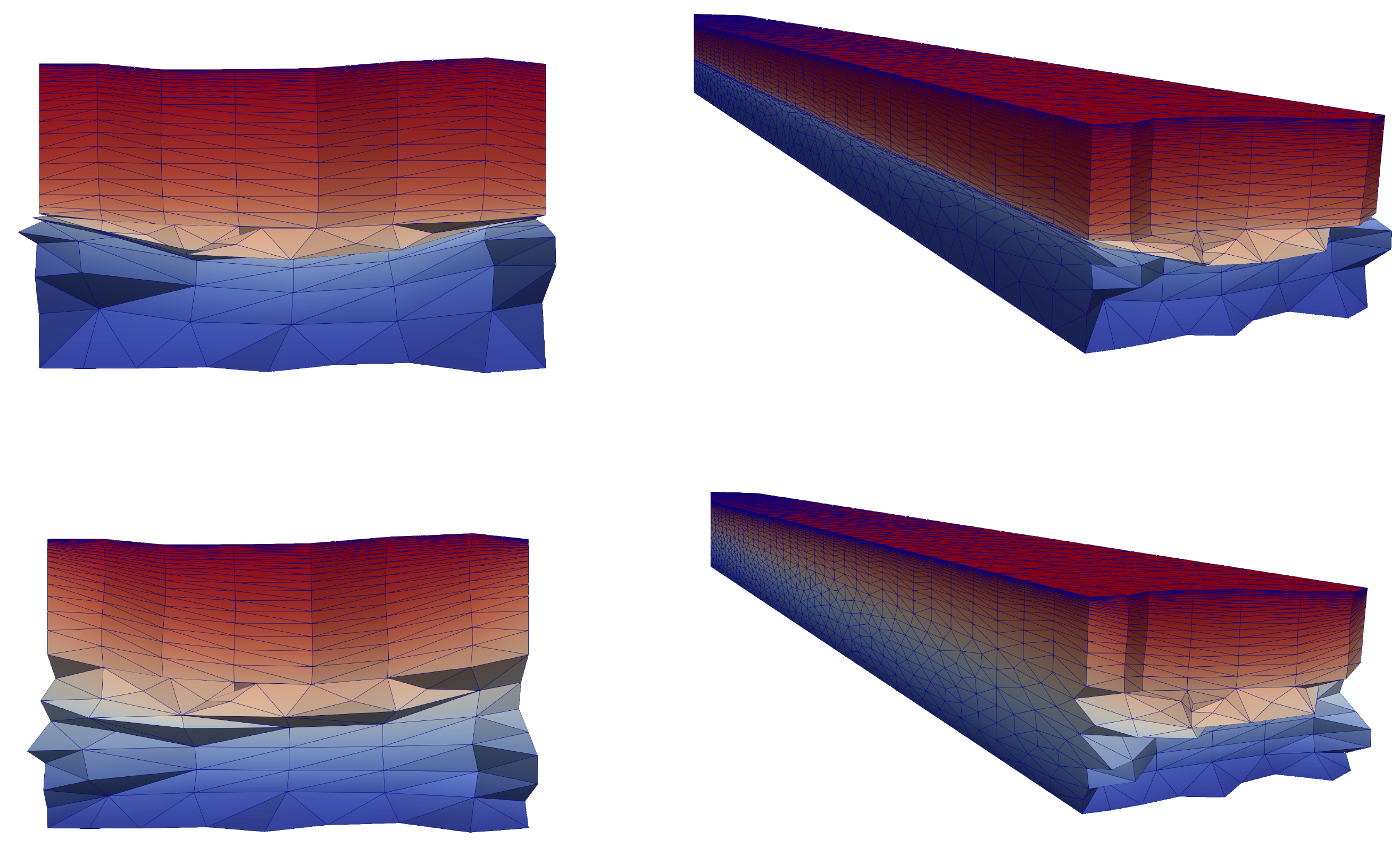}
  \caption{A crinkle-cut cross section was taken at the mid-section of the channel after 70\% compression using a biharmonic (top) and a harmonic (bottom) surface deformation.}
  \label{fig:duct70CompressedXSection}
\end{figure}

\begin{table*}[t!]
\centering
\caption{ Normalized Scaled Jacobian Mesh Quality Statistics for Compressed Boundary Layer Mesh Under Harmonic Surface Deformation. }
\begin{tabular}{|| c | c | c | c ||}
  \hline
  \% Compressed & Mean Normalized Quality & Min Normalized Quality & Max Normalized Quality \\
  \hline
  \hline
  0 & 1.000000 &  1.000000 &  1.000000 \\
  \hline
  10 & 0.958131 &  0.897114 &  1.225868 \\
  \hline
  20 & 0.906009 &  0.792184 &  1.514773 \\
  \hline
  30 & 0.845514 &  0.681641 &  1.883110 \\
  \hline
  40 & 0.777266 &  0.567598 &  2.289681 \\
  \hline
  50 & 0.701735 &  0.440565 &  2.549398 \\
  \hline
  60 & 0.619149 &  0.304348 &  2.868457 \\
  \hline
  70 & 0.530068 &  0.141035 &  2.476053 \\
  \hline
  80 & 0.435559 & -0.053814 &  2.005864 \\
  \hline
\end{tabular}
\label{table:BL_Harmonic_Compression}
\end{table*}

\begin{table*}[t!]
\centering
\caption{ Normalized Scaled Jacobian Mesh Quality Statistics for Compressed Boundary Layer Mesh Under Biharmonic Surface Deformation. }
\begin{tabular}{|| c | c | c | c ||}
  \hline
  \% Compressed & Mean Normalized Quality & Min Normalized Quality & Max Normalized Quality \\
  \hline
  \hline
  0  & 1.000000 &  1.000000 &  1.000000 \\
  \hline
  10 & 0.971301 &  0.865901 &  1.315387 \\
  \hline
  20 & 0.931247 &  0.727356 &  1.750157 \\
  \hline
  30 & 0.881689 &  0.585910 &  2.300805 \\
  \hline
  40 & 0.823444 &  0.442486 &  2.915207 \\
  \hline
  50 & 0.756937 &  0.293090 &  2.488440 \\
  \hline
  60 & 0.682633 &  0.132583 &  2.332846 \\
  \hline
  70 & 0.601878 & -0.107032 &  2.067191 \\
  \hline
\end{tabular}
\label{table:BL_biharmonic_Compression}
\end{table*}

\subsection{Compressing a Boundary Layer Mesh} \label{sec:compressBL}
In this subsection, we examine the mesh quality response of a rectangular channel with a boundary layer grown from the top face downward into the volume undergoing severe compression. The mesh of this domain consists of 103,325 tetrahedral elements. The channel is $1.05 m$ long with a cross-section $0.1 m$ tall by $0.05 m$ wide. The boundary layer contains 20 layers, extending into the domain $0.04 m$. The channel was compressed by employing the Translate Feature action to fix the bottom face and move the upper face downward. In this test, harmonic and biharmonic deformations were employed to propagate the surface deformation through the vertical sides of the geometry. The deformation was applied in $10\%$ increments up to $80\%$ compression. As before, each deformation was performed separately in one step. Under both surface deformation strategies, the Jacobian stiffened linear elasticity solver was employed for volume deformation.

Table \ref{table:BL_Harmonic_Compression} summarizes the normalized mesh quality statistics for the compressed channel under harmonic surface deformation where we see element inversion at $80\%$ compression. Table \ref{table:BL_biharmonic_Compression} summarizes the mesh quality statistics normalized by the undeformed mesh quality for the compressed channel under biharmonic surface deformation where we see element inversion at $70\%$ compression. Despite experiencing element inversion sooner, the biharmonic deformation did demonstrate better average normalized quality reduction throughout each compression with an overall mean reduction in quality of $39.8\%$ at $70\%$ compression. Under harmonic deformation, the mean reduction in quality at $70\%$ compression was $47.0\%$. However, the biharmonic deformation led to nonuniform deformation throughout the thickness of the domain including a pinched region through the midspan of the domain where element quality degraded rapidly. This is expressed in the far reduced minimum normalized mesh quality of the biharmonic deformation scheme compared to the harmonic deformation scheme.

In Figure \ref{fig:duct70Compressed}, we see the dramatic difference in deformation behavior the selection of polyharmonic order has on the deformation of a mesh. In the biharmonic case, the boundary layer is nearly rigidly translated, leading to preservation of the boundary layer. However, the elements close to the bottom face are also held nearly rigid, so only the elements between these two regions are allowed to be compressed, leading to a layer of extreme compression and eventual element inversion. In contrast, the harmonic deformation strategy spreads the deformation across the entire free domain, leading to a smoother overall deformation and avoidance of element inversion. However, the boundary layer experiences deformation outside rigid body translation meaning the properties of the boundary layer mesh present in the initial mesh are not preserved. Figure \ref{fig:duct70CompressedXSection} shows that this behavior roughly propagates into the domain. The boundary layer post-compression of the channel which underwent biharmonic surface deformation has a significantly less curved boundary layer mesh internally than the mesh which underwent harmonic surface deformation. Though some curving is evident in the layers of the boundary layer mesh after the linear elasticity solver computes the volumetric deformation, as a boundary condition for the volumetric solver, the biharmonic surface deformation was able to hold taut the boundary layer mesh significantly better than the harmonic surface deformation.

\section{Solution Convergence Under Domain Modification} \label{sec:solCvg}
In this section, we investigate solution convergence under mesh refinement when the geometric domain is deformed. In order to probe convergence, we require a problem definition with predictable and well behaved physics which can be confidently simulated in the absence of applied geometric deformation as well as a geometric modification which does not adversely affect mesh quality. To accomplish these goals, our experiment is to model pressure drop due to internal flow in a pipe and then measure how the pressure drop changes when the outlet end of the pipe is widened. Widening the outlet end of the pipe will have the effect of diffusing the flow, and as such, the pressure drop across the length of the modified pipe geometry will be decreased relative to the initial geometric configuration. In this experiment, we can expect that the pressure drop through the initial pipe geometry will converge under mesh refinement, and we desire to show that the pressure drop through the pipe post-modification also converges. Further, we seek to examine the duration of the transient response to the geometric deformation.

\subsection{Experimental Setup}

The initial geometry is specified by a circular cross-section pipe with radius of $R = 0.025 m$ and a length of $ L = 0.5 m$. We target Poiseuille flow with a pressure drop of $\Delta p = 10 Pa$ across the length of the pipe. In cylindrical coordinates, the velocity in the $z$ direction takes the form
\begin{equation}
  u_z(r) = \frac{\Delta p}{4 \mu L}(R^2 - r^2).
\end{equation}
Taking the dynamic viscosity to be $\mu = 0.01 Ns/m^2$, we use this formulation to set the inlet velocity profile as well as the initial conditions for the flow throughout the pipe. Computing explicitly for the initial velocity and pressure profiles, we arrive at
\begin{equation} \label{eqn:cvgFlowProfile}
  \begin{split}
  u_z(r) &= 0.3125 - 500 r^2, \\
  p(z)   &= 10 + 10(1-2z),
\end{split}
\end{equation}
where we have a maximum velocity in the $z$ direction of $0.3125 m/s$ at the center of the pipe, and a maximum pressure at the inlet of $20 Pa$ with an analytical drop in pressure from inlet to outlet of $10 Pa$. Using the length of the pipe divided by the maximum velocity at the inlet, we can compute a \textit{flow through time} for this flow configuration of $T_{ft} = 1.6 s$.

\begin{figure}[!b]
  \centering
  \includegraphics[width=\linewidth]{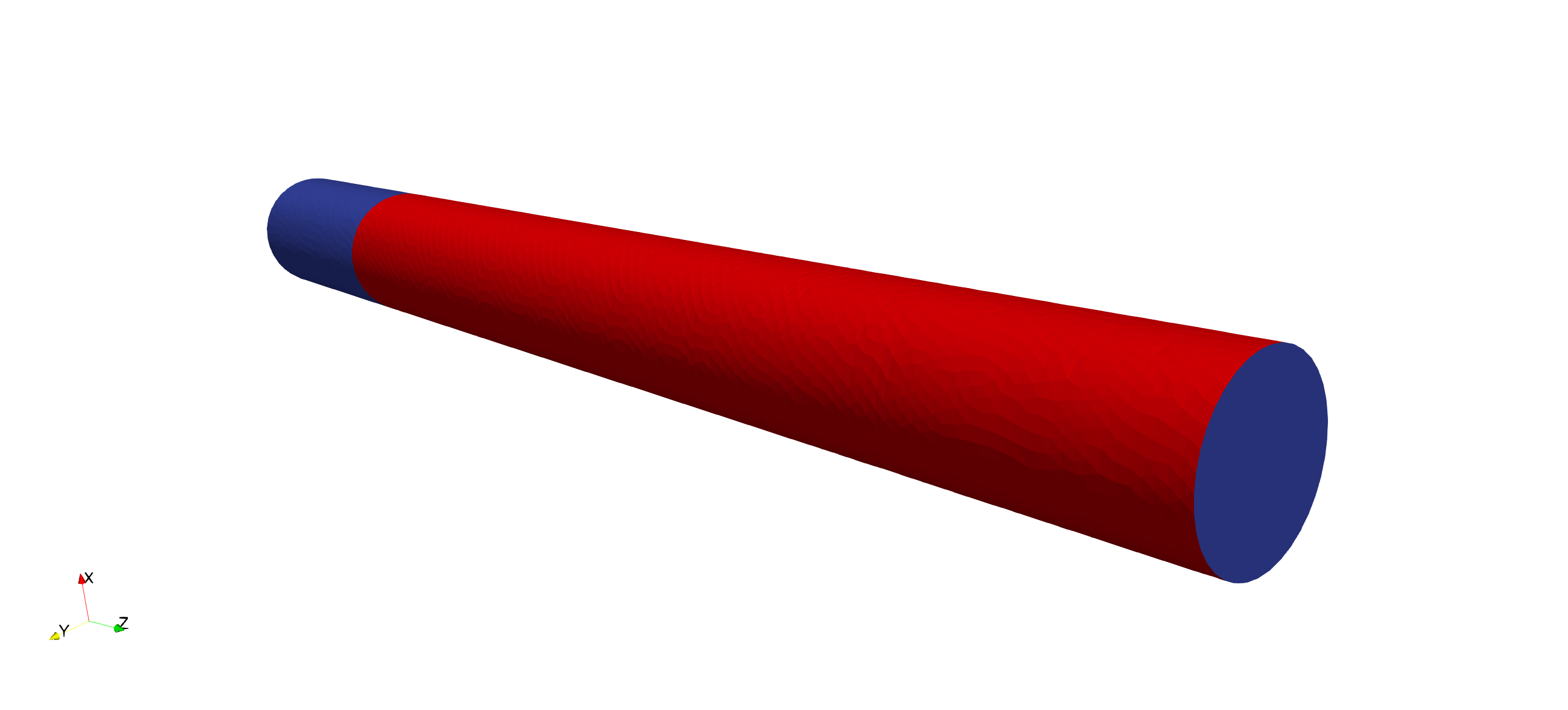}
  \caption{The red area within $z=[0.1, 0.5]m$ is specified to scale in the $x$ and $y$ directions. The blue regions are set to freely deform under a biharmonic surface deformation. The inlet face (hidden) is held fixed. }
  \label{fig:convergenceGeomSpec}
\end{figure}

\begin{table*}[t!]
\centering
\caption{ Pressure Drop Convergence Before and After Geometry Modification. }
\begin{tabular}{|| c | c | c ||}
  \hline
  Mesh Elements & Pressure Drop Pre-Modification (Pa) & Pressure Drop Post-Modification (Pa) \\
  \hline
  \hline
  123,494 & 9.694 &  5.011 \\
  \hline
  238,037 & 9.756 & 5.094 \\
  \hline
  471,293 & 9.762 & 5.105 \\
  \hline
\end{tabular}
\label{table:pressureDropConvergence}
\end{table*}

\begin{figure*}[!ht]
  \centering
  \includegraphics[width=0.75\linewidth]{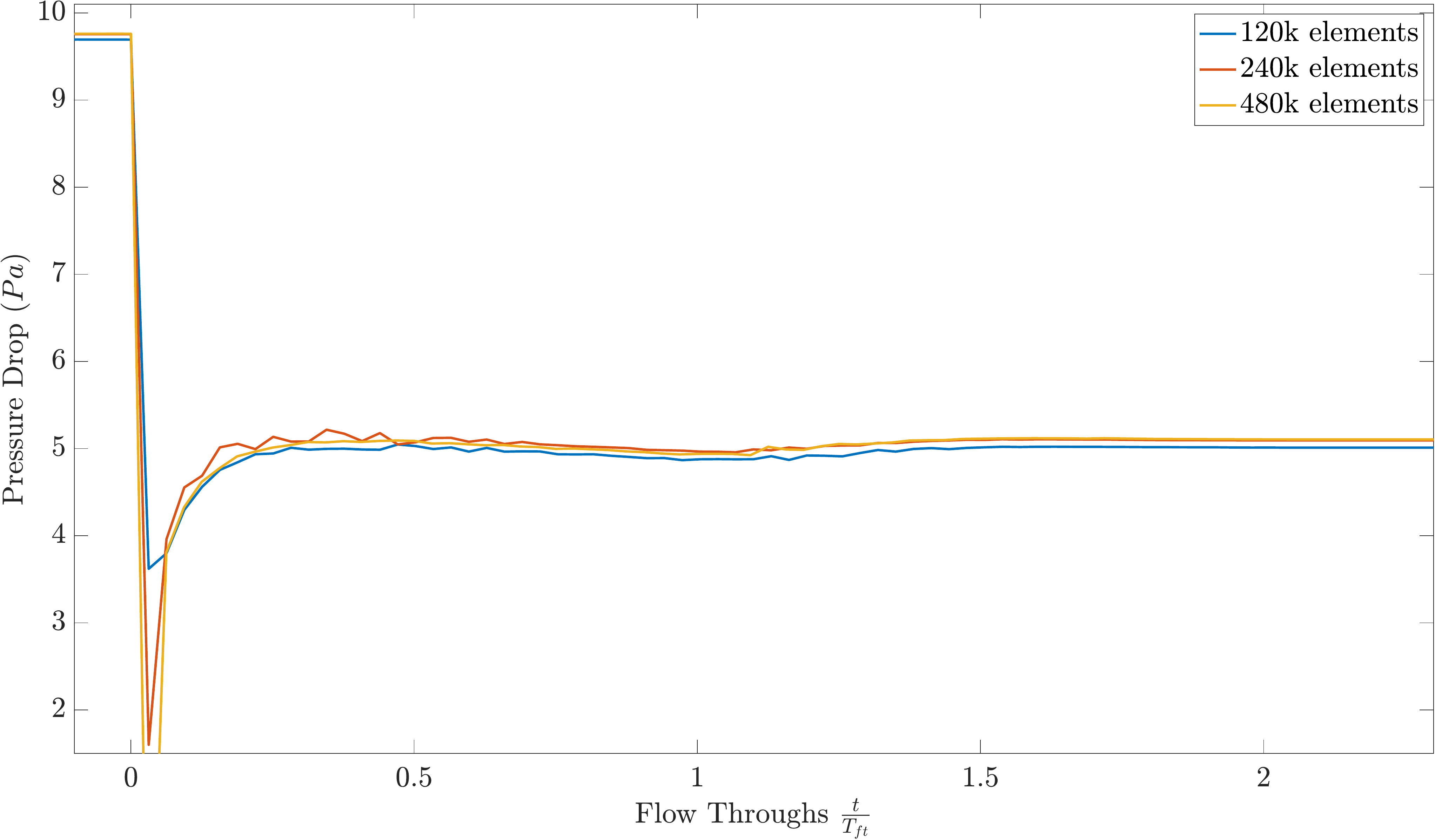}
  \caption{The average drop in pressure from the inlet to the outlet of the pipe is plotted with respect to flow-throughs past the geometry modification event which takes place at $\frac{t}{T_{ft}} = 0$.}
  \label{fig:cvgPressureResponse}
\end{figure*}

At the inlet of the pipe, we set the velocity using Equation \eqref{eqn:cvgFlowProfile}. At the outlet, we set a zero viscous traction boundary condition and we set the pressure to $10 Pa$. The most correct boundary condition would be to constrain tangential velocity to zero at the outlet, though we opted for the weaker zero tangential traction boundary condition in order to not overly constrain the PDE in anticipation for the prescribed geometry deformation.

The flow is first simulated until the pressure field reaches a steady state. The relaxed boundary condition choice of zero-tangential traction at the outlet led to a slightly reduced initial pressure drop from the theoretically expected drop. The observed steady state pressure drop in the pipe post deformation is approximately $\Delta p = 9.76 Pa$. Then, the outlet of the pipe is widened such that the cross-sectional area of the outlet is $10\%$ greater than the inlet cross-sectional area. The pipe is widened by applying a Scale by Direction action, specifying scaling of the region of the pipe $0.1 m$ from the inlet onward to the outlet as shown in Figure \ref{fig:convergenceGeomSpec}.

\subsection{Results}
Expanding the outlet region of the pipe by $10\%$ of its initial cross-sectional area resulted in a change in pressure drop from approximately $9.76 Pa$ to approximately $5.11 Pa$. The least refined mesh poorly represented the analytical parabolic flow profile, and thus was farthest from capturing the pressure drop. Table \ref{table:pressureDropConvergence} details the average pressure drop before and after deformation for each of the three meshes studied.

Figure \ref{fig:cvgPressureResponse} shows the temporal response of the pressure drop through the pipe across the geometry modification event. We see an initially steady pressure drop immediately respond to the artificial instantaneous change in the pipe's volume and then rapidly recover to a physically realizable pressure field though with some noise and a low frequency oscillation which rapidly decays away to a steady solution. Approximately 1.6 flow throughs are required for the pressure field to recover from the event.

\begin{figure*}[!ht]
  \centering
  \includegraphics[width=0.9\linewidth]{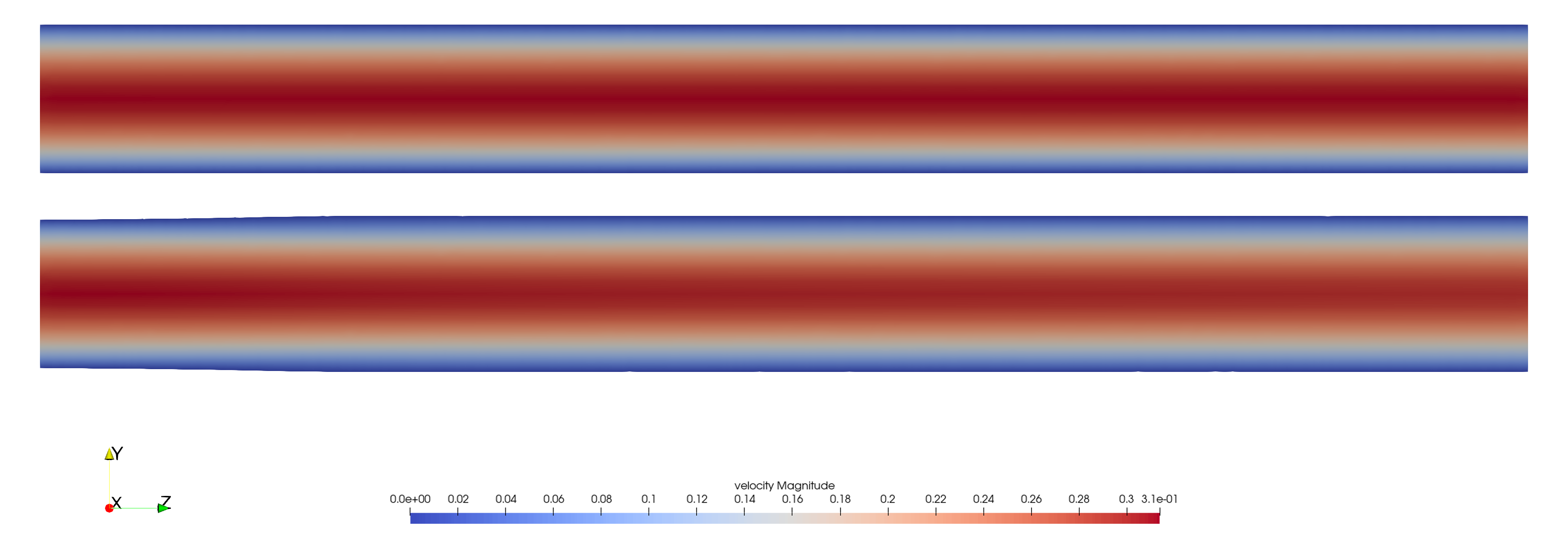}
  \caption{The velocity magnitude is plotted at the mid-plane of the pipe before geometry modification at $\frac{t}{T_{ft}} = 0.0$ (top) immediately before deformation. Then, the deformation occurs and the second, third and fourth pressure field are taken at $\frac{t}{T_{ft}} = 2$ (bottom). Both images are captured from the finest mesh.}
  \label{fig:convergenceTestVelocityProfile}
\end{figure*}

Figure \ref{fig:convergenceTestVelocityProfile} depicts the velocity field magnitude along the mid-plane of the pipe before and after geometry modification. We see an expectedly well behaved flow profile before modification at $\frac{t}{T_{ft}} = -0.05$ and a fully recovered flow profile shortly after reaching steady state post-modification at $\frac{t}{T_{ft}} = 2$. Though the small change in radius is not visually obvious, the change in cross-sectional area is significant enough to halve the pressure drop through the pipe.

\begin{figure*}[!ht]
  \centering
  \includegraphics[width=0.9\linewidth]{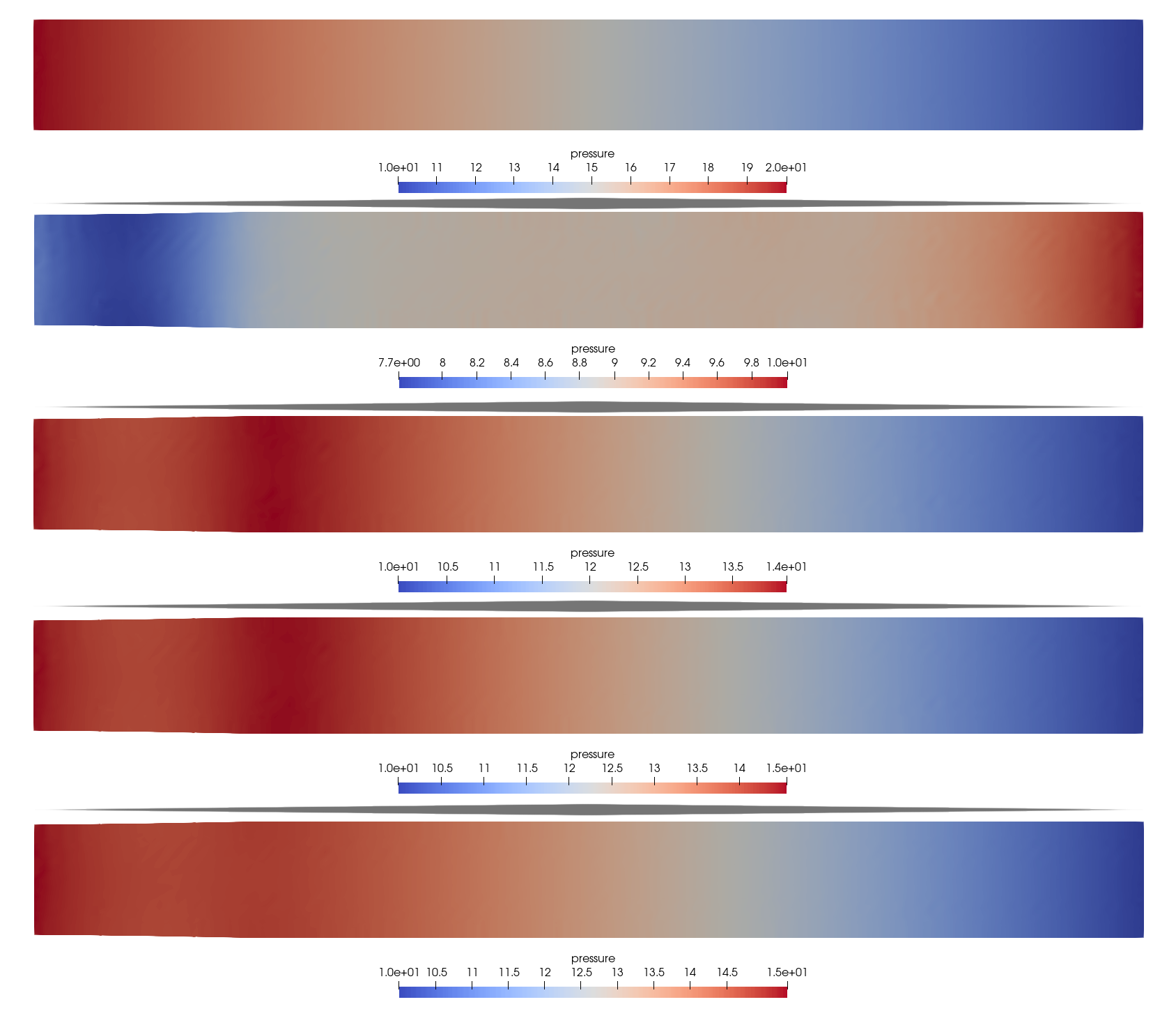}
  \caption{The pressure field is plotted at the mid-plane of the pipe before geometry modification at $\frac{t}{T_{ft}} = -0.05$ (top) and after geometry modification at $\frac{t}{T_{ft}} = 0.0313, 0.0625, 0.0938$ respectively. The bottom profile shows the converged pressure field at $\frac{t}{T_{ft}} = 2.250$. All pressure fields are captured from the finest mesh. Note that each pressure profile has its own scale as denoted by the color bar directly below each image in order to better highlight the pressure gradient at each snapshot.}
  \label{fig:cvgpressurePlots}
\end{figure*}

Finally, Figure \ref{fig:cvgpressurePlots} shows snapshots of the transient behavior of the pressure field before, during and long after the geometry deformation event. Before the geometry deformation event, the pressure profile is expectedly linear through the length of the pipe. Then, a strong adverse pressure gradient develops serving to rapidly decelerate the flow. The reversal of the pressure gradient is quickly relieved, leaving behind a mild adverse pressure gradient downstream of the inlet at the point where the diameter was expanded. As the flow attains steady state, this adverse pressure gradient decays to a mild adverse pressure gradient through the expansion just past the inlet . Note that despite the strong adverse pressure gradient that appears following the geometric deformation, no flow reversal is observed throughout the length of the simulation.

\section{Interactivity Scaling} \label{sec:interactiveScaling}
Thus far, we have presented a novel workflow for geometric design iteration through interactively manipulating a simulation domain as the simulation is running. We then characterized the system's behavior with respect to affecting mesh quality and incompressible flow solution convergence. In this section, we aim to estimate the human time cost associated with this system. Our system circumvents the typical design iteration loop of CAD modeling, meshing, partitioning, simulating and analyzing which incurs many human time costs along the way (including the oft dreaded HPC queue wait time). These hidden time costs are difficult to estimate in general, and vary wildly from user to user. As such, we will not attempt to compare directly to an estimate of time investment for existing workflows, and instead report the approximate human time cost of this presented workflow. Further, we investigate the time cost growth with respect to the resolution of the computational domain. Because the interactivity loop is implemented serially, rather than demonstrating parallel scaling, we demonstrate acceptable interactive performance to reasonably large scale simulation domains. As \texttt{PHASTA} has been demonstrated to scale favorably with problem size, and the \texttt{Catalyst} visualization system employed here has also been shown to scale well with problem size, we will measure here the time cost of the new interactivity loop, the requisite message passing system interfacing the interactivity loop with the server application, and the parallel volume mesh deformation routine running on the server. 

\begin{figure}[!t]
  \centering
  \includegraphics[width=\linewidth]{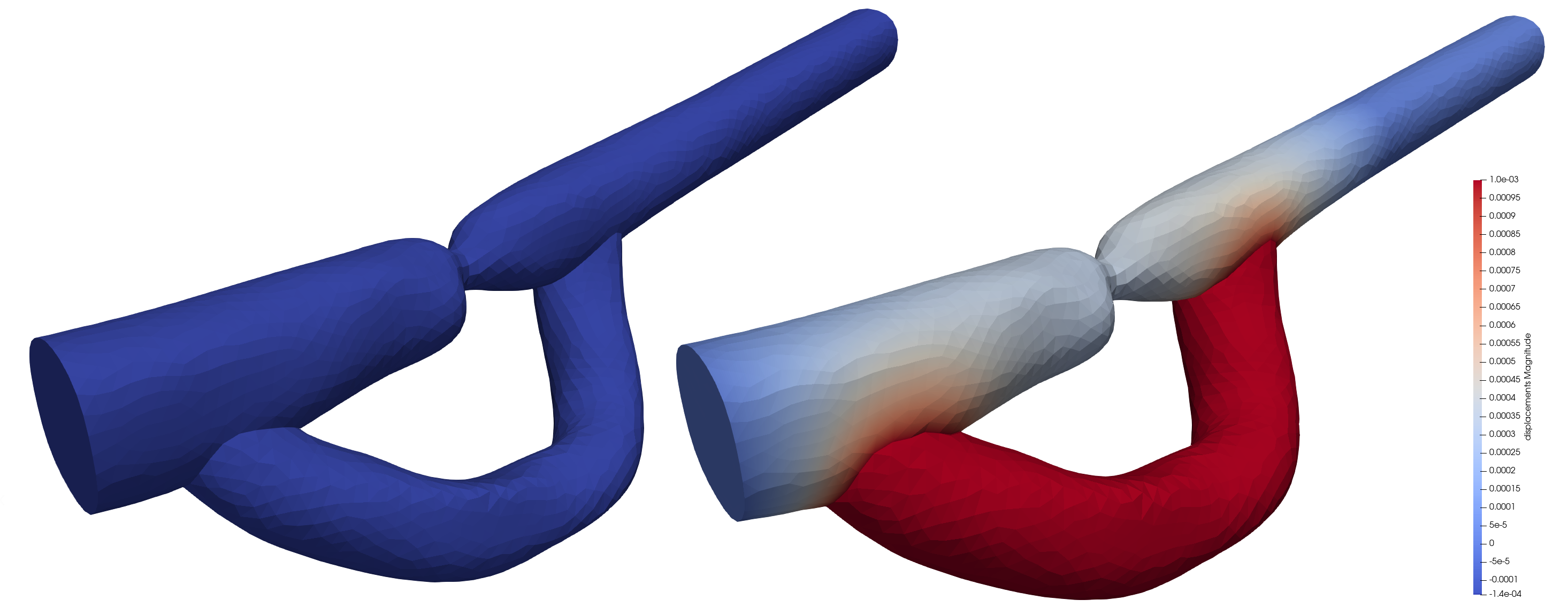}
  \caption{The bypass section of the geometry was scaled in the direction of the outward surface normal, thus increasing the diameter of the bypass by approximately 2 mm. Mesh 2 is depicted here before (left) and after (right) the modification was applied.  The meshes are colored by displacement magnitude.}
  \label{fig:meshScaled}
\end{figure}

\subsection{Experimental Setup}
In this test, we evaluate the simulation of flow through an arterial bypass. In this incompressible fluid dynamics simulation, the inlet velocity was set to a constant velocity in the direction of the inlet face normal vector into the domain of $0.15 m/s$, and the outlet was set to a zero gauge pressure condition along with a zero viscous traction condition. No-slip wall conditions were applied across the arterial wall and bypass wall. The kinematic viscosity of the fluid was set to $3.3e-6 m^2/s$. With the inlet diameter of approximately $0.014 m$, this leads to an inlet Reynolds number of approximately 636. The time step size for the simulation was set to $0.001 s$. A second order generalized-alpha time integration scheme was employed to progress the simulation \cite{jansen2000generalized}.

For each simulation, a single geometric modification was provided. Here, we apply a Scale by Normals filter to the bypass, widening the bypass diameter by approximately $2 mm$. Figure \ref{fig:meshScaled} depicts the displacement field before and after applying the modification to the surface mesh. This modification was applied with a deformation schedule of just one step. The Server was set to check for new deformations every 10 time steps of the simulation.

All simulations were run on the RMACC Summit Supercomputer at the University of Colorado Boulder. The general allocation was used, providing access to compute nodes equipped with 2 Intel Xeon E5-2680 v3 at 2.50GHz per node for a total of 24 cores per node \cite{anderson2017deploying}. The Client application and \texttt{ParaView}-based geometry manipulation were run on a 13-inch 2018 MacBook Pro equipped with a quad-core Intel Core i5 processor at 2.3GHz.

\begin{figure}[!t]
  \centering
  \includegraphics[width=\linewidth]{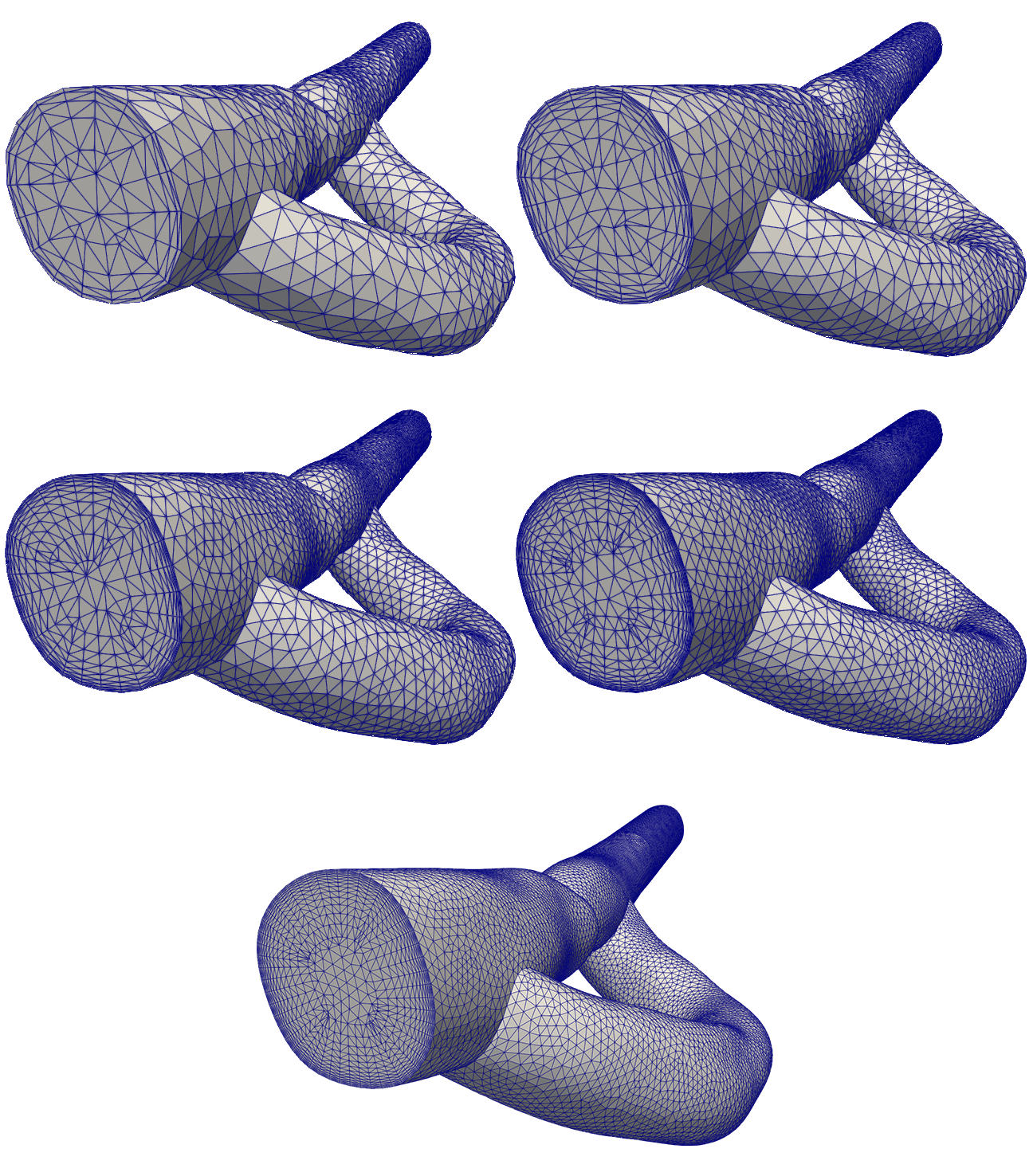}
  \caption{Representative images of the five meshes used in the timing study. Top-Left: Mesh 0, Top-Right: Mesh 1, Middle-Left: Mesh 2, Middle-Right: Mesh 3, Bottom, Mesh 4.}
  \label{fig:meshRefinement}
\end{figure}

Table \ref{table:meshes} details the mesh statistics of the particular meshes used in this study. We chose to produce meshes partitioned with approximately 5000 elements per process. With the exception of Mesh 4, we targeted job sizes with multiples of 12 MPI processes to line up with the node layout of the Summit Supercomputer. Figure \ref{fig:meshRefinement} depicts the five meshes used in this timing study with the boundary layer meshes visible at the inlet. The targeted number of elements for each mesh was achieved through a combination of boundary layer refinement along with isotropic $h$-refinement. All meshes were generated with the Simmetrix \texttt{Simmodeler} meshing application.

\begin{table*}[t!]
\centering
\caption{Arterial Bypass Mesh Resolutions Used in Interactivity Study. }
\begin{tabular}{||l || c | c | c | c | c ||}
 \hline
                  & Mesh 0& Mesh 1 & Mesh 2 & Mesh 3 & Mesh 4\\ [0.5ex]
 \hline\hline
 Volume Elements  & 59850 & 120147 & 238553 & 481545 & 1294658 \\
 \hline
 Volume Vertices  & 11241 & 21533  & 42600  & 84566  & 225361 \\
 \hline
 Surface Faces    & 3806  &  5028  & 9340   & 16008  & 32768 \\
 \hline
 Surface Vertices & 1903  & 2514   & 4670   & 8004   & 20165 \\
 \hline
 MPI Processes Utilized by Server  & 12    & 24     & 48     & 96     & 256 \\
 \hline
\end{tabular}
\label{table:meshes}
\end{table*}

In order to characterize the performance of our freeform computational steering system, we collected elapsed wall-time intervals for each major operation of the Client and the Server applications.

On the Client system, file I/O was employed to interface with \texttt{ParaView}. This approach was rather antithetical to the otherwise file I/O abjuring design used by the Server application. As such, we time each file I/O required by the Client application. We measured time to export a surface mesh to a \texttt{VTK} file which could be read by \texttt{ParaView}. We also measure the time to export a displacement field from \texttt{ParaView}. We then time the communication time to send that displacement field back to the  Server application. In between, we recorded the time interval of the displacement field computation performed in \texttt{ParaView}.

On the Server application, our key interest lies in how much computational overhead or \textit{tax} is incurred by our system on the otherwise unfettered operations of the simulation. To do this, we took wall-time intervals for each component of the Server application involved in mesh manipulation and communication of data. First, the one-time expense of extracting and consolidating the surface mesh was measured. Second, the time required to send the extracted surface mesh to the Client application was measured. Next, in order to assess the performance of the volume deformation system with some granularity, the runtimes of each component of the volume deformation routine were measured independently. These components include matrix allocation (including sparsity pattern computation), formation and assembly of the linear elasticity system, and solving the system for the volumetric displacements. Finally, as a basis of comparison, the time required to progress the simulation 10 time steps was computed by measuring the duration of three independent sets of 10 time steps and taking the average over those three sets. Three separate simulation runs were conducted over each of the five meshes for a total of 15 simulation runs.

\subsection{Results}
In this section, we report results of the timing study. In Table \ref{table:clientOps}, we report the average wall-time intervals for Client application operations, and in Table \ref{table:serverOps}, we report the average wall-time intervals for Server application operations.

As shown in Table \ref{table:clientOps}, performing the modification of the surface mesh scales like $\mathcal{O}(n^{2/3})$ where $n$ is the total number of degrees of freedom of the full volumetric simulation mesh. The wall-time interval for various File I/O operations on the Client device appears to grow faster than the $\mathcal{O}(n^{2/3})$ demonstrated for the surface mesh manipulation computation. This is evident especially with the average time to export the surface mesh of Mesh 3 and 4 to a \texttt{ParaView} file format for transmission to \texttt{ParaView}.

\begin{table*}[t!]
\centering
\caption{ Client Operation Timing Interval Averages. }
\begin{tabular}{||l || c | c | c | c | c ||}
 \hline
                  & Mesh 0& Mesh 1 & Mesh 2 & Mesh 3 & Mesh 4 \\ [0.5ex]
 \hline\hline
 Export to \texttt{ParaView} (ms)                      & 51   & 107   & 373  & 1243 & 5652\\
 \hline
 Perform Modification (ms)                    & 207  & 319   & 569  & 955 & 2175\\
 \hline
 Export Displacement Field from \texttt{ParaView} (ms) & 5    &  6    &  12  &  19 & 41 \\
 \hline
 Send Displacement Field to Server (ms)       & 29   &  28   &  29  & 27  & 28 \\
 \hline
\end{tabular}
\label{table:clientOps}
\end{table*}

\begin{table*}[t!]
\centering
\caption{ Server Operation Timing Interval Averages. }
\begin{tabular}{||l || c | c | c | c | c ||}
 \hline
                  & Mesh 0& Mesh 1 & Mesh 2 & Mesh 3 & Mesh 4\\ [0.5ex]
 \hline\hline
 Extract and Consolidate Surface Mesh (ms)   & 2   & 3    & 6      & 5    & 272\\
 \hline
 Send Surface Mesh to Client (ms)            & 483 & 386  & 377    & 607  & 545\\
 \hline
 Compute Volume Deformation  & & & & &\\ 
 \enspace Allocate matrix (ms)               &  40 & 71  & 154    & 102  & 579 \\
 \enspace Formation \& Assembly (ms)         & 441 & 573  & 943    & 1163 & 3925 \\
 \enspace Solve (ms)                         &  23 &  22  &  30    & 46   & 131\\
 \enspace GMRES Iterations                   & 34  &  35  &  46    & 78   & 173\\
 \hline
 Progress Flow Solve 10 steps (ms)  & 1230 & 1375 & 1201 & 1132  & 1993 \\
 \hline
\end{tabular}
\label{table:serverOps}
\end{table*}

\begin{table*}[t!]
\centering
\caption{ Volume Deformation Overhead Measured in Added Simulation Time Steps. }
\begin{tabular}{||l || c | c | c ||}
 \hline
                & Best Case Overhead & Average Overhead & Worst Case Overhead \\ [0.5ex]
 \hline\hline
 Mesh 0            & 4.0 & 4.1  & 4.2  \\
 \hline
 Mesh 1            & 4.1 & 4.8  & 6.1  \\
 \hline
 Mesh 2            & 2.9 & 9.4  & 21.8 \\
 \hline
 Mesh 3            & 4.4 & 11.6 & 23.0 \\
 \hline
 Mesh 4            & 9.6 & 23.2 & 38.3 \\
 \hline
\end{tabular}
\label{table:overheadInSteps}
\end{table*}

In Table \ref{table:serverOps}, we see a weak dependence on problem size in timing the various operations on the Server application.  There is a jump in wall-time average for some operations from the Mesh 1 case to the Mesh 2 case.  This is to be expected, however, as the Mesh 0 and Mesh 1 cases involve just one node on Summit while the Mesh 2, Mesh 3, and Mesh 4 cases involve multiple nodes and thus are subject to interconnect bandwidth and latency.  Also as expected, PHASTA demonstrated competent weak scaling performance, with an average wall-time interval for 10 time steps across the first four mesh cases of 1234 ms, with a standard deviation of 102.3 ms.

The Mesh 4 case exhibited reduced weak scaling performance across all parallel operations on the Server application. This includes an anomalously slow surface mesh extraction and consolidation operation recorded in Run 1. Even \texttt{PHASTA}, which for the first four mesh cases demonstrated consistent weak scaling, exhibited deteriorated performance, averaging 62\% slower runtime over 10 time steps compared to the average performance of the first four mesh cases.  Recall however that the Mesh 4 case does not line up with the node layout of Summit, and this could be responsible for loss of performance.

It is difficult to directly compare the overhead our framework imposes on a typical simulation. Performing a modification of the mesh provides imposes some overhead on the simulation, however, it provides insights akin to running two simulations. That being said, if the overhead of our freeform computational steering framework was so great as to impose an entire simulation's worth of overhead, this framework would be intractable. Fortunately, in Table \ref{table:overheadInSteps}, we record the best, average and worst case overhead as measured in added simulation time steps. We see that in the best case scenario our system imposes the equivalent of just 2.9 simulation time steps and in the worst case scenario our system imposes just 38.3 time steps worth of overhead. 

To get a sense of how our framework's imposed overhead relates to physically simulated time, we can compute an intrinsic bulk time scale for this particular simulation. We can define the \textit{bypass time scale} $t_L$ as a physical time scale related to the time required for a disturbance to propagate through the entire bypass. We define it as the approximate distance traveled by a fluid particle through the bypass $L$, divided by the inlet velocity $U$,
\begin{equation}
  t_L = L / U.
\end{equation}
For the simulation studied in this section, we set $L = 0.1198 m$ as the approximate distance from the inlet, through the curved bypass, and then through the outlet, and we set $U = 0.15 m/s$ as the prescribed velocity inlet boundary condition. The bypass time scale was then computed to be approximately 0.799 seconds. Our simulation time steps were statically set to 0.001 seconds. Thus, in order to simulate one bypass time scale, we require running 799 time steps. If we consider this physical time scale as the minimum required length of a simulation in order to get some insight into the simulated dynamics, then imposing a mesh modification using our framework amounts to overhead in the best case of just 0.36\%. In the worst case scenario, this overhead inflates to just 4.7\%.

One further observation should be made about the conducting of this timing evaluation. Several of the simulation runs conducted in this test sat dormant in the Summit queue waiting to run for over twelve hours. That wait time was orders of magnitude longer than the runtime of every one of the simulations conducted in this study combined. That human time cost should not go unmentioned. Minimizing the total number of jobs required to explore a design is a great side effect of our freeform computational steering system. Even exploring a single variation in a design during an \textit{in situ} design space exploration session saves the user potentially hours of downtime waiting to gather results which will provide more insight into their design challenge.

\section{Conclusions} \label{sec:conclusions}
In this work, we designed a new computational steering workflow for leveraging high performance simulations to explore how geometry may affect the dynamics of a simulated system. A client-server architecture was employed to provide accessible user-interaction on a user's workstation while also leveraging high performance computing resources in order to enable exploration of detailed, highly resolved finite element simulations. A communication layer was developed to isolate the user-interaction loop occurring on a user's workstation from the simulation loop occurring on a high performance compute resource. Custom interactivity plugins were written using \texttt{ParaView}'s plugin system to transform the visualization software into a computational geometry steering interface, tethering visualization and user interaction to a single platform. We implemented all of this capability in the open source software suite \texttt{Shoreline} \cite{wetterernelson2021}.

The high performance CFD application \texttt{PHASTA} was utilized as the demonstration finite element solver in our workflow. \texttt{PHASTA} was chosen for its well-established track record as a scalable tool targeting large computing resources, coupled with its history of \textit{in situ} visualization and computational steering. We demonstrated our workflow with \texttt{PHASTA} with a walled channel which the user can pinch into a converging-diverging nozzle. The deformation of the domain was transferred into the \texttt{PHASTA} runtime which thus transitioned the flow from attached channel flow to detached separated flow. 

Based on the constricting nozzle demonstration, a rapid domain modification schedule is appropriate in some contexts. This result is encouraging in the context of computational steering, where rapid response to user input is a desirable feature. However, for more complex flow scenarios, a deformation schedule which eases the deformation over many simulation time steps could be beneficial to numerical stability.

We performed a full freeform design exploration, studying the behavior of an arterial bypass implanted at two different angles. This particular design problem demonstrated the utility of our freeform computational steering system for problems with organic geometry. In such contexts, geometric parameters suitable for parametric design optimization or ensemble simulations are nonexistent or at least exceedingly difficult to define. However, with our tools, a practitioner can probe the design space of these organic systems directly with the simulation domain. The interactive curve skeleton was intuitive to use in controlling the shape of the bypass.

We investigated how freeform manipulation of the surface of a geometry may affect the quality of the enclosed volumetric mesh. We found that under modest deformation, mesh quality can remain well within acceptable levels, and only under extreme deformation will individual elements begin to invert leading to a totally invalid mesh. As noted before, each simulation solver will have its own tolerance for what constitutes poor mesh quality. As such, any individual user will need to have an understanding of how far they may be able to push their prescribed deformations given their simulation solver.

In order to characterize solution convergence, we took a well studied incompressible flow solution and modified its geometry to see how it responds. By instantaneously increasing the volume of the pipe, the pressure field instantaneously responded with a massive pressure spike through the upstream region extending to the inlet, leading momentarily to an overall drastically reduced pressure drop across the pipe. The pressure drop drops precipitously after the applied geometric deformation. This response is expected, as the pre-deformation  velocity field is interpolated onto the deformed geometry and the pressure field in turn must slow the flow down to maintain constant mass flow-rate through the entire pipe. The pressure field rapidly recovers however, and the velocity field behaves physically after a short transient is blown through the pipe. Further, we see that with mesh refinement, we are able to better resolve that \textit{water hammer} effect caused by the instantaneous volume change, as shown by a progressively more severe pressure change at the geometry modification event. 

Finally, we profiled the freeform computational steering system's capacity to scale with problem size. The experimental setup mimicked that of a typical weak scaling analysis where the problem is partitioned such that each process maintains approximately the same computational load, and as problem size is increased, the number of processor cores is increased in tandem. This was done in order to isolate the various computational taxes our system imposes on a typical simulation. We found that as expected, \texttt{PHASTA} is quite capable at weak scaling, as the average time for the simulation to step in time was quite consistent for each simulation size. For each mesh size, a set of 10 time steps averaged approximately 1320 ms. We found that the volume deformation components of the Server application ran on the order of 40-116\% the wall-time of a set of 10 simulation time steps. Since the cadence of volume deformation and simulation time steps is completely user-specified, these results are not completely generalizable, and performance will vary from user to user. However, the magnitude of these timings in this collection of simulations is certainly marginal, meaning the computations did not impede interactivity. Further, we were conservative in measuring overhead against just 10 simulation time steps. In expected use cases, it would be unlikely to perform a geometric modification every 10 time steps, so that 40-116\% tax would never be applied to the whole simulation run. In the worst case found in this study, performing a geometric manipulation amounted to adding approximately 38 simulation time steps to the total simulation run. In the best case found in this study, a geometric modification amounted to around four added time-steps to the total simulation runtime.

\section{Future Work} \label{sec:futureWork}
Moving forward, several paths exist for extending this work. First, projecting simulation solutions onto the updated mesh is currently done using a naive interpolation approach. Though our deformation scheduling results presented in Section \ref{sec:nozzle} and convergence results from Section \ref{sec:solCvg} were quite encouraging, our method does not guarantee numerical stability of any given flow when met with arbitrary geometric manipulation. As such, drawing from fluid structure interaction, an Arbitrary Lagrangian Eulerian (ALE) formulation of the fluid-geometry interaction ought be investigated in the context of freeform computational steering \cite{donea1982arbitrary, bazilevs2008isogeometric, nobile1999stability}. Our current system does not respect mass conservation or momentum conservation of the fluid flow between deformation steps. In compressible flow especially, this can lead to spurious oscillations which can be potentially unstable \cite{lesoinne1996geometric}. As such, incorporating techniques such as ALE would improve numerical stability and widen the space of problems which our freeform computational steering system could interface with.

We hope to see this work applied to novel systems and enable new procedures for probing geometric modification response in fluid dynamical systems across the incompressible and compressible regimes, laminar and turbulent regimes, single phase and multiphase flows, etc. However, this system is not uniquely designed for studying fluid flow systems, and as such, future applications should extend this workflow to problems in solid mechanics, wave propagation or plasma physics. The benefit of analyzing dynamic changes to a system as the geometric design is modified is one of the key strengths of our framework, and as such, studying systems with transient physical processes have the most to gain from our framework.

The clearest path to dramatically improve robustness under most freeform manipulation would be to include adaptive mesh refinement in the volume deformation algorithm. By adaptively re-meshing regions of reduced quality, a target mesh quality could be maintained across the geometry. However, most adaptive mesh refinement algorithms rely on a mesh association with a static CAD model \cite{shephard2007parallel, smith2018memory}. By performing geometric manipulation directly on the finite element mesh, we explicitly disassociate the mesh from its progenitor CAD model. This leads to the question of what geometry ought the mesh be adapted to? The user-modified discrete surface mesh is the obvious candidate, but it was already just a discrete approximation of the initial CAD model before deformation, and geometric error may rapidly accumulate if multiple deformation and refinement steps were performed. 

\section{Acknowledgements}
The authors would like to thank Cameron Smith for his continued assistance with SCOREC tools. This material is based upon work supported by the National Science Foundation under Grant Number 1740330.
\vspace{12pt}

\bibliography{mybibfile}

\end{document}

%% file: background.tex
\section{Background} \label{sec:background}
Computational steering takes many forms and varies significantly in levels of interactivity across a wide spectrum of applications. The domain spans applications from monitoring tools with basic means of changing simulation parameters to fully immersive virtual reality environments with complex controls for real-time interaction with the simulated domain. The proceeding sections provide some context for the work presented in this article.

\subsection{Notification and Monitoring Systems} \label{sec:notAndMonitor}
One of the most common design goals of computational steering systems is to be as minimally invasive to the simulation as possible. In an effort to minimally probe typical \enquote*{black box} simulations, efforts have been made to employ monitoring and notification systems which query data generated by simulation systems during runtime \cite{santos2009enabling, tchoua2010collaborative, pintas2013}. These monitoring systems can provide basic readouts and inform a user of undesirable simulation behavior such as solver divergence, but they do not enable live configuration of the simulation as it is on-going. These systems wrap around existing simulation systems, and provide some basic information that a simulation may not output by default. Though information such as solver diagnostics is useful for simulation practitioners concerned with the health of their simulation, they provide no interactivity.

\subsection{\textit{In Situ} Visualization}
In the 2009 Sandia National Laboratory report by Thompson et al. \cite{thompson2009design}, a clear and targeted case was made for systems which mitigate or eliminate some or all analysis data which would be written to disk for long term storage and future analysis. Even then, difficulties were rapidly arising with simulations capacity for data output completely eclipsing existing storage capacity. The strategy they proposed was to process the simulation data in between simulation time steps and present data insight to the user, without user intervention. Since then, many solutions to this problem have been proposed and successfully implemented.

It is becoming a common practice to render video of system solutions on the compute resource or a separate visualization resource. This system is exemplified by \texttt{ParaView Catalyst} \cite{ayachit2015}. This software developed by Kitware reflects the state of the art in \textit{in situ} visualization frameworks. \texttt{Catalyst} exposes a lightweight application programming interface (API) which can be added to existing simulation software, allowing the user to run \texttt{ParaView} visualization pipelines within their simulation, and send that result to an attached \texttt{ParaView} server which can transmit visual data to a user operating a \texttt{ParaView} client application. Utilizing the full extent of the \texttt{ParaView} visualization capabilities allows for complex visualization and analysis to be performed during a simulation's runtime. Further, these pipelines are implemented as Python scripts which can be reconfigured during runtime as well, enabling flexible and dynamic insight gathering. \texttt{Catalyst} provides a profound capability for \textit{in situ} visualization, though on its own is not capable of manipulating the simulation. \texttt{Catalyst} does provide a pause button which temporarily halts a simulation's progression, allowing a user to manipulate their visualization script before proceeding with their simulation.

Related systems have enabled \textit{in situ} visualization such as \cite{kuhlen2011parallel} which employed \texttt{Lib-Sim} in the visualization software \texttt{VisIt} and \cite{lofstead2008flexible} which employed \texttt{ADIOS} also in \texttt{VisIt}. Custom solutions also exist, such as for combustion dynamics \cite{yu2010situ}. These various solutions all address the rising tide of data proliferation caused by simulation data output capabilities far outpacing our capacity to store data for future analysis.

\subsection{Steering with Parameters in \texttt{PHASTA}}
Throughout this article, we demonstrate our freeform computational steering system with the high performance CFD software \texttt{PHASTA}. \texttt{PHASTA} is a scalable, unstructured grid finite element solver for compressible and incompressible CFD with a suite of turbulence modeling capabilities. As we make significant use of this solver, it is appropriate to discuss some of the computational steering work which has been explored using \texttt{PHASTA}. In \cite{Yi2014}, the goal was to modify simulation parameters such as solver coefficients and time step size while a simulation was running, and to do so informed by \textit{in situ} visualization provided by \texttt{ParaView}. The ability to adjust simulation parameters on the fly allowed a user to start up a large simulation without needing to know all the parameters up front, and then avoid restarting the simulation over and over (avoiding long queue times on high performance compute resources) by simply modifying parameters which lead to better time step convergence. This capability is convenient for very large simulations requiring tens or hundreds of thousands of computer cores. However, the goals of this previous work targeted a single flow configuration, and it did not allow for significant changes to the problem definition as a truly interactive experience was not their intention. Further, this system lacks any capacity for modifying the geometric domain of the simulation.

Recently, Newberry et al. demonstrated utilizing the high performance software adaptor \texttt{SENSEI} to study the effect of modifying physical and geometric parameters of an on-going simulation \textit{in situ} within \texttt{PHASTA} \cite{newberry2021software}. However, the approach to geometry modification was fully parametric, that is, an underlying geometry controller was employed that was custom tailored to the specific problem. That is, little flexibility was available for manipulating the geometry. 

\texttt{PHASTA} was also used as one of the first demonstrators of the \texttt{ParaView} co-processing library for \textit{in situ} visualization of parallel CFD simulations \cite{fabian2011paraview}. This history of pioneering new simulation workflows is one of the many reasons \texttt{PHASTA} was selected as the key simulation technology on which to demonstrate our freeform computational steering framework presented in this work.

\subsection{Interactive Simulation Geometry} \label{sec:interactivity}
Interacting with the simulation domain is a key component of the work presented in this document. Domain modification takes two main forms in the literature: rigid body motion of objects and deformation of the computational mesh. In this section, we will discuss various efforts to enable geometric domain interactivity of simulation domains. First, we will discuss a branch of the field concerned with real-time fluid dynamics simulation and user interactivity with those simulations. Then, we will discuss new work in simulated surgery, where a surgeon can perform virtual surgery in preparation or practice for real surgery.

\subsubsection{Real-Time Interactive Fluid Dynamics}
Recently, multiple papers have been published on interactive fluid dynamics simulation which employ a compute resource to handle the fluid simulation, while a user operates interactively from their workstation  \cite{linxweiler2010, harwood2018, harwood2019, gobbert2018, wang2019}. Most of the interactivity was attained through custom solutions. One group went so far as to use the video game engine \texttt{Unity} as their interactivity driver \cite{wang2019}. The choice of a video game engine gave them significantly more flexibility to develop custom rendering techniques. They mentioned \texttt{ParaView} as their foil where custom rendering techniques are more difficult to implement. In \cite{wang2019}, a communication layer was in place and a server ran a 3D Lattice-Boltzmann simulation, and a client pinged the server for a volume fraction which represented their solution variable for rendering in \texttt{Unity}. Then, they were able to \enquote*{manipulate} their simulation domain by employing \texttt{Unity} features for moving geometry around. The only significant example provided was moving a simple dam wall (represented by a rectangle) up and down. These interactive simulation environments are limited to low order Lattice-Boltzmann solvers because of their speed and efficient implementation on GPGPU systems. The approaches taken in the works cited in this section target real-time simulation which necessitates low-fidelity simulations, fast solvers, and low order methods, leaving much to be desired in physical accuracy of the fluid state. In contrast, our interactive system targets large, high-fidelity simulations running on high performance computing clusters where scientific analysis is the goal.

Interactive simulations targeting real-time performance have in recent years employed simple controls for moving around objects within a domain. These systems rely on particle methods where colliders can be moved without mapping or modifying a solution field defined over the domain \cite{linxweiler2010, harwood2018, harwood2019, gobbert2018, wang2019}. This methodology has the advantage of being computationally inexpensive, and when coupled with inexpensive particle based fluid simulations, this can lead to intuitive and interesting user interaction with the simulated fluid domain.

Harwood et al in 2018 devised a virtual reality environment using the video game engine \texttt{Unreal Engine 4} to allow a user to walk around a virtual wind tunnel. The fluid simulation here was a low-order 3D Lattice-Boltzmann simulation targeting real-time performance. Their system allowed for arbitrary objects represented by triangulated surfaces to be placed into a wind tunnel, and it allowed a user to walk around the virtual space, observing the simulated incompressible fluid flow. In this work, they note the difficulty in achieving real-time simulation, recognizing the significant reduction in accuracy, stability and fluid domain size required to meet targeted real-time performance \cite{harwood2018}. Unfortunately, results from these admittedly low-accuracy simulations is not usable for detailed scientific discovery or precise design exploration. They do however provide a stepping stone toward truly immersive simulation experiences, virtually placing the user into the simulated domain.

It should be noted in this work we target \textit{useful-time simulation} rather than real-time simulation.  We say a simulation is run in useful-time when the data-output time-scale is synchronized appropriately with the human interactivity time-scale. When presented with data too quickly or too slowly, a human may have difficulties interpreting how the simulation is evolving in time.  Fortunately, while highly-accurate real-time simulation is currently out of reach for many scientific and engineering systems of interest, physically realistic useful-time simulation is possible with today’s existing high performance computing systems in many simulation contexts.

\begin{figure*}[t!]
  \centering
  \includegraphics[width=0.60\linewidth]{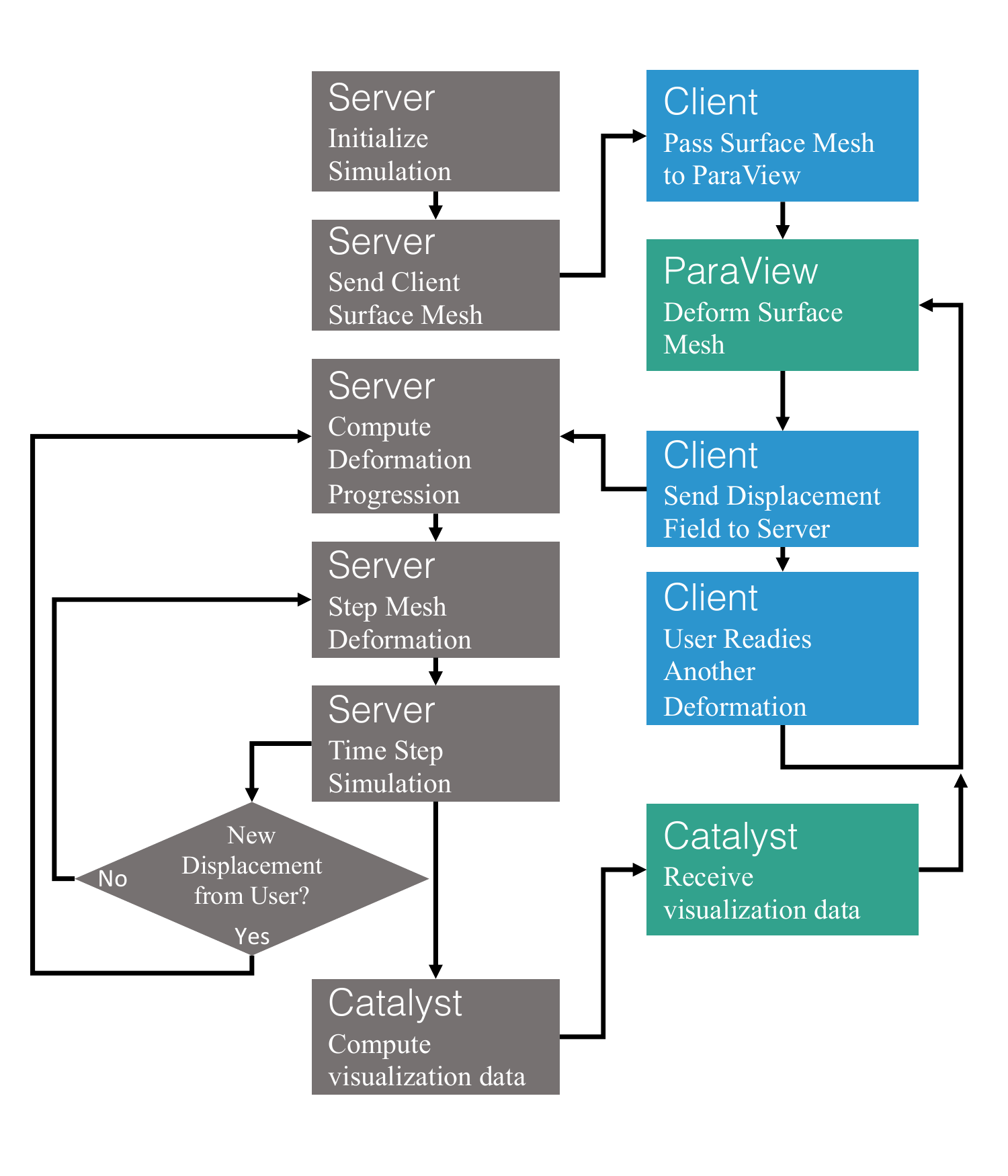}
  \caption{The mesh deformation steering system operates in a server-client configuration where a user interacts directly with a client application, and the server system responds to the information received from the client.}
  \label{fig:Flowchart}
\end{figure*}

\subsubsection{Virtual Surgery}
Currently, significant research efforts have been made to create immersive interactive biomedical simulations for use as teaching tools and virtual surgical planning and training systems. These systems employ finite element soft-body simulations of biological tissues to model the tissue's response to user interaction. These interactions range from dynamically modifying state variables and simulation parameters \cite{lloyd2012artisynth}, to complex interactive systems employing virtual reality headsets and interactive controllers that approximate the user's touch \cite{martins2019interactive, jayasudha2020soft}. In the context of surgical planning and training, these systems typically employ linear elasticity models for soft tissue, trading physical accuracy for computational performance.

These virtual surgery systems provide an incredible degree of interactivity with a simulation. We draw some broad inspiration from these for our work presented here such as from their approach to human-computer interaction and how these systems enable physical interaction with virtual domains. However, these systems are by no means design tools. They are intended to train users to understand how physiological processes respond to medical intervention, and are as such completely focused on modeling physiological systems. In this context, the simulated system undergoes physically realizable responses to interactively applied external stimuli, where in the software framework described in this article, the simulated system undergoes responses to a change in geometric design.